\documentclass[namedreferences]{solarphysics}
\usepackage[optionalrh]{spr-sola-addons} 
\usepackage{graphicx}        
\usepackage{amssymb}        
\usepackage{color}           
\usepackage{url}             

\begin{document}

\begin{article}

\begin{opening}

\title{Solar Orbiter\\ {\it Exploring the Sun--heliosphere connection}}

\author{D.~\surname{M\"uller}$^{1}$\sep
        R.G.~\surname{Marsden}$^{1}$\sep
        O.C.~\surname{StCyr}$^{2}$\sep
        H.R.~\surname{Gilbert}$^{2}$
        for the Solar Orbiter Team
       }
\runningauthor{M\"uller et al.}
\runningtitle{Solar Orbiter -- Exploring the Sun--heliosphere connection}

   \institute{$^{1}$ European Space Agency, ESTEC, Noordwijk,\\ The Netherlands\\
                     email: \url{Daniel.Mueller@esa.int} email: \url{Richard.Marsden@esa.int}\\ 
              $^{2}$ NASA Goddard Space Flight Center, Greenbelt, MD, USA\\
                     email: \url{Chris.StCyr@nasa.gov} email: \url{Holly.R.Gilbert@nasa.gov}\\
             }

\begin{abstract}
The heliosphere represents a uniquely accessible
domain of space, where fundamental physical processes common to solar,
astrophysical and laboratory plasmas can be studied under conditions
impossible to reproduce on Earth and unfeasible to observe from astronomical distances.
Solar Orbiter, the first mission of ESA's Cosmic Vision 2015--2025 programme, will address the central question of heliophysics: How does the
Sun create and control the heliosphere? In this paper, we present the scientific goals of the mission and provide an overview of the mission
implementation.
\end{abstract}

\end{opening}


\section{Introduction}
     \label{S-Introduction} 

We live in the extended atmosphere of the Sun, a region of space known as the heliosphere. Understanding the connections and the coupling between the Sun and the heliosphere is of fundamental importance to understanding how our solar system works. The results from current and past solar and heliospheric missions such as Helios \cite{Porsche:1977aa,Schwenn:1990aa,Schwenn:1991ab}, Voyager \cite{Stone:1977aa}, Ulysses \cite{Wenzel:1992aa}, Yohkoh \cite{Acton:1992aa}, SOHO \cite{Domingo:1995aa}, TRACE \cite{Handy:1999aa}, RHESSI \cite{Lin:2002aa}, Hinode \cite{Kosugi:2007aa}, STEREO \cite{Kaiser:2008aa} and SDO \cite{Pesnell:2012aa} have formed
the foundation of our understanding of the solar corona, the solar wind, and the three-dimensional heliosphere. Each of these missions had a specific focus, being part of an overall strategy of coordinated solar and heliospheric research. However, none of these missions have been able to fully explore the interface region where the solar wind is born and heliospheric structures are formed with sufficient instrumentation to link solar wind structures back to their source regions at the Sun (Helios 1 and 2, e.g., carried no imaging instruments). This is the goal of Solar Orbiter, a mission of collaboration between ESA and NASA that was recently selected as the first medium (M)-class mission of ESA's Cosmic Vision 2015--2025 programme.

With a combination of in-situ and remote-sensing instruments and its inner-heliospheric mission design, Solar Orbiter will address the central question of heliophysics: How does the Sun create and control the heliosphere? This primary, overarching scientific objective can be expanded into four interrelated top-level scientific questions that will be addressed by Solar Orbiter:

\begin{itemize}
\item What drives the solar wind and where does the coronal magnetic field originate from?
\item How do solar transients drive heliospheric variability?
\item How do solar eruptions produce energetic particle radiation that fills the heliosphere?
\item How does the solar dynamo work and drive connections between the Sun and the heliosphere?
\end{itemize}

These questions represent fundamental challenges in solar and heliospheric physics today. By addressing them, we expect to make major breakthroughs in our understanding of how the inner solar system works and is driven by solar activity. 
To answer these questions, it is essential to make in-situ measurements of the solar wind plasma, fields, waves, and energetic particles close enough to the Sun that they are still relatively pristine and have not had their properties modified by subsequent transport and propagation processes. This is one of the fundamental drivers for the Solar Orbiter mission, which will approach the Sun to as close as 0.28\,AU. 

Relating these in-situ measurements back to their source regions and structures on the Sun requires simultaneous, high-resolution imaging and spectroscopic observations of the Sun in and out of the ecliptic plane. The resulting combination of in-situ and remote-sensing instruments on the same spacecraft, together with the new, inner-heliospheric perspective, distinguishes Solar Orbiter from all previous and current missions, enabling science which can be achieved in no other way. 

The following section introduces the science payload and mission design. Section~\ref{S-Science} describes the mission's science objectives in detail: The present state of knowledge is presented for all major science questions of the mission, followed by descriptions of how Solar Orbiter will advance our understanding. Section~\ref{S-Spacecraft} introduces the Solar Orbiter spacecraft, followed by an overview of the science operations in  Section~\ref{S-Operations}. Table \ref{T-MissionSummary} gives a one-page mission summary.


\begin{table}
\caption{Solar Orbiter Mission Summary.}

\label{T-MissionSummary}
\begin{tabular}{lp{7cm}} 
\hline              
{\bf Top-level Science Questions} & 
\begin{itemize}
\item What drives the solar wind and where does the coronal magnetic field originate from?
\item How do solar transients drive heliospheric variability?
\item How do solar eruptions produce energetic particle radiation that fills the heliosphere?
\item How does the solar dynamo work and drive connections between the Sun and the heliosphere?
\end{itemize}\vspace{-6mm}\\
\hline
{\bf Science Payload} &
{\bf In-Situ Instruments:} 
\begin{itemize}
\item Energetic Particle Detector (EPD) 
\item Magnetometer (MAG)
\item Radio and Plasma Wave analyser (RPW)
\item Solar Wind Analyser (SWA)
\end{itemize}
{\bf Remote-Sensing Instruments:}
\begin{itemize}
\item EUV full-Sun and high-resolution Imager (EUI)
\item Coronagraph (METIS)
\item Polarimetric and Helioseismic Imager (PHI)
\item Heliospheric Imager (SoloHI)
\item EUV spectral Imager (SPICE)
\item X-ray spectrometer/telescope (STIX)
\end{itemize}\vspace{-6mm}\\
 \hline
{\bf Mission Profile} & 
\vspace{-6mm}
\begin{itemize}
\item Launch on NASA-provided Evolved Expendable Launch Vehicle (Ariane 5 as back-up)
\item Interplanetary cruise with chemical propulsion and gravity assists at Earth and Venus
\item Venus resonance orbits with multiple gravity assists to increase inclination
\end{itemize}
\vspace{-6mm}\\
\hline
{\bf Closest Perihelion} & 0.28\,AU\\
\hline
{\bf Max. Heliolatitude} & 25$^\circ$ (nominal mission) / 34$^\circ$--36$^\circ$ (extended mission)\\
 \hline
{\bf Spacecraft} &
3-axis stabilized platform, heat shield, two adjustable, single-sided solar arrays, dimensions: 2.5 $\times$ 3.0 $\times$ 2.5 ${\rm m}^3$ (launch configuration)
\\
 \hline
{\bf Telemetry Band} & Dual X-band\\
 \hline
{\bf Data Downlink} & 150 kbit/s at 1\,AU spacecraft--Earth distance\\
 \hline
{\bf Launch Date} & Jan-2017 (Mar-2017 and Sep-2018 back-ups)\\
 \hline
{\bf Nominal Mission Duration} & 7 years (incl.\ cruise phase) \\
 \hline
 {\bf Extended Mission Duration} & 3 years\\
 \hline
\end{tabular}
\end{table}


\newpage

\section{Instruments and Mission Design}

\subsection{Scientific Payload}
\label{S-Payload}
The scientific payload elements of Solar Orbiter will be provided by ESA member states, NASA and ESA and have been selected and funded through a competitive selection process. These are:
 
\subsubsection*{The in-situ instruments:} 
\begin{itemize}
\item The Energetic Particle Detector (EPD) experiment (J.\ Rodriguez-Pacheco, PI, Spain) will measure the properties of suprathermal ions and energetic particles in the energy range of a few keV/n to relativistic electrons and high-energy ions (100\,MeV/n protons, 200\,MeV/n heavy ions). 
\item The Magnetometer (MAG) experiment (T.S.\ Horbury, PI, UK) will provide detailed in-situ measurements of the heliospheric magnetic field. 
\item The Radio and Plasma Waves (RPW) experiment (M.\ Maksimovic, PI, France) will measure magnetic and electric fields at high time resolution and determine the characteristics of electromagnetic and electrostatic waves in the solar wind from almost DC to 20\,MHz.
\item The Solar Wind Analyser (SWA) instrument suite (C.J.\ Owen, PI, UK) will fully characterize the major constituents of the solar wind plasma (protons, alpha particles, electrons, heavy ions) between 0.28 and 1.2\,AU.
\end{itemize}

\subsubsection*{The remote-sensing instruments:}
\begin{itemize}
\item The Extreme Ultraviolet Imager (EUI, P.\ Rochus, PI, Belgium) will provide image sequences of the solar atmospheric layers from the photosphere into the corona. 
\item The Multi Element Telescope for Imaging and Spectroscopy (METIS) Coronagraph (E.\ Antonucci, PI, Italy) will perform broad-band and polarized imaging of the visible K-corona, narrow-band imaging of the UV and EUV corona and spectroscopy of the most intense lines of the outer corona. 
\item The Polarimetric and Helioseismic Imager (PHI, S.K.\ Solanki, PI, Germany) will provide high-resolution and full-disk measurements of the photospheric vector magnetic field and line-of-sight velocity as well as the continuum intensity in the visible wavelength range. 
\item The Solar Orbiter Heliospheric Imager (SoloHI, R.A.\ Howard, PI, USA) will image both the quasi-steady flow and transient disturbances in the solar wind over a wide field of view by observing visible sunlight scattered by solar wind electrons. 
\item A European-lead extreme ultraviolet imaging spectrograph (SPICE) with contributions from ESA member states and ESA. This instrument will remotely characterize plasma properties of regions at and near the Sun.
\item The Spectrometer/Telescope for Imaging X-rays (STIX) (S.\ Krucker, PI, Switzerland) provides imaging spectroscopy of solar thermal and non-\\thermal X-ray emission from $\sim4-150$\,keV. 
\end{itemize}

\begin{figure}   
   \centerline{\includegraphics[width=\textwidth,clip=true]{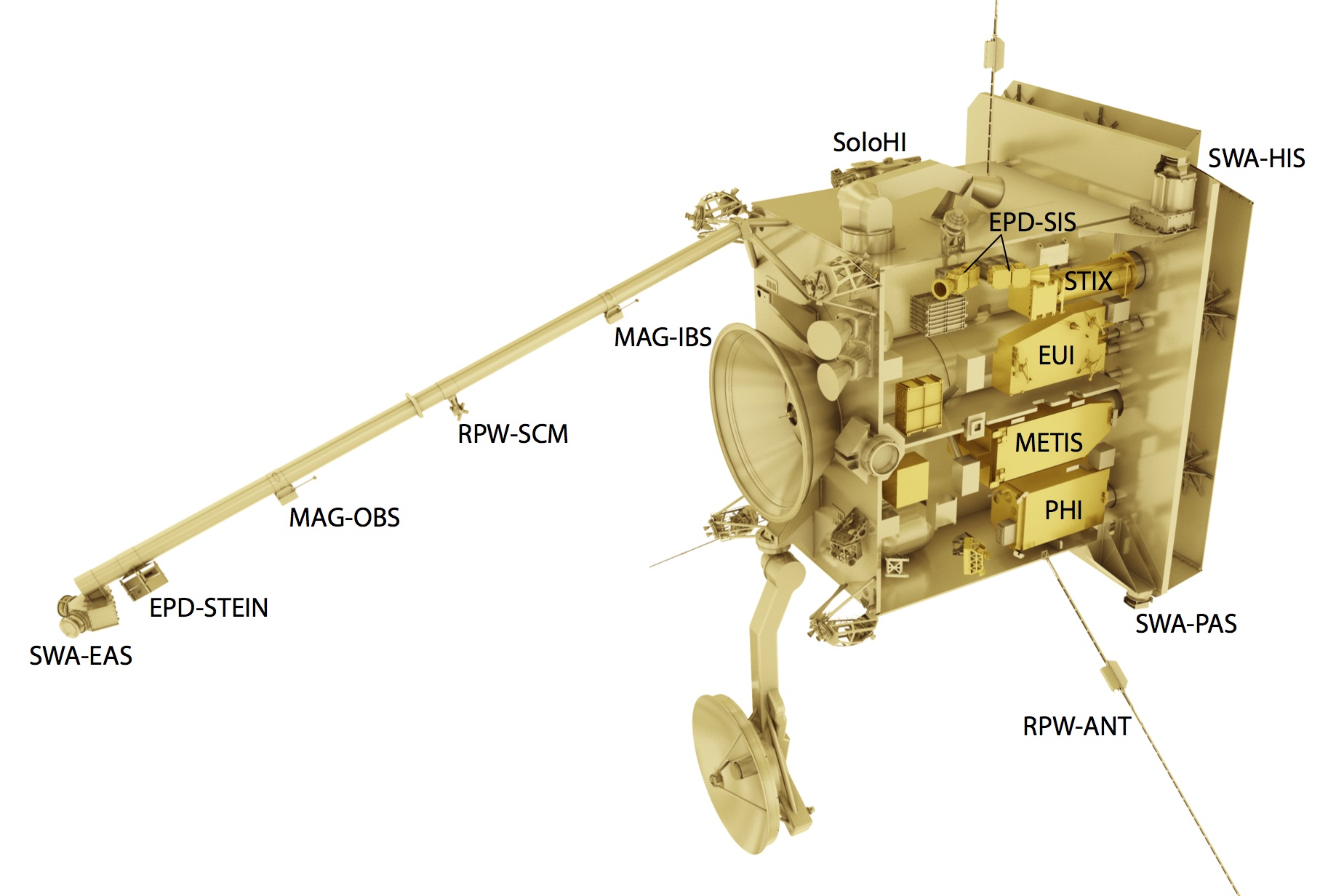}}
              \caption{Payload accommodation onboard Solar Orbiter. In this rendering, one side wall has been removed to expose the remote-sensing instruments mounted on the payload panel. The SPICE instrument (not visible) is mounted to the top panel from below. See Section~\ref{S-Payload} for a payload description and acronyms.}
   \label{F-SC_payload}
   \end{figure}

\noindent
The accommodation of the science payload onboard the spacecraft is illustrated in Figure~\ref{F-SC_payload}. A detailed description of the payload elements, as well as traceability matrices of the science goals are given in
\inlinecite{Marsden:2011aa}.

\subsection{Mission Design}
\label{S-MissionDesign}
\begin{figure}   
   \centerline{\includegraphics[width=\textwidth,clip=true]{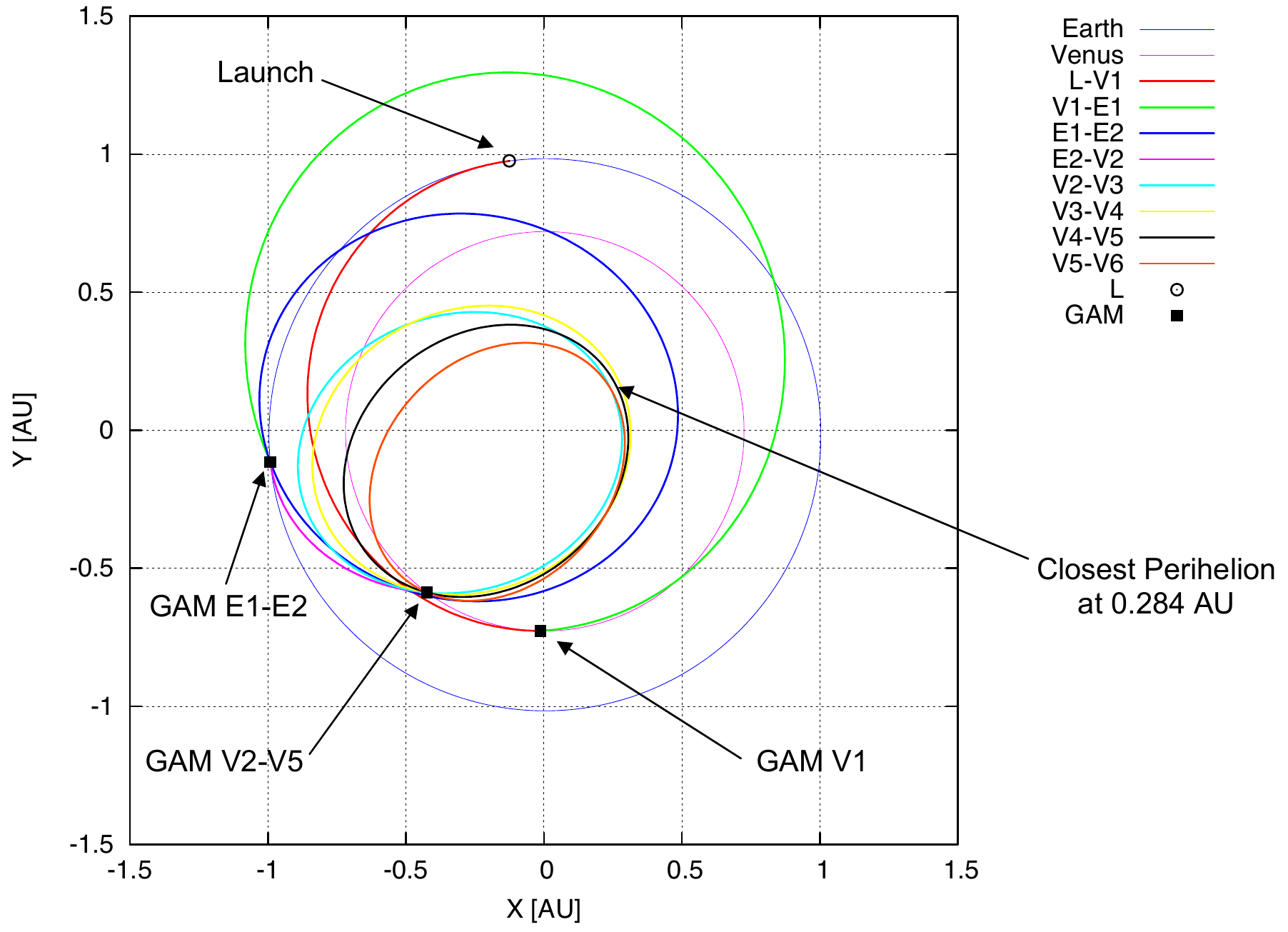}}
              \caption{Solar Orbiter's trajectory viewed from above the ecliptic (January 2017 launch). The gravity assist maneuvers (GAM) at Earth (E) and Venus (V) are indicated, along with the orbits of these two planets.}
   \label{F-Jan2017_ecliptic}
   \end{figure}

\begin{figure}   
                 \centerline{\includegraphics[width=\textwidth,clip=true]{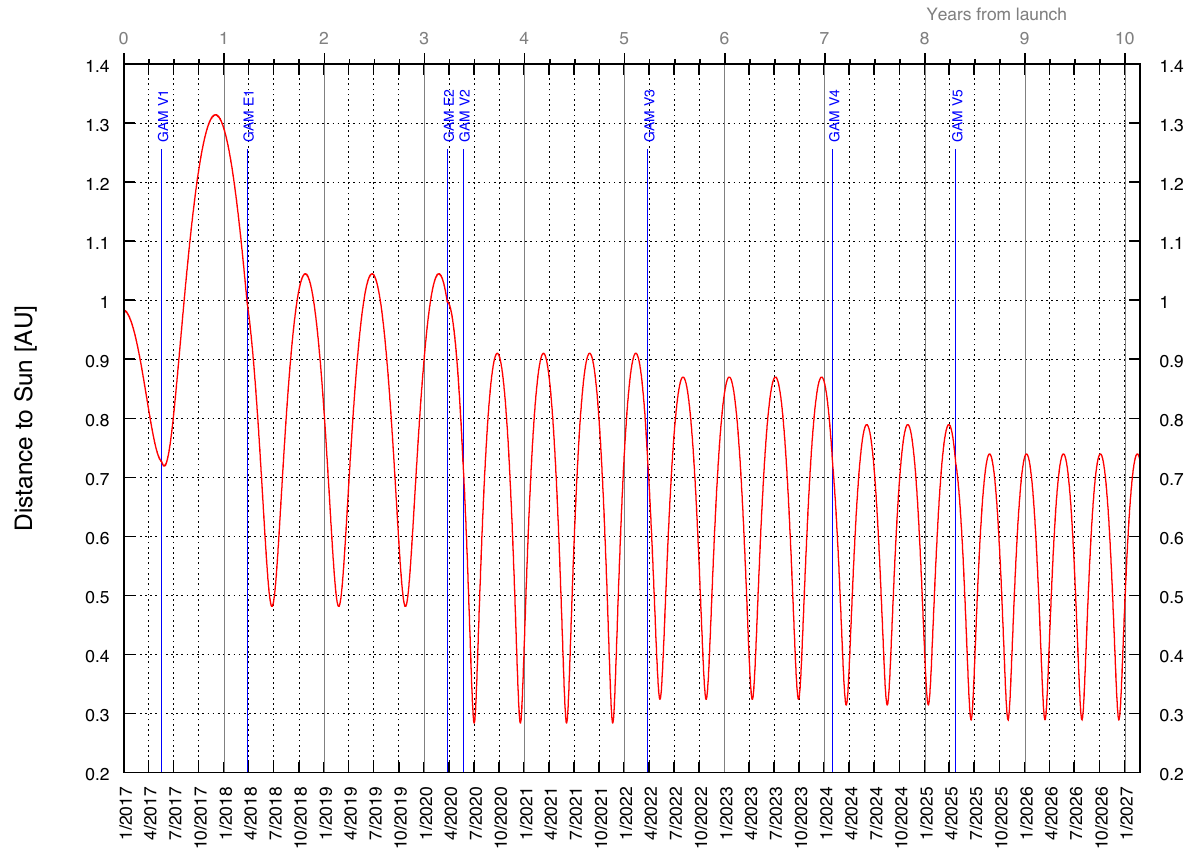}}
                  \centerline{\includegraphics[width=\textwidth,clip=true]{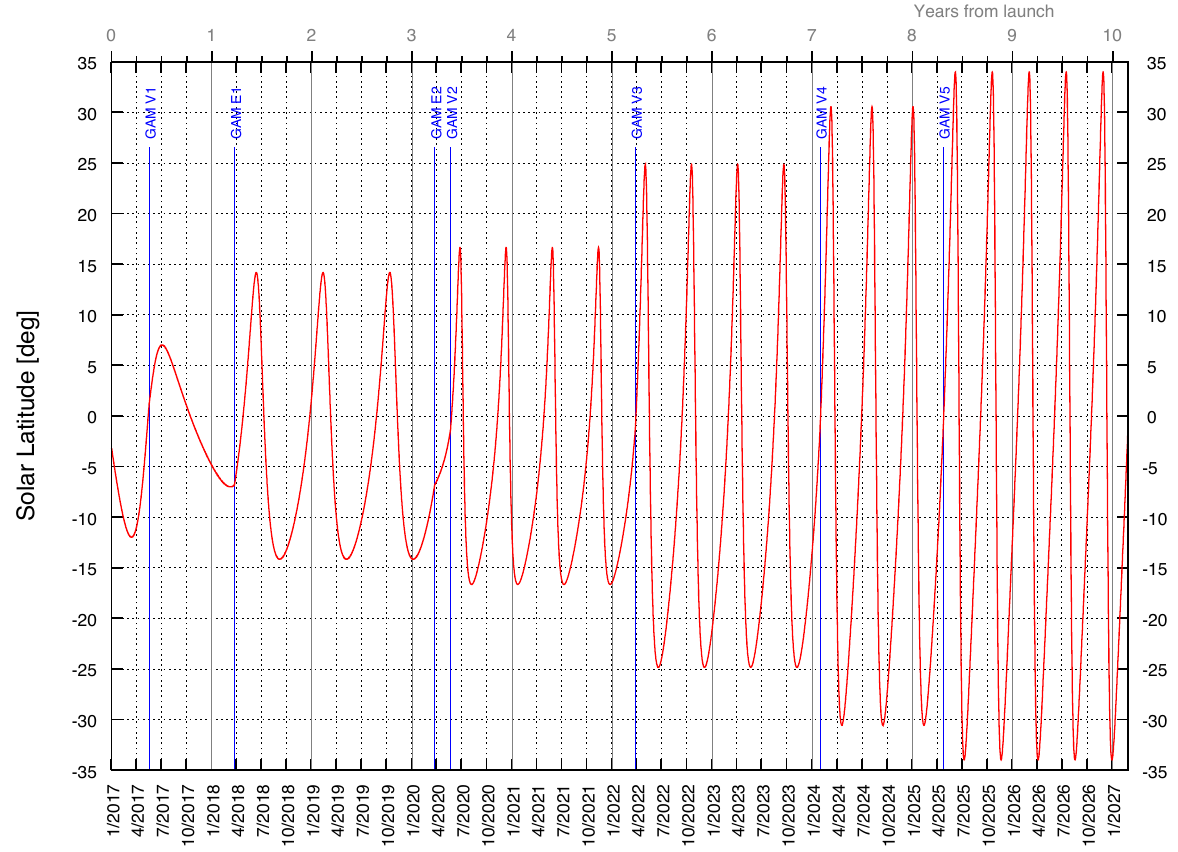}}
              \caption{Mission profile for a January 2017 launch, showing heliocentric distance (top) and latitude (bottom) of Solar Orbiter as a function of time. Also indicated are the times at which gravity assist maneuvers at Venus and Earth occur (blue).
}
   \label{F-Jan2017_latdist}
   \end{figure}

The baseline mission is planned to start in January 2017 with a launch on a NASA-provided launch vehicle from Cape Canaveral, placing the spacecraft on a ballistic trajectory that will be combined with planetary gravity assist maneuvers (GAM) at Earth and Venus (Figure~\ref{F-Jan2017_ecliptic}). The second Venus GAM places the spacecraft into a 4:3 resonant orbit with Venus at a perihelion radius of 0.284\,AU. The first perihelion at this close distance to the Sun is reached 3.5 years after launch. This orbit is the start of the sequence of resonances 4:3-4:3-3:2-5:3 that is used to raise gradually the solar inclination angle at each Venus GAM (Figure~\ref{F-Jan2017_latdist}). The resulting operational orbit has a period of 168 days during the nominal mission with a minimum perihelion radius of 0.28\,AU. The end of the nominal mission occurs 7 years after launch, when the orbit inclination relative to the solar equator reaches $25^\circ$. The inclination may be further increased during an extended mission phase using additional Venus GAMs, to reach a maximum of $34^\circ$ for the January 2017 baseline and $36^\circ$ for a launch in March 2017.

\newpage
\section{Science Objectives}
\label{S-Science}

The solar corona continuously expands and develops into a supersonic wind that extends outward, interacting with itself and with Earth and other planets, to the heliopause boundary with interstellar space, far beyond Pluto's orbit, as measured by the Voyager spacecraft \cite{Stone:1977aa}. The solar wind has profound effects on planetary environments and on the planets themselves --- for example, it is responsible for many of the phenomena in Earth's magnetosphere and is thought to have played a role in the evolution of Venus and Mars through the erosion of their upper atmospheres. 

\begin{figure}   
   \centerline{\includegraphics[width=\textwidth,clip=true]{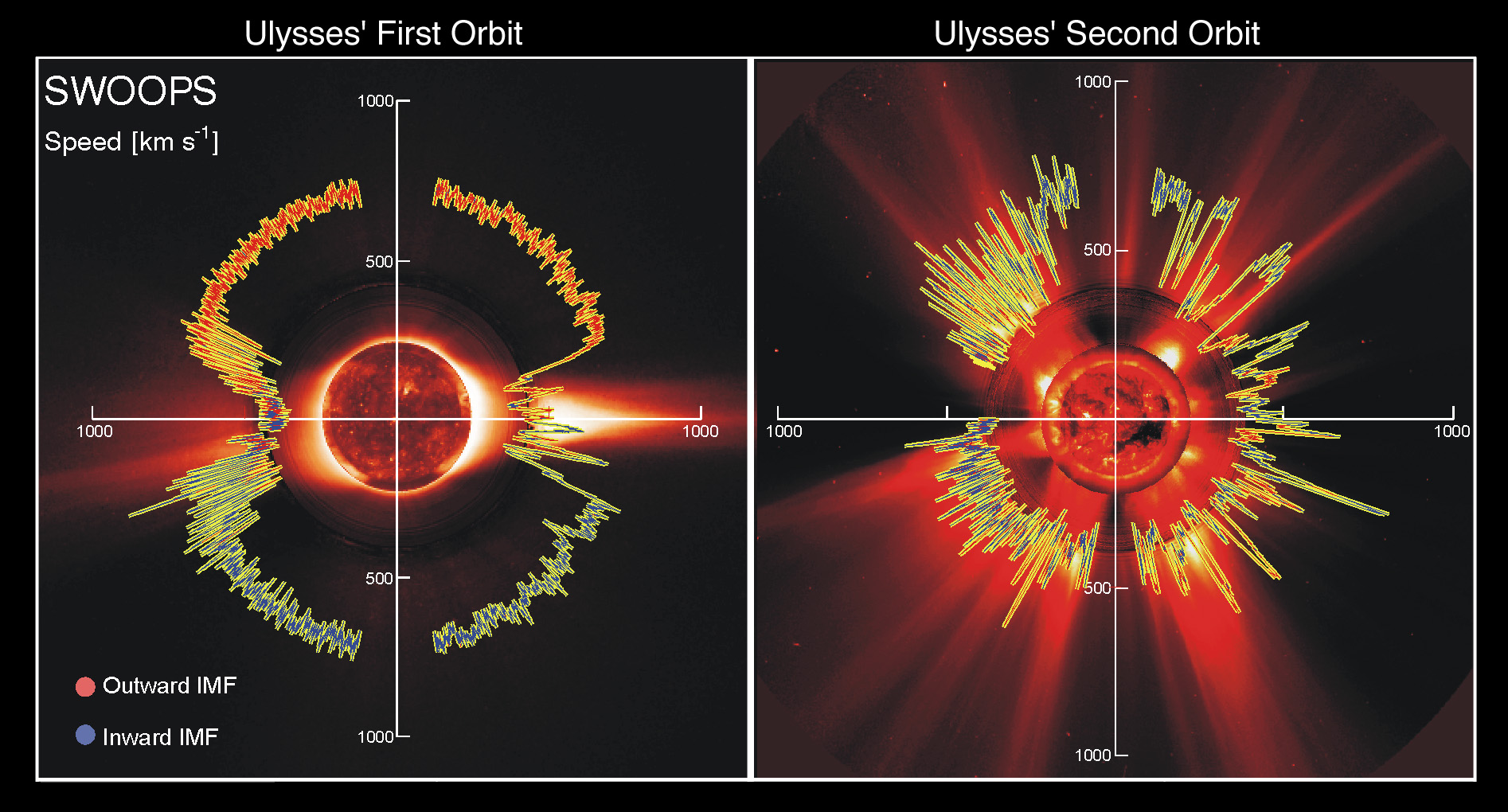}}
              \caption{Polar plots of the solar wind speed, colored by the interplanetary magnetic field (IMF) polarity for Ulysses' first two polar orbits. The earliest times are on the left (nine o'clock position) and progress around counterclockwise. The characteristic solar images for solar minimum for cycle 22 (left), solar maximum for cycle 23 (right) are from SOHO EIT and C2 coronagraph and the Mauna Loa K coronameter. Through a combination of remote-sensing and in-situ measurements, Solar Orbiter will map structures measured in the inner heliosphere to features observed in the corona. (From \protect\opencite{McComas:2008aa})}
             \label{F-McComas+al2008}
   \end{figure}

Two classes of solar wind --- 'fast' and 'slow'  --- fill the heliosphere, and the balance between them is modulated by the 11-year solar cycle (Figure \ref{F-McComas+al2008}). The fast solar wind ($\sim700$\,km/s and comparatively steady) is known to arise from coronal holes. The slow solar wind ($\sim300 - 500$\,km/s) permeates the plane of the ecliptic during most of the solar cycle so it is important to Earth's space environment. The slow solar wind shows different mass flux and composition than the fast wind, consistent with confined plasma in the solar corona.

\begin{figure}   
   \centerline{\includegraphics[width=\textwidth,clip=true]{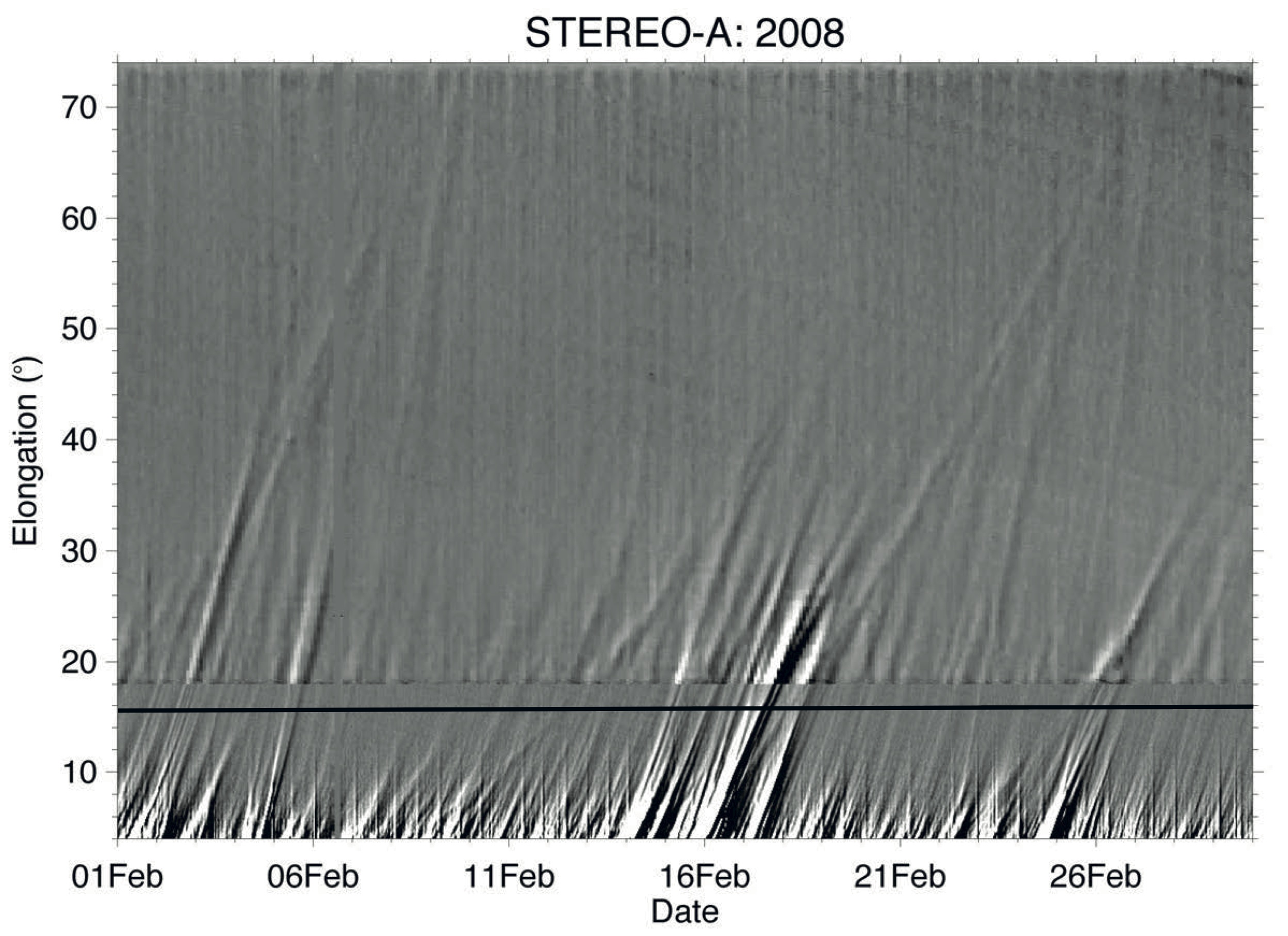}}
              \caption{Small-scale structures flowing in the solar wind are visible as diagonal lines in this time-distance plot from the STEREO Heliospheric Imager instrument, moving further from the Sun (elongation) with time. Structures moving at different speeds collide and merge, smoothing out the flow and removing information about their relative origins. Solar Orbiter will travel to 0.28\,AU, corresponding to an elongation of $15.6^{\circ}$ (black horizontal line), making it possible to measure unevolved small-scale solar wind structures for the first time. (Courtesy J.\ Davies, Rutherford Appleton Laboratory, UK)
}
   \label{F-STEREO-HI}
   \end{figure}

The specific escape mechanism through the largely closed magnetic field is not known since candidate sites and mechanisms cannot be resolved from 1\,AU as much of the crucial physics in the formation and activity of the heliosphere takes place much closer to the Sun. By the time magnetic structures, shocks, energetic particles and solar wind pass by Earth they have already evolved and in many cases mixed so as to blur the signatures of their origin (Figure~\ref{F-STEREO-HI}). It is clear that our understanding can be advanced by flying a spacecraft combining remote and in-situ observations into the inner solar system. From this inner heliospheric vantage point, solar sources can be identified and studied accurately and combined with in-situ observations of solar wind, shocks, energetic particles, etc., before they evolve significantly. Solar Orbiter can therefore be seen as the next step in our exploration of the Sun and heliosphere. In this section, we will expand the four overarching science questions of Solar Orbiter into subquestions and describe how Solar Orbiter will address them.

\subsection{What drives the solar wind and where does the coronal magnetic field originate from?}
\label{S-Goal1}
Hot plasma in the Sun's atmosphere flows radially outward into interplanetary space to form the solar wind, filling the solar system and blowing a cavity in the interstellar medium known as the heliosphere. During solar minimum, large-scale regions of a single magnetic polarity in the Sun's atmosphere  --- polar coronal holes  --- open into space and are the source of high speed, rather steady solar wind flows (Figure \ref{F-McComas+al2008}). The slow wind emanates from magnetically complex regions at low latitudes and the periphery of coronal holes. It is highly variable in speed, composition, and charge state. The origin of the slow wind is not known. At solar maximum, this stable bimodal configuration gives way to a more complex mixture of slow and fast streams emitted at all latitudes, depending on the distribution of open and closed magnetic regions and the highly tilted magnetic polarity inversion line. The fast wind from the polar coronal holes carries magnetic fields of opposite polarity into the heliosphere, which are then separated by the heliospheric current sheet (HCS) embedded in the slow wind. {\it Ulysses} and {\it Wind} measurements over a range of latitudes far from the Sun show that this boundary is not symmetric around the Sun's equator, but is on average displaced southward \cite{Smith:2000aa}. \inlinecite{Wang:2011aa} demonstrate that this southward displacement follows from Joy's law and the observed hemispheric asymmetry in the sunspot numbers, with activity being stronger in the southern (northern) hemisphere during the declining (rising) phase of cycles 20--23. They find that during the last four cycles, the polarity of the interplanetary magnetic field (IMF) around the equator tended to match that of the north polar field both before and after polar field reversal, while during cycle 19, the HCS showed in fact an average northward displacement, when the northern hemisphere became far more active than the southern hemisphere during the declining phase of the cycle.

The energy that heats the corona and drives the wind comes from the mechanical energy of convective photospheric motions, which is converted into magnetic and/or wave energy. In particular, both turbulence and magnetic reconnection are implicated theoretically and observationally in coronal heating and acceleration  (for reviews, see, e.g., \opencite{Klimchuk:2006aa} and \opencite{Reale:2010ab}).

However, existing observations have not been able to adequately constrain these theories, and the identity of the mechanisms that heat the corona and accelerate the solar wind remains one of the unsolved mysteries of solar and heliospheric physics. How the coronal plasma is generated, energized, and the way in which it breaks loose from the confining coronal magnetic field are fundamental physical questions with crucial implications for predicting our own space environment, as well as for the understanding of the physics of other astrophysical objects, from other stars, to accretion disks and their coronae, to energetic phenomena such as jets, X- and gamma-ray bursts, and cosmic-ray acceleration. 

The solar wind contains waves and turbulence on scales from millions of kilometres to below the electron gyroradius. The turbulence scatters energetic particles, affecting the flux of particles that arrives at the Earth; local kinetic processes dissipate the turbulent fluctuations and heat the plasma. Properties of the turbulence vary with solar wind stream structure, reflecting its origins near the Sun, but the turbulence also evolves as it is carried into space with the solar wind, blurring the imprint of coronal conditions and making it difficult to determine its physical origin. The inner heliosphere, where Solar Orbiter will conduct its observations, provides the ideal laboratory for understanding magnetohydrodynamic turbulence, which is expected to be ubiquitous in astrophysical environments.

Below we discuss in more detail three interrelated questions which flow down from this top-level question: What are the source regions of the solar wind and the heliospheric magnetic field? What mechanisms heat and accelerate the solar wind? What are the sources of turbulence in the solar wind and how does it evolve?

\subsubsection{What are the source regions of the solar wind and the heliospheric magnetic field?}
\paragraph*{Present state of knowledge.} At large scales, the structure of the solar wind and heliospheric magnetic field and their mapping to the solar corona are reasonably well understood. However, extending this global understanding of the overall connection between the corona and the solar wind deeper into the solar atmosphere and to the photosphere where the magnetic field can be measured has been difficult due to the dynamically evolved state of the plasma measured in situ at 1\,AU and to the lack of simultaneous in-situ measurements and high-cadence, high-resolution remote sensing of solar plasma. A number of fundamental questions remain unanswered both about the source of the fast and slow solar wind and about the source of the magnetic field that the solar wind carries into the heliosphere.

\vspace{3mm}
\noindent
{\it(a) Source regions of the solar wind.} The speed of the solar wind is empirically anti-correlated with the (modelled) expansion rate of the magnetic field with radial distance close to the Sun \cite{Wang:2006aa}, where central areas of polar coronal holes give rise to the fastest solar wind streams, while regions closer to the coronal hole boundary give rise to progressively slower wind. Within coronal holes, strong outflows are well correlated with the intense flux elements found at the intersection of the photospheric supergranular cells; these expand into the corona as 'funnels,' preferentially from regions dominated by flux of the dominant hole polarity \cite{Tu:2005aa,McIntosh:2006aa}.

\begin{figure}   
   \centerline{\includegraphics[width=\textwidth,clip=true]{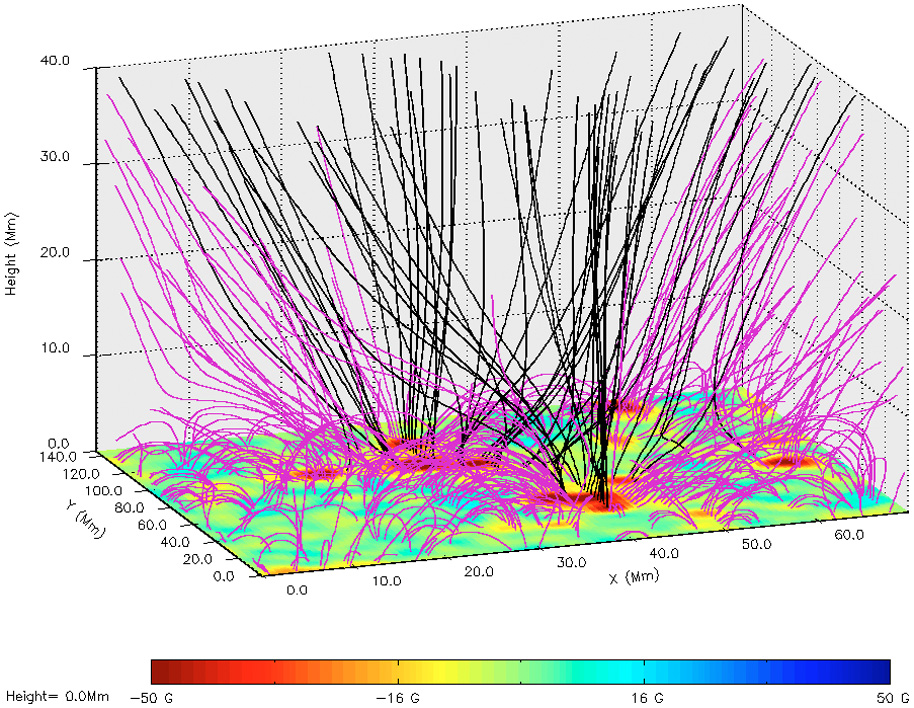}}
              \caption{The modelled magnetic field of the transition region and lower corona in a polar coronal hole based on measurements of the photospheric magnetic field strength. The figure illustrates the complex connections between the solar surface and space:  only the black field lines extend far from the surface. A central goal of Solar Orbiter is to establish the links between the observed solar wind streams and their sources back on the Sun. Understanding the dynamics of the Sun's magnetic atmosphere and its signatures in the measured solar wind holds the key to understanding the origin of all solar wind flows. (From \protect\opencite{Marsch:2006aa})}
     \label{F-Marsch+al2006}
   \end{figure}

The source region of the wind, at chromospheric and transition region heights, is extremely structured and dynamic (Figure~\ref{F-Marsch+al2006}). The chromosphere is permeated by spicules, cool and dense jets of chromospheric plasma. Spicules have been thought to be too slow and cold to contribute significantly to the solar wind, but a more dynamic type of spicule, with shorter lifetimes, faster motions, and a hotter plasma component has recently been discovered by Hinode. Such spicules also support waves, possibly with sufficient energy to accelerate fast wind streams in coronal holes \cite{De-Pontieu:2009aa,De-Pontieu:2011aa}.

Hinode has also observed the frequent occurrence of very small-scale X-ray jets in polar coronal holes \cite{Cirtain:2007aa}. Given the high velocities and frequency of these events, it has been suggested that they contribute to the fast solar wind. Their relation to the photospheric magnetic field, however, has not been established as the high latitudes at which they are observed hamper the accurate determination of their photospheric footpoints from the ecliptic plane. Other fine-scale ray-like structures --- coronal plumes --- permeate coronal holes and are correlated with small-scale bipolar structures inside the hole. Ultraviolet measurements show that these structures are cooler than the surrounding background hole plasma, and have slower, but denser outflows. In-situ measurements reveal the existence of faster and slower microstreams within the fast wind \cite{Neugebauer:1995aa} as well as other fine-scale structures \cite{Thieme:1990aa}, but the two have not been unambiguously linked to coronal features. 

The anti-correlation of expansion/wind speed suggests that the slow wind is accelerated along those open field lines with the greatest expansion rate, notably corresponding to the bright rays at the coronal hole-streamer interface (e.g., \opencite{Wang:2007aa}) and to outflows from coronal hole boundaries \cite{Antonucci:2005aa}. However, composition measurements tend to call this notion into question: a significant elemental fractionation is observed in the solar wind plasma relative to that of the photosphere (e.g., \opencite{Geiss1982SSRv}), which scales with the first ionization potential (FIP). Metallic ions, with low FIP, are more abundant in the solar wind than mid- or high-FIP elements, when compared with their photospheric compositions \cite{von-Steiger:1997aa}. Ulysses has revealed a systematic difference in the degree of fractionation depending on the solar wind type. Fast wind associated with coronal holes has a composition similar to that of the photosphere, whereas the slow solar wind is characterized by a substantially larger degree of fractionation. 

\begin{figure}   
   \centerline{\includegraphics[width=\textwidth,clip=true]{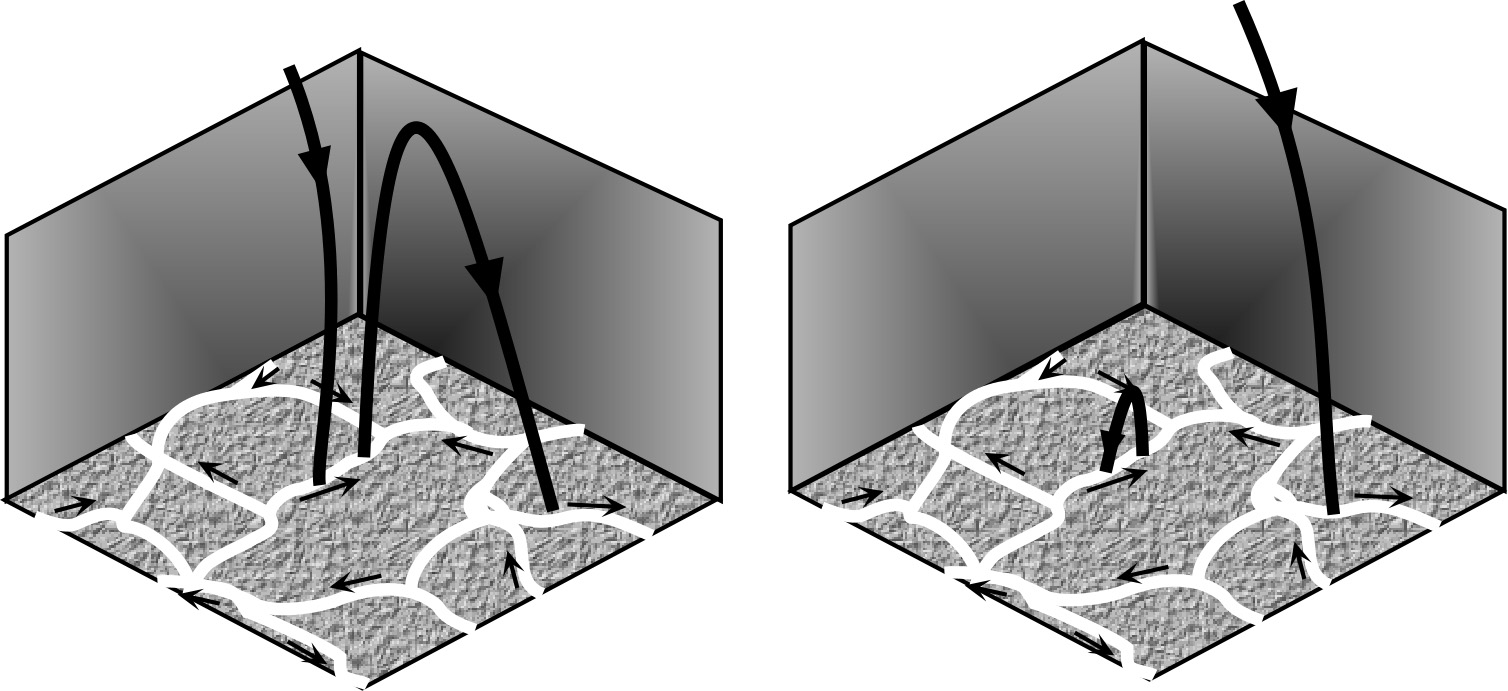}}
              \caption{Convective cells (white lines) in the photosphere can bring together oppositely directed magnetic field lines (left). These can undergo interchange reconnection, altering which field lines open into space and which close back to the surface (right). This process is thought to be important in generating the slow solar wind flow, as well as moving magnetic flux over the solar cycle, but observational evidence for it is currently weak. Solar Orbiter will combine high resolution observations of photospheric motion and magnetic fields with measurements of the solar wind and magnetic field flowing outward to determine the quantitative effects of interchange reconnection. (From \protect\opencite{Fisk:2006aa})
}
     \label{F-Convective_cells_sketch}
   \end{figure}

Recently, \inlinecite{Antiochos:2011aa} have presented a new model, which can account both for the observed large angular width (up to $\approx 60^\circ$) of the slow wind as well as its FIP-enhanced coronal composition. They argue that the most likely source for the slow wind is a network of narrow, possibly singular, open-field corridors in the surrounding closed-field corona that map to a web of separatrices (termed 'S-web') and quasi-separatrix layers in the heliosphere. In this case, the process that releases the coronal plasma to the wind would have to be either the opening of closed flux or interchange reconnection between open and closed magnetic field lines. 
Closed magnetic fields lines close to the Sun confine the plasma in loops, where the compositional differentiation occurs, but these are continuously destroyed when neighbouring open field lines are advected into them. Interchange reconnection between the open and closed field allows the plasma to flow outwards into space (Figure~\ref{F-Convective_cells_sketch}). This process should occur predominantly at the coronal hole boundary, but may also be active in the intermediate areas of quiet Sun, and is the underlying mechanism invoked by Fisk and coworkers (\opencite{Fisk:1998aa}; \opencite{Fisk:2003aa}; \opencite{Fisk:2006aa}; \opencite{Fisk:2009aa}) in their model for the heliospheric magnetic field.

\begin{figure}   
   \centerline{\includegraphics[width=\textwidth,clip=true]{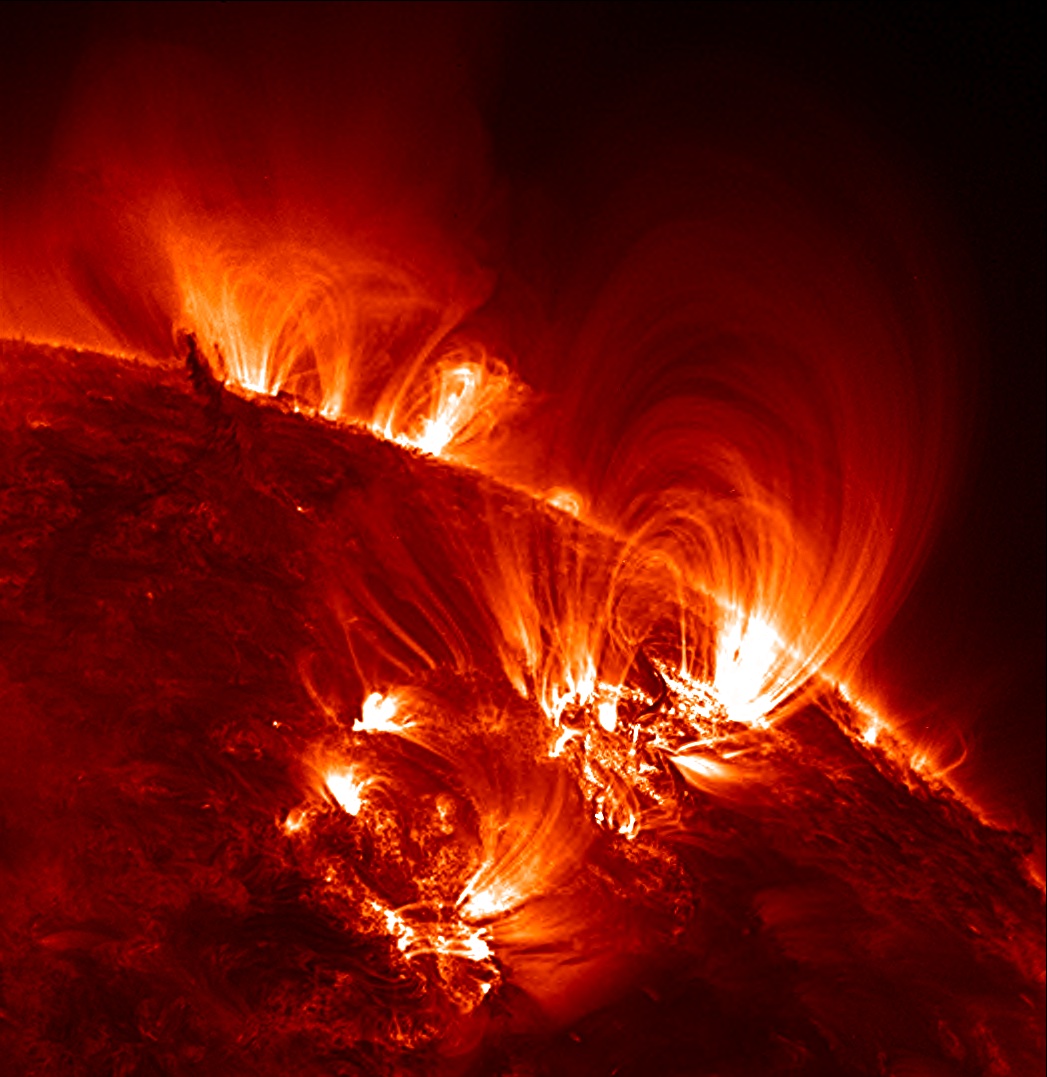}}
              \caption{Ultraviolet emission from plasma in the Sun's atmosphere, revealing the complex magnetic field structures around active regions. (SDO AIA 17.1\,nm image)
}
     \label{F-AR_AIA}
   \end{figure}

Additional contributions to the slow wind could arise from the opening of previously closed field lines in the middle and lower corona, from the tops of helmet streamers or the complex magnetic fields around active regions (Figure~\ref{F-AR_AIA}), for example, releasing plasma blobs or plasmoids into the heliosphere. White-light coronagraph observations show streamer blobs that might be plasmoids or might be pile-up from reconnection high in the corona. Finally, there might be a continuous outward leakage of plasma from high in the solar corona where the plasma pressure becomes comparable to the magnetic pressure in the weak field at the apex of closed loops. 

\vspace{3mm}
\noindent
{\it (b) Source regions of the heliospheric magnetic field.} Our current knowledge of the surface magnetic field of the Sun and its extension into the solar atmosphere and interplanetary space is based on measurements of the photospheric line-of-sight (and recently, vector) magnetic field, coupled with spacecraft measurements of the field in situ. The vast majority of the magnetic flux from the Sun closes in the lower layers of the solar atmosphere, within the chromosphere and lower corona, in multiple small scale bipolar regions with strong local fields, and it is only a small fraction which extends high enough in the solar atmosphere to be dragged out into the heliosphere by the solar wind. In addition, the intense magnetic fields in the lower atmosphere are highly variable and dynamic at scales extending down to instrument resolution limits in both time and space, continuously reconnecting and contributing to the intense activity, spicules and jets in the chromosphere and lower corona. The magnetic connection between the solar wind and the solar source therefore hinges on understanding what determines the amount of open flux from the Sun, how open field lines are distributed at the solar surface at any given time, and how these open field lines reconnect and change their connection across the solar surface in time, processes which are controlled by interchange reconnection (\opencite{Wang:2000aa}; \opencite{Fisk:2001aa}).

The HCS is embedded in slow solar wind and, like the slow wind, is full of small-scale structure. The origin of fine-scale structure in the magnetic field is therefore directly related to the origin of the slow solar wind. As mentioned in section~\ref{S-Goal1}, one of the most surprising results regarding the heliospheric current sheet is that it is not symmetric around the equator, but appears to be displaced southward by around $10^{\circ}$ \cite{Smith:2000aa} during solar minima, causing a difference in cosmic ray fluxes between hemispheres. Similar asymmetries exist in the Sun's polar magnetic fields and even sunspot numbers \cite{Wang:2011aa}, but what is the origin of this asymmetry, and how does the Sun produce it in space? 

\paragraph*{How Solar Orbiter will address the question.}
 Solar Orbiter will measure the solar wind plasma and magnetic field in situ while simultaneously performing remote-sensing measurements of the photosphere and corona, thus allowing the properties of the solar wind measured in situ to be correlated with structures observed in the source regions at the Sun. During its perihelion passages, Solar Orbiter will determine the plasma parameters and compositional signatures of the solar wind, which can be compared directly with the spectroscopic signatures of coronal ions with differing charge-to-mass ratios and FIP. 

Solar Orbiter will determine magnetic connectivity by measuring energetic electrons and the associated X-rays and radio emissions and using these measurements to trace the magnetic field lines directly to the solar source regions. Photospheric magnetic field measurements, together with those made in situ, will allow the coronal magnetic field to be reconstructed by extrapolation with well-defined boundary conditions. Extreme ultraviolet imaging (EUV) imaging and spectroscopy will provide the images and plasma diagnostics needed to characterize the plasma state in the coronal loops, which can erupt and deliver material to solar wind streams in the outer corona. As Solar Orbiter observes different source regions, from active regions to quiet Sun to coronal holes, hovering for substantial amounts of time over each during the near-corotation periods, it will be able to provide insight into the origin of the solar wind.

EUV spectroscopy and imaging are needed to detect magnetic reconnection in the transition region and corona, e.g., by the observation of plasma jets or of explosive events as seen in the heavy-ion Doppler motions believed to mark the reconnection-driven plasma outflow. These events appear to be associated with impulsive energetic particle bursts observed near 1\,AU. The study of the time evolution of such events, and of their particle and radiation outputs, can reveal whether reconnection is quasi-steady or time-varying, and a comparison with magnetic field data will indicate the locations of the reconnection sites with respect to the overall magnetic field structure and topology. Solar Orbiter's coronagraph will construct global maps of the H and He outflow velocity and measure the degree of correlation of wind speed and He fraction. 

Within the solar wind, the in-situ instruments will measure the radial, latitudinal, and longitudinal gradients of plasma and field parameters in the inner heliosphere, providing information fundamental to diagnosing the connection of the solar wind with the coronal structure. Combining Solar Orbiter data from the in-situ and remote-sensing instruments taken at different intervals will make it possible to determine the relative contributions of plumes, jets, and spicules to the fast wind.

\subsubsection{What mechanisms heat and accelerate the solar wind?}
\paragraph*{Present state of knowledge.} Despite more than a half-century of study, the basic physical processes responsible for heating the million-degree corona and accelerating the solar wind are still not known. Identification of these processes is important for understanding the origins and impacts of space weather and to make progress in fundamental stellar astrophysics. 

Ultimately, the problem of solar wind acceleration is a question of the transfer, storage and dissipation of the abundant energy present in the solar convective flows. The key challenge is to establish how a small fraction of that energy is transformed into magnetic and thermal energy above the photosphere. Both emerging magnetic flux and the constant convective 'braiding' of magnetic field lines contribute to the processing of the energy in what is an extremely structured, highly dynamic region of the solar atmosphere, the route to dissipation involving cascading turbulence, current sheet collapse and reconnection, shocks, high-frequency waves, and wave-particle interactions. The advent of high-cadence high-resolution observations has demonstrated the extremely complex phenomenology of the energy flux in the lower atmosphere, including many types of transient events discovered and classified by Yohkoh, SOHO, TRACE, RHESSI, Hinode and SDO. 

Energy deposited in the corona is lost in the form of conduction, radiation (negligible in coronal holes), gravitational enthalpy, and kinetic energy fluxes into the accelerating solar wind plasma. Transition region pressure, coronal densities, temperature and the asymptotic solar wind speed are sensitive functions of the mode and location of energy deposition. The mass flux is not, however, as it depends only on the amplitude of the energy flux \cite{Hansteen:1995aa}. A relatively constant coronal energy flux therefore explains the small variations in mass flux between slow and fast solar wind found by Ulysses during its first two orbits, although the dramatic decrease in mass flux over the present cycle points also to a decreased efficiency of coronal heating and therefore to its dependence on the solar magnetic field \cite{McComas:2008aa,Schwadron:2008aa}.
 
\begin{figure}   
   \centerline{\includegraphics[width=\textwidth,clip=true]{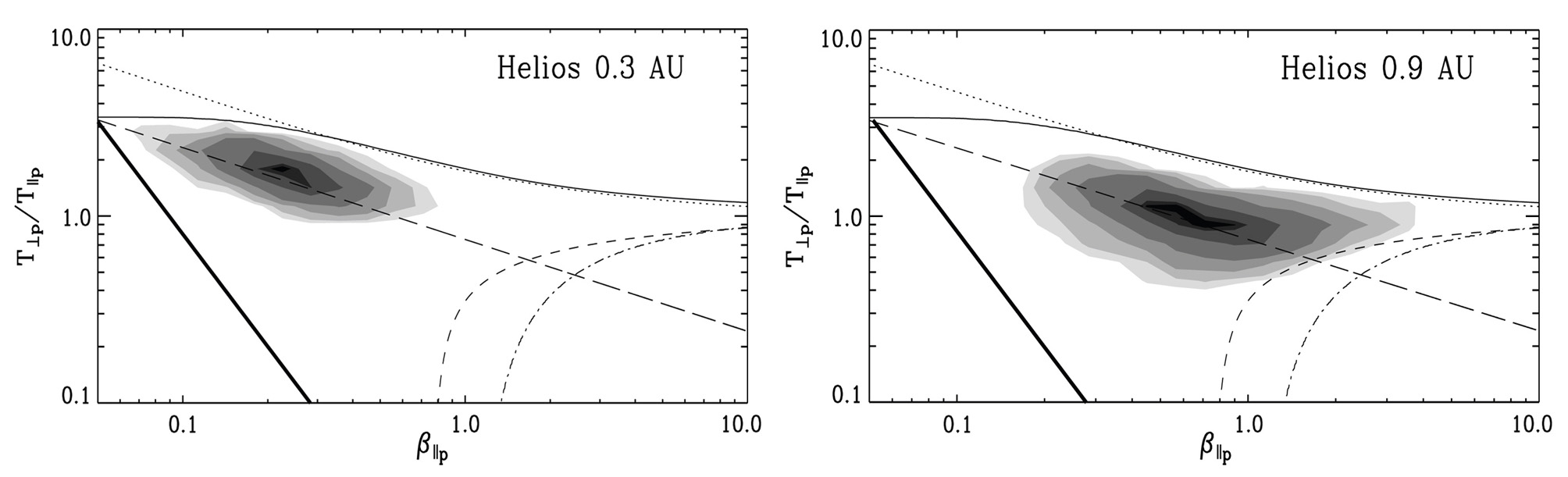}}
              \caption{Histograms of the solar wind proton temperature anisotropy (ratio of perpendicular to parallel temperatures) versus the plasma pressure parallel to the field (parallel plasma beta) in the fast solar wind measured at two different radial distances by Helios. The dark line shows the decrease of anisotropy expected if the wind were expanding adiabatically without heating (dark continuous), and the actual distribution function contours with best fit of the run of anisotropy. Instability threshold conditions for the ion-cyclotron (solid), the mirror (dotted), the parallel (dashed) and oblique (dash-dotted) fire hose  instabilities are also shown. Distribution functions display perpendicular heating and evolve towards marginal stability with distance from the Sun. Solar Orbiter will determine initial conditions for the perpendicular anisotropies and help determine the nature of the plasma-wave interactions responsible for this heating. (From \protect\opencite{Matteini:2007aa})
}
     \label{F-SolarWind_histo}
   \end{figure}

One of the fundamental experimental facts that has been difficult to account for theoretically is that the fast solar wind originates in regions where the electron temperature and densities are low, while the slow solar wind comes from hotter regions of the corona. The anti-correlation of solar wind speed with electron temperature is confirmed by the anti-correlation between wind speed and 'freezing in' temperature of the different ionization states of heavy ions in the solar wind \cite{Geiss:1995aa} and implies that the electron pressure gradient does not play a major role in the acceleration of the fast wind. On the other hand, the speed of the solar wind is positively correlated with the in-situ proton temperature, and the fastest and least collisionally coupled wind streams also contain the largest distribution function anisotropies. Observations of the very high temperatures and anisotropies of coronal heavy ions suggest that other processes such as magnetic mirror and wave-particle interactions should also contribute strongly to the expansion of the fast wind \cite{Li:1998aa,Kohl:1997aa,Kohl:1998aa,Kohl:2006aa,Dodero:1998aa}. In particular, either the direct generation of high-frequency waves close to the cyclotron resonance of ions or the turbulent cascade of energy to those frequencies should play an important role (Figure~\ref{F-SolarWind_histo}). For an extensive review of the kinetic physics of the solar corona and solar wind, see \inlinecite{Marsch:2006ab}.

Theoretical attempts to develop self-consistent models of fast solar wind acceleration have followed two somewhat different paths. First, there are models in which the convection-driven jostling of magnetic flux tubes in the photosphere drives wave-like fluctuations that propagate up into the extended corona. The waves partially reflect back toward the Sun, develop into strong turbulence, and/or dissipate over a range of heights.
These models also tend to attribute the differences between the fast and slow solar wind not to any major differences in the lower boundary conditions, but to the varying expansion factor of magnetic field lines in different areas of coronal holes (\opencite{Cranmer:2007aa}, and references therein).

In the second class of models, the interchange reconnection models, the energy flux usually results from magnetic reconnection between closed, loop-like magnetic flux systems (which are in the process of emerging, fragmenting, and being otherwise jostled by convection) and the open flux tubes that connect to the solar wind. Here the differences between fast and slow solar wind result from qualitatively different rates of flux emergence, reconnection, and coronal heating in different regions on the Sun \cite{Axford:1992aa,Fisk:1999aa,Schwadron:2003aa}. It has been difficult to evaluate competing models of fast wind acceleration and to assess observationally the relative contributions of locally emerging magnetic fields and waves to the heat input and pressure required to accelerate the wind  largely because of the absence of measurements of the solar wind close to the Sun where they can be mapped with sufficient precision to a solar source region.

\paragraph*{How Solar Orbiter will address the question.} Solar Orbiter's combination of high-resolution measurements of the photospheric magnetic field together with images and spectra at unprecedented spatial resolution will make it possible to identify plasma processes such as reconnection/shock formation and wave dissipation in rapidly varying surface features, observe Doppler shifts of the generated upflows, and determine compositional signatures. Whatever the scale, magnetic reconnection leads to particle dissipative heating and acceleration and wave generation, which have the net effect  of a local kinetic energy increase in the lower solar atmosphere that can be revealed through high-resolution extreme ultraviolet (EUV) imaging and spectroscopy. Wave propagation will be traced from the source site to the region of dissipation through observations of EUV-line broadening and Doppler shifts.

Global maps of the H and He outflow velocity, obtained by applying the Doppler dimming technique to the resonantly scattered component of the most intense emission lines of the outer corona (H\,I 121.6\,nm and He\,II 30.4\,nm), will provide the contours of the maximum coronal acceleration for the two major components of the solar wind, and the role of high-frequency cyclotron waves will be comprehensively assessed by measuring spectroscopically the particle velocity distribution across the field and determining the height where the maximum gradient of outflow velocity occurs \cite{Telloni:2007aa}. 

Solar Orbiter's heliospheric imager will measure the velocity, acceleration, and mass density of structures in the accelerating wind, allowing precise comparison with the different acceleration profiles of turbulence-driven and interchange reconnection-driven solar wind models.

As it is performing imaging and spectroscopic observations of the corona and photosphere, Solar Orbiter will simultaneously measure in situ the properties of the solar wind emanating from the source regions. The in-situ instrumentation will determine all of the properties predicted by solar wind acceleration models: speed, mass flux, composition, charge states, and wave amplitudes. Moving relatively slowly over the solar surface near perihelion, Solar Orbiter will measure how properties of the solar wind vary depending on the changing properties of its source region, as a function of both space and time, distinguishing between competing models of solar wind generation.

\subsubsection{What are the sources of turbulence in the solar wind and how does it evolve?}
\paragraph*{Present state of knowledge.} The solar wind is filled with turbulence and instabilities. At large scales, the fast solar wind is dominated by anti-sunward propagating Alfv{\'e}n waves thought to be generated by photospheric motions. At smaller scales, these waves decay and generate an active turbulent cascade, with a spectrum similar to the Kolmogorov hydrodynamic scaling of $f^{-5/3}$. In the slow solar wind, turbulence does not have a dominant Alfv{\'e}nic component, and it is fully developed over all measured scales. There is strong evidence that the cascade to smaller scales is anisotropic, but it is not known how the anisotropy is generated or driven \cite{Horbury:2008aa}. What do the differences between the turbulence observed in the fast wind and that observed in the slow wind reveal about the sources of the turbulence and of the wind itself?

Little is known about what drives the evolution of solar wind turbulence. Slow-fast wind shears, fine-scale structures, and gradients are all candidate mechanisms \cite{Tu:1990aa,Breech:2008aa}. To determine how the plasma environment affects the dynamical evolution of solar wind turbulence it is essential to measure the plasma and magnetic field fluctuations in the solar wind as close to the Sun as possible, before the effects of mechanisms such as velocity shear become significant, and then to observe how the turbulence evolves with heliocentric distance. 
The dissipation of energy in a turbulent cascade contributes to the heating of the solar wind plasma. However, while measurements of the properties of solar wind turbulence in near-Earth orbit largely agree with observed heating rates \cite{Smith:2001aa,Marino:2008aa}, the details are controversial and dependent on precise models of turbulent dynamics. In order to establish a full energy budget for the solar wind, the heating rates as a function of distance and stream properties must be determined, including turbulence levels before the cascade develops significantly.
The statistical analysis of the fluctuating fields also reveals pervasive fine-scale structure (e.g., discontinuities and pressure balanced structures). The origin of these structures is uncertain: are they the remnant of complex coronal structuring in the form of strands of small-scale flux tubes advected by the solar wind flow \cite{Borovsky:2008aa,Bruno:2001aa}, or are they generated locally by turbulent fluctuations? 

At scales around the proton gyroradius and below, turbulent fluctuations interact directly with the solar wind ions. The precise nature of the turbulent cascade below the proton gyroradius is poorly understood and might even vary depending on local plasma conditions. Below the electron gyroradius, conditions are even less certain and the partitioning of turbulent energy into electron or ion heating is unknown at this time. In addition, solar wind expansion constantly drives distribution functions toward kinetic instabilities, where fluctuations with characteristic signatures are generated (e.g., \opencite{Marsch:2006aa}). What physical role do kinetic effects play with distance from the Sun? What role do wave-particle interactions play in accelerating the fast solar wind? What contribution do minor ions make to the turbulent energy density in near-Sun space?

\paragraph*{How Solar Orbiter will address the question.}
Solar Orbiter will measure waves and turbulence in the solar corona and solar wind over a wide range of latitudes and distances, including closer to the Sun than ever before, making it possible to study turbulence before it is significantly affected by stream-stream interactions. By travelling over a range of distances, the spacecraft will determine how the turbulence evolves and is driven as it is carried anti-sunward by the solar wind. 
Detailed in-situ data will make it possible to distinguish between competing theories of turbulent dissipation and heating mechanisms in a range of plasma environments and are thus of critical importance for advancing our understanding of coronal heating and of the role of turbulence in stellar winds. 

By entering near-corotation close to the Sun, Solar Orbiter will be able to distinguish between the radial, longitudinal, and temporal scales of small-scale structures, determining whether they are the signatures of embedded flux tubes or are generated by local turbulence.

Solar Orbiter's magnetic and electric field measurements, combined with measurement of the full distribution functions of the protons and electrons will fully characterize plasma turbulence over all physically relevant time scales from very low frequencies to above the electron gyrofrequency. Because Solar Orbiter is a three-axis stabilized spacecraft, it can continuously view the solar wind beam with its proton instrument, measuring proton distributions at the gyroperiod and hence making it possible directly to diagnose wave-particle interactions in ways that are not possible on spinning spacecraft. By travelling closer to the Sun than ever before, it will measure wave-particle interactions before the particle distributions have fully thermalized, studying the same processes that occur in the corona. By measuring how the distributions and waves change with solar distance and between solar wind streams with different plasma properties, Solar Orbiter will make it possible to determine the relative effects of instabilities and turbulence in heating the plasma.

The solar wind is the only available plasma 'laboratory' where detailed studies of magnetohydrodynamic (MHD) turbulence can be carried out free from interference with spatial boundaries, and in the important domain of very large magnetic Reynolds numbers. Detailed comparison between experimental in-situ data and theoretical concepts will provide a more solid physical foundation for MHD turbulence theory, which will be of critical importance for understanding the solar (stellar) coronal heating mechanism and the role or turbulence in the solar (a stellar) wind. 

\subsection{How do solar transients drive heliospheric variability?}

The Sun exhibits many forms of transient phenomena, such as flares, coronal mass ejections (CMEs), eruptive prominences, and shock waves. Many directly affect the structure and dynamics of the outflowing solar wind and thereby also eventually affect Earth's magnetosphere and upper atmosphere. Understanding these impacts, with the ultimate aim of predicting them, has received much attention during the past decade and a half under the banner of 'space weather.' However, many fundamental questions remain about the physics underpinning these phenomena and their origins, and these questions must be answered before we can realistically expect to be able to predict the occurrence of solar transients and their effects on geospace and the heliosphere. These questions are also pertinent, within the framework of the 'solar-stellar connection,' to our understanding of other stellar systems that exhibit transient behaviour such as flaring (e.g., \opencite{Getman:2008aa}).

Solar Orbiter will provide a critical step forward in understanding the origin of solar transient phenomena and their impact on the heliosphere. Located close to the solar sources of transients, Solar Orbiter will be able both to determine the inputs to the heliosphere and to measure directly the heliospheric consequences of eruptive events at distances close enough to sample the fields and plasmas in their pristine state, prior to significant processing during their propagation to 1\,AU. Solar Orbiter will thus be a key augmentation to the chain of solar-terrestrial observatories in Earth orbit and at the libration points, providing a critical perspective from its orbit close to the Sun and out of the ecliptic. 

Below we discuss in more detail three interrelated questions which flow down from this top-level question: How do CMEs evolve through the corona and inner heliosphere? How do CMEs contribute to solar magnetic flux and helicity balance? How and where do shocks form in the corona and inner heliosphere?

\subsubsection{How do CMEs evolve through the corona and inner heliosphere?}
\begin{figure}   
   \centerline{\includegraphics[width=\textwidth,clip=true]{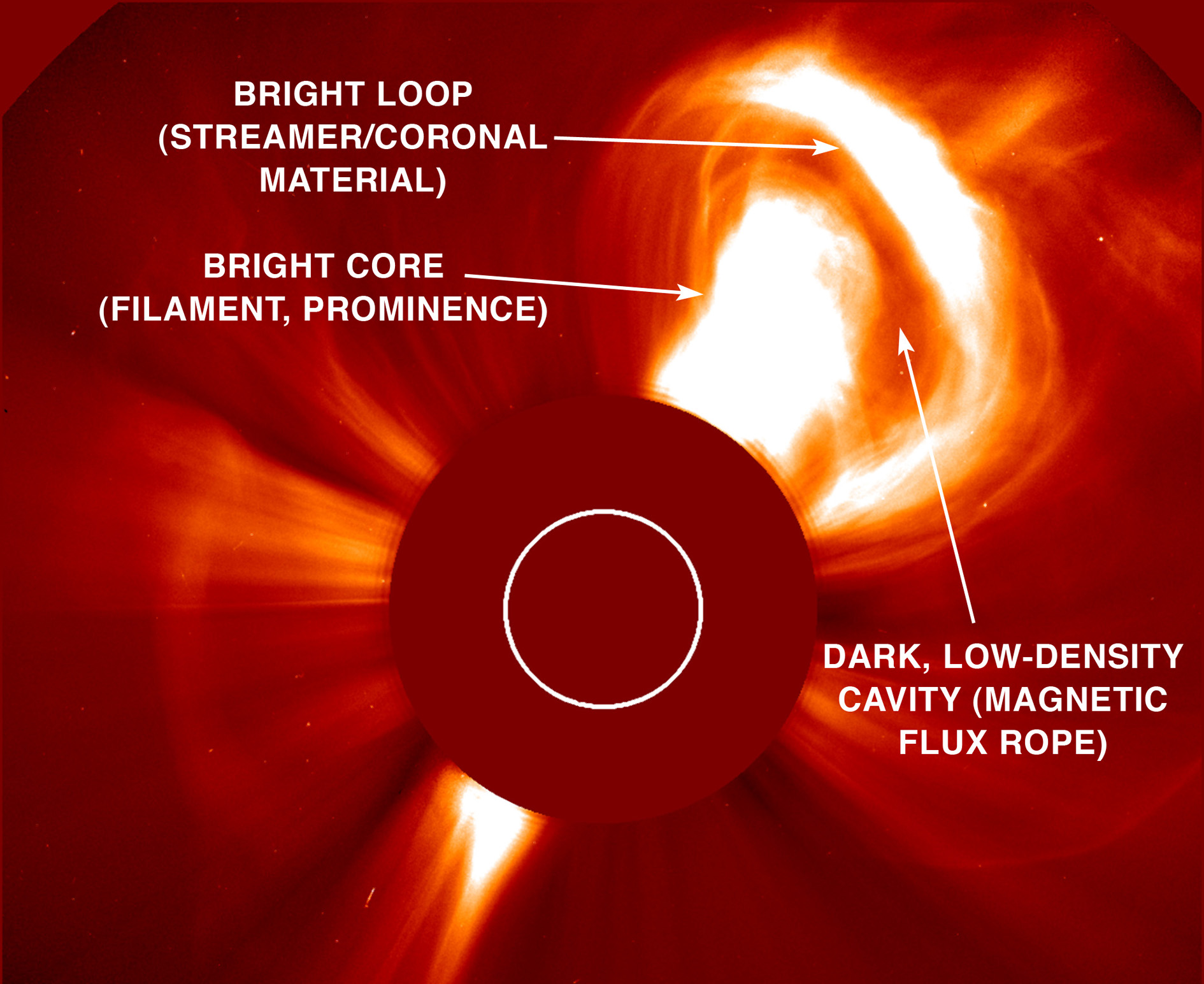}}
              \caption{A coronal mass ejection (CME) on the solar limb as viewed by LASCO on SOHO in December 2002. The dark, low-density region inside the structure formed by the bright loop of streamer material is thought to be the magnetic flux rope predicted by current CME initiation models. CMEs are believed to originate from prominence eruptions, yet in interplanetary coronal mass ejections (ICMEs) observed at 1\,AU prominence plasma (the bright core in this image) is very rarely detected. Solar Orbiter will enable in-situ measurements of the ejecta and their radial (and out-of-the-ecliptic) evolution in more detail than possible from Earth orbit, where many features have been washed out.}
     \label{F-CME_3parts}
   \end{figure}

\paragraph*{Present state of knowledge.} Following earlier observations by space-based white-light coronagraphs, considerable progress in understanding CMEs has been achieved using data from the ESA-NASA SOHO mission, which provides continuous coverage of the Sun and combines coronagraphs with an EUV imager and off-limb spectrometer. Other spacecraft, such as ACE, WIND and Ulysses, which carried comprehensive in-situ instrumentation, have contributed significantly to our understanding of the interplanetary manifestation of these events. With a full solar cycle of CME observations, the basic features of CMEs are now understood. CMEs often appear to originate from highly-sheared magnetic field regions on the Sun known as filament channels, which support colder plasma condensations known as prominences. Eruptions are frequently impulsively accelerated in the low corona within 10-15\,minutes (the initial phase can take significantly longer, see \opencite{Liu:2010ab}), while the associated shocks cross the solar disk within 1\,hour. CMEs reach speeds of up to 3000\,km/s and carry energies (kinetic, thermal and magnetic) of $\sim10^{25}$\,J ($= 10^{32}$\,ergs). They can also accelerate rapidly during the very early stages of their formation, with the CME velocity being closely tied, in time, to the associated flare's soft X-ray light profile \cite{Zhang:2006aa}. SOHO's coronagraph images (Figure~\ref{F-CME_3parts}) have provided evidence for a magnetic flux rope structure in some CMEs as well as for post-CME current sheets. Both features are predicted by CME initiation models (e.g., \opencite{Lin:2000aa,Lynch:2004aa}).

STEREO observations are making it possible to chart the trajectories of CMEs in the corona and heliosphere in three dimensions, thereby improving our understanding of CME evolution and propagation. STEREO data have supported detailed comparison both of in-situ measurements with remote-sensing observations and of MHD heliospheric simulations with observations. The combination of high-cadence coronagraphic and EUV imaging simplifies the separation of the CME proper from its effects in the surrounding corona \cite{Patsourakos:2009aa} and allows a more accurate determination of its dynamics.

Despite the advances in our understanding enabled by SOHO and STEREO, very basic questions remain unanswered. These concern the source and initiation of eruptions, their early evolution, and the heliospheric propagation of CMEs. All current CME models predict that the topology of interplanetary coronal mass ejections (ICMEs) is that of a twisted flux rope as a result of the flare reconnection that occurs behind the ejection. Observations at 1\,AU, however, find that less than half of all ICMEs, even those associated with strong flares, have a flux rope structure (\opencite{Richardson:2004aa}; \opencite{Richardson:2010aa}; \opencite{Kilpua:2011aa}). Many ICMEs at 1\,AU appear to have a complex magnetic structure with no clearly-defined topology. Moreover, for ICMEs that do contain flux ropes, the orientation is often significantly different from that expected on the basis of the orientation of the magnetic fields in the prospective source region. CMEs are believed to originate from prominence eruptions, yet in ICMEs observed at 1\,AU prominence plasma is very rarely detected. These major disconnects between theoretical models of prominence eruption and CME propagation and observations need to be resolved if any understanding of the CME process is to be achieved.

\paragraph*{How Solar Orbiter will address the question.} To advance our understanding of the structure of ICMEs and its relation to CMEs at the Sun beyond what has been achieved with SOHO and STEREO requires a combination of remote-sensing and in-situ measurements made at close perihelion and in near-corotation with the Sun. Through combined observations with its magnetograph, imaging spectrograph, and soft X-ray imager, Solar Orbiter will provide the data required to establish the properties of CMEs at the Sun and to determine how coronal magnetic energy is released into CME kinetic energy, flare-associated thermal/non-thermal particle acceleration, and heating. Observations with the imaging spectrograph will be used to determine the composition of CMEs in the low corona and to establish how they expand and rotate and will also provide important clues to the energy partition within a CME once it is released. Solar Orbiter will make comprehensive in-situ measurements of the fields and plasmas (particularly composition) of ICMEs during the early phases of their propagation through the heliosphere. 

These measurements will allow the properties of an ICME to be related to those of the CME at the Sun and to the conditions in the CME source region as observed by Solar Orbiter's remote-sensing instruments and will make it possible to examine the evolution of CMEs in the inner heliosphere. Solar Orbiter's combination of remote-sensing and in-situ observations will also establish unambiguously the magnetic connectivity of the ICME and reveal how the magnetic energy within flux ropes is dissipated to heat and accelerate the associated particles. Solar Orbiter data will also reveal how the structure of the magnetic field at the front of a CME evolves in the inner heliosphere  --- a critical link in understanding, and eventually predicting, the geoeffective potential of transient events on the Sun.

To fully understand the physical system surrounding CME ejection, the temporal evolution of active regions and CME-related shocks and current sheets must be tracked from their formation in the corona to their expulsion in the solar wind. During the mission phases when the spacecraft is in near-corotation with the Sun, Solar Orbiter will continuously observe individual active regions, free from projection complications, over longer periods than are possible from Earth orbit. Solar Orbiter will thus be able to monitor the development of sheared magnetic fields and neutral lines and to trace the flux of magnetic energy into the corona. Observations from this vantage point will make it possible to follow the evolution of the current sheet behind a CME with unprecedented detail and to clarify the varying distribution of energy in different forms (heating, particle acceleration, kinetic).

\subsubsection{How do CMEs contribute to the solar magnetic flux and helicity balance?}
\label{CME_balance}
\begin{figure}   
   \centerline{\includegraphics[width=\textwidth,clip=true]{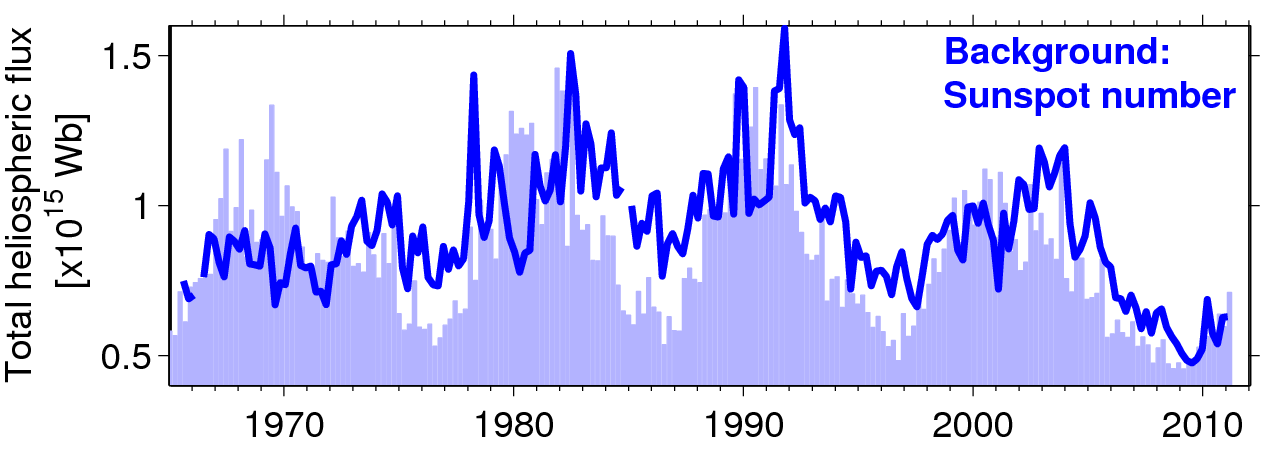}}
              \caption{Near-Earth interplanetary magnetic field strength (thick line) and sunspot number (background filled values) for the last 45 years. The magnetic field strength varies over the solar cycle, but was lower during the recent minimum than at any time since the beginning of the space age. The depth, as well as the length of this minimum was not predicted and is not understood. Solar Orbiter will investigate the evolving links between solar and interplanetary magnetic fields. (From M.\ Owens, University of Reading)
}
     \label{F-HeliosphericFlux_Jun11}
   \end{figure}

\paragraph*{Present state of knowledge.} Magnetic flux is transported into the heliosphere both by the solar wind, in the form of open flux carried mostly by the fast wind from polar coronal holes, and by coronal mass ejections, which drag closed flux with them as they propagate into the heliosphere. 
At some point the closed flux introduced by CMEs must be opened to avoid an unsustainable build-up of magnetic flux in the heliosphere. Measurements of the magnetic flux content of the heliosphere from near Earth, covering more than 40 years, show that the total amount of magnetic flux in the solar system changes over the solar cycle (\opencite{Owens:2008aa} and Figure~\ref{F-HeliosphericFlux_Jun11}). In addition, there is evidence that the heliospheric magnetic flux has increased substantially in the last hundred years, perhaps by as much as a factor of two \cite{Lockwood:1999aa,Rouillard:2007aa}, possibly due to a long-term change in the Sun's dynamo action. Surprisingly, however, during the recent solar minimum the IMF strength was lower than at any time since the beginning of the space age. 

The relative contribution of the solar wind and CMEs to the heliospheric magnetic flux budget is an unresolved question, as is the process by which the flux added by the CMEs is removed. Models to explain the solar cycle variation assume a background level of open flux, to which CMEs add extra flux during solar maximum, increasing the intensity of the IMF. The exceptionally low intensity of the IMF during the current minimum has been attributed to the low rate of CME occurrence \cite{Owens:2008aa}. Alternatively, there may simply be no 'background' open flux level. 

There is evidence that the flux introduced into the heliosphere by CMEs may be removed by magnetic reconnection within the trailing edges of CMEs, which disconnects the CME from the Sun or by interchange reconnection closer to the solar surface (e.g., \opencite{Owens:2006aa}). Recent observations show that reconnection processes occur quite often in the solar wind, even when the magnetic field is not being compressed. However, the rate and/or locations at which reconnection generally removes open flux are not at present known. 

Together with magnetic flux, the solar wind and CMEs carry magnetic helicity away from the sun. Helicity is a fundamental property of magnetic fields in plasmas, where it plays a special role because it is conserved not only by the ideal dynamics, but also during the relaxation which follows instabilities and dissipation. Helicity is injected into the corona when sunspots and active regions emerge, via the twisting and braiding of magnetic flux. During the coronal heating process the overall helicity is conserved and tends to accumulate at the largest possible scales. It is natural to assume that critical helicity thresholds may be involved in the triggering of CMEs, but how solar eruptions depend on the relative amounts of energy and helicity injection during active region emergence and evolution is unknown. Yet this understanding could be a crucial element in the prediction of large solar events.

\paragraph*{How Solar Orbiter will address the question.} Fundamental to the question of contribution of CMEs to the heliospheric flux budget is the flux content of individual events. Encountering CMEs close to the Sun before interplanetary dynamics affects their structure, Solar Orbiter will measure their magnetic flux content directly; comparisons with remote-sensing measurements of their source regions will clarify the relation between CME flux and the eruption process. As Solar Orbiter moves through the inner heliosphere, it will encounter CMEs at different solar distances, making it possible to quantify the effect of interplanetary dynamics on their apparent flux content.
The flux carried outwards by CMEs must eventually disconnect completely from the Sun, or undergo interchange reconnection with existing open field lines. Solar Orbiter will diagnose the magnetic connectivity of the solar wind and CME plasma using suprathermal electron and energetic particle measurements. These particles, which stream rapidly along the magnetic field from the Sun, indicate whether a magnetic flux tube is connected to the Sun at one end, at both ends, or not at all. Suprathermal particles disappear when the field is completely disconnected, or may reverse their flow direction as a result of interchange reconnection. However, scattering and reflection due to curved, tangled, or compressed magnetic field lines act to smear out these signatures with increasing solar distance, leading to ambiguity in connectivity measurements. Around perihelion, Solar Orbiter will be able to determine the original level of magnetic connectivity; covering a wide range of distances in the inner heliosphere, the spacecraft will measure how the connectivity changes as field lines are carried away from the Sun.

Solar Orbiter will also directly sample reconnection regions in the solar wind as they pass the spacecraft, determining their occurrence rates in the inner heliosphere as a function of distance and testing theories of CME disconnection by searching for reconnection signatures in the tails of CMEs.

The contribution to the heliospheric magnetic flux of small scale plasmoids, ejected from the tops of streamers following reconnection events, is unclear. Solar Orbiter, slowly moving above the solar surface during perihelion passes, will determine the magnetic structure, connectivity, and plasma properties including composition of these ejecta, using spectroscopic imaging observations to unambiguously link them to their source regions.
To assess the role of CMEs in maintaining the solar magnetic helicity balance, Solar Orbiter will compare the helicity content of active regions as determined from remote sensing of the photospheric magnetic field with that of magnetic clouds measured in situ. Such a comparison requires both extended remote-sensing observations of the same active region over the region's lifetime and in-situ measurements of magnetic clouds from a vantage point as close to the solar source as possible. Around its perihelia, Solar Orbiter will 'dwell' over particular active regions and observe the emergent flux for a longer interval (more than 22 days) than is possible from 1\,AU, where perspective effects complicate extended observations. The resulting data will be used to calculate the helicity content of an active region, track its temporal variation, and determine the change in helicity before and after the launch of any CMEs. Should a magnetic cloud result from an eruptive event in the active region over which Solar Orbiter is dwelling, the relatively small distance between the solar source and the spacecraft will make it probable that Solar Orbiter will directly encounter the magnetic cloud soon after its release. Determination of the cloud's properties and connectivity through Solar Orbiter's in-situ measurements will enable the comparison of a magnetic cloud in a relatively unevolved state with the properties of the solar source, an impossibility with measurements made at 1\,AU. The comparison of the helicity change in the source region with the value measured in the magnetic cloud will provide insight into the role of CMEs in the helicity balance of the Sun.

\subsubsection{How and where do shocks form in the corona and inner heliosphere? }
\paragraph*{Present state of knowledge.} The rapid expulsion of material during CMEs can drive shock waves in the corona and heliosphere. Shocks in the lower corona can also be driven by flares, and in the case of CME/eruptive flare events it may be difficult to unambiguously identify the driver \cite{Vrsnak:2008aa}. CME-driven shocks are of particular interest because of the central role they play in accelerating coronal and solar wind particles to very high energies in solar energetic particles (SEP) events (see Section~\ref{SEP}). 

Shocks form when the speed of the driver is super-Alfv{\'e}nic. The formation and evolution of shocks in corona and the inner heliosphere thus depend (1) on the speed of the driving CME and (2) on the Alfv{\'e}n speed of the ambient plasma and its spatial and temporal variations. According to one model of the radial distribution of the Alfv{\'e}n speed in the corona near active regions, for example, shocks can form essentially in two locations, in the middle corona ($1.2 - 3$\,R$_{\rm Sun}$), where there is an Alfv{\'e}n speed minimum, and distances beyond an Alfv{\'e}n speed maximum at $\sim 4$\,R$_{\rm Sun}$ \cite{Mann:2003aa}. A recent study of CMEs with and without type II radio bursts (indicative of shock formation) has shown that some of the fast and wide CMEs observed produced no shock or only a weak shock because they propagated through tenuous regions in the corona where the Alfv{\'e}n velocity exceeded that of the CME \cite{Gopalswamy:2008aa}. CME shock formation/evolution can also be affected by the interaction between an older, slower-moving CME and a faster CME that overtakes it. Depending on the Alfv{\'e}n speed in the former, the interaction may result in the strengthening or weakening of an existing shock driven by the overtaking CME or, if there is no existing shock, the formation of one \cite{Gopalswamy:2001aa,Gopalswamy:2002aa}. 

Recent studies of LASCO images obtained during the rising phase of solar cycle 23 have demonstrated the feasibility of detecting CME-driven shocks from a few to $\sim 20$\,R$_{\rm Sun}$ and of measuring their density compression ratio and propagation direction \cite{Vourlidas:2003aa,Ontiveros:2009aa}. This development has opened the way for the investigation of shock formation and evolution in the lower corona and heliosphere through Solar Orbiter's combination of remote-sensing observations and in-situ measurements. 

\paragraph*{How Solar Orbiter will address the question.}Understanding shock generation and evolution in the inner heliosphere requires knowledge of the spatial distribution and temporal variation of plasma parameters (density, temperature, and magnetic field) throughout the corona. Solar Orbiter's remote-sensing measurements  --- in particular electron density maps derived from the polarized visible-light images and maps of the density and outflow velocity of coronal hydrogen and helium  --- will provide much improved basic plasma models of the corona, so that the Alfv{\'e}n speed and magnetic field direction can be reconstructed over the distance range from the Sun to the spacecraft. Remote sensing will also provide observations of shock drivers, such as flares (location, intensity, thermal/non-thermal electron populations, time-profiles) and manifestations of CMEs (waves, dimmings, etc.) in the low corona with spatial resolution of a few hundred kilometres and cadence of a few seconds. It will measure the acceleration profile of the latter and then track the CMEs through the crucial heights for shock formation ($2 - 10$\,R$_{\rm Sun}$) and provide speed, acceleration, and shock compression ratio measurements.

Type II bursts, detected by Solar Orbiter, will indicate shock-accelerated electron beams produced by the passage of a CME and thus provide warning of an approaching shock to the in-situ instruments. These in-situ plasma and magnetic field measurements will fully characterize the upstream and downstream plasma and magnetic field properties and quantify their microphysical properties, such as turbulence levels and transient electric fields (while also directly measuring any solar energetic particles (SEPs) --- cf.\ Section~\ref{SEP}). Spacecraft potential measurements also allow for rapid determinations of the plasma density, and of electric and magnetic field fluctuations, on microphysical scales, comparable to the Doppler-shifted ion scales, which are characteristic of the spatial scales of shocks. The evolution of such parameters will provide insight into the processes dissipating shock fronts throughout the parameter space. Because of Solar Orbiter's close proximity to the Sun, the measurements of the solar wind plasma, electric field, and magnetic field will be unspoiled by the dynamical wind interaction pressure effects due to solar rotation and will provide the first reliable data on the magnetosonic speed, the spatial variation of the plasma pressure and magnetic field in the inner heliosphere. A recent MHD modelling study has shown that interactions among recurring CMEs and their shocks occur typically in the distance range around $0.2 - 0.5$\,AU \cite{Lugaz:2005aa}. Solar Orbiter will spend significant time in the regions of recurring CME interactions and so will be able to investigate the effects of such interactions on the evolution of CME-driven shocks.

\subsection{How do solar eruptions produce energetic particle radiation that fills the heliosphere?}
\label{SEP}
Astrophysical sites that have the ability to accelerate ions and electrons to high speeds, forming energetic particle radiation, exist throughout the solar system and beyond. Detected remotely from radio and light emission around supernovae remnants, the Sun and planets, or directly from particles that reach our detectors, this radiation arises from the explosive release of stored energy that can cause magnetic fields to rearrange, or can launch shock waves which accelerate particles by repeatedly imparting many small boosts to their speed. The nearly universal occurrence of energetic particle radiation, along with the effects it can have on planetary environments, evolution of life forms, and space systems has fostered a broad interest in this phenomenon that has long made it a high priority area of investigation in space science. Since remote sites in the galaxy cannot be studied directly, solar system sources of energetic particles give the best opportunity for studying all aspects of this complex problem. 

The Sun is the most powerful particle accelerator in the solar system, routinely producing energetic particle radiation at speeds close to the speed of light, sufficiently energetic to be detected at ground level on Earth even under the protection of our magnetic field and atmosphere. SEP events can severely affect space hardware, disrupt radio communications, and cause re-routing of commercial air traffic away from polar regions. In addition to large events, which occur roughly monthly during periods of high sunspot count, more numerous, smaller solar events can occur by the thousands each year, providing multiple opportunities to understand the physical processes involved. Below we discuss in more detail three interrelated questions that flow down from this top-level question: How and where are energetic particles accelerated at the Sun? How are energetic particles released from their sources and distributed in space and time? What are the seed populations for energetic particles?

\subsubsection{How and where are energetic particles accelerated at the Sun?}
\begin{figure}   
   \centerline{\includegraphics[width=\textwidth,clip=true]{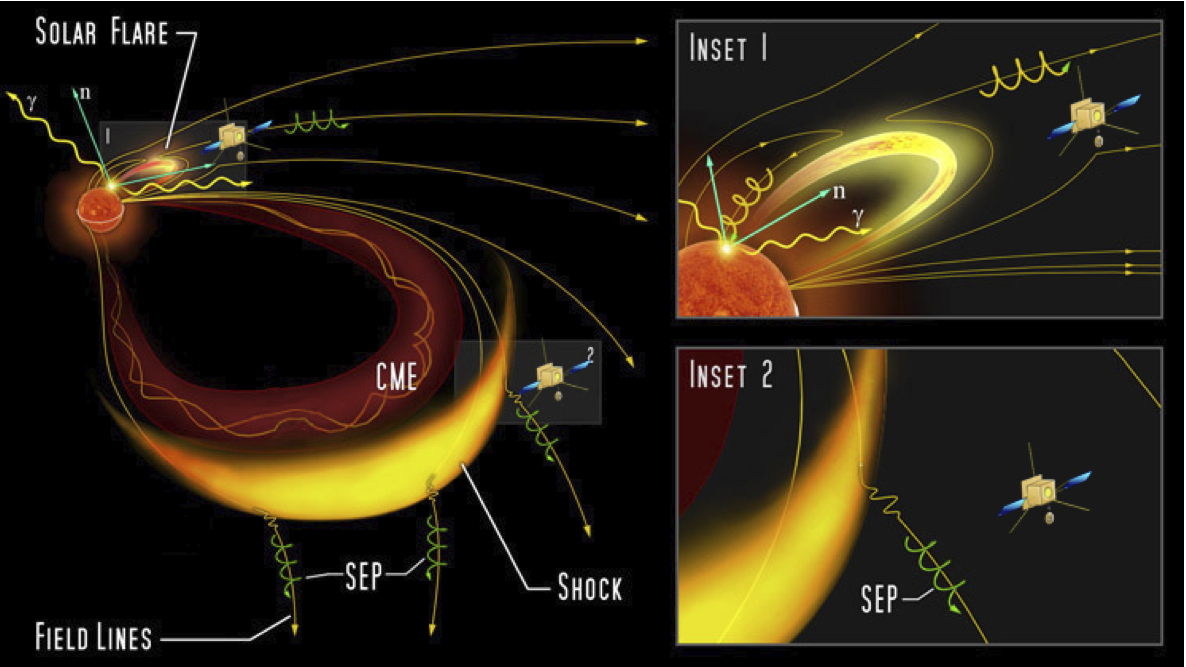}}
              \caption{Sketch showing a solar flare and CME driving an interplanetary shock. Both the flare source and shock may contribute to the interplanetary energetic particle populations. However, the relative importance of acceleration processes due to flares and CME-driven shocks cannot be determined at 1\,AU because of particle mixing. Solar Orbiter will allow tests of the relative importance of the different acceleration mechanisms around perihelion. There, the shock will pass over Solar Orbiter while still in the early phases of particle acceleration, making it possible to directly compare the energetic particles with shock properties such as mach number, turbulence level, and with the local seed population. Simultaneous in-situ observations of magnetic field lines connecting back to flare sites and to shock fronts driven by CMEs, together with concurrent remote imaging of flares, wide field-of-view imaging of CMEs and spectroscopic identification of the CME-driven shocks from Solar Orbiter, will help to determine the relative importance of the associated acceleration processes. (Adapted from NASA's Solar Sentinels STDT report)}
     \label{F-SEP_gfx}
   \end{figure}

\begin{figure}   
   \centerline{\includegraphics[width=\textwidth,clip=true]{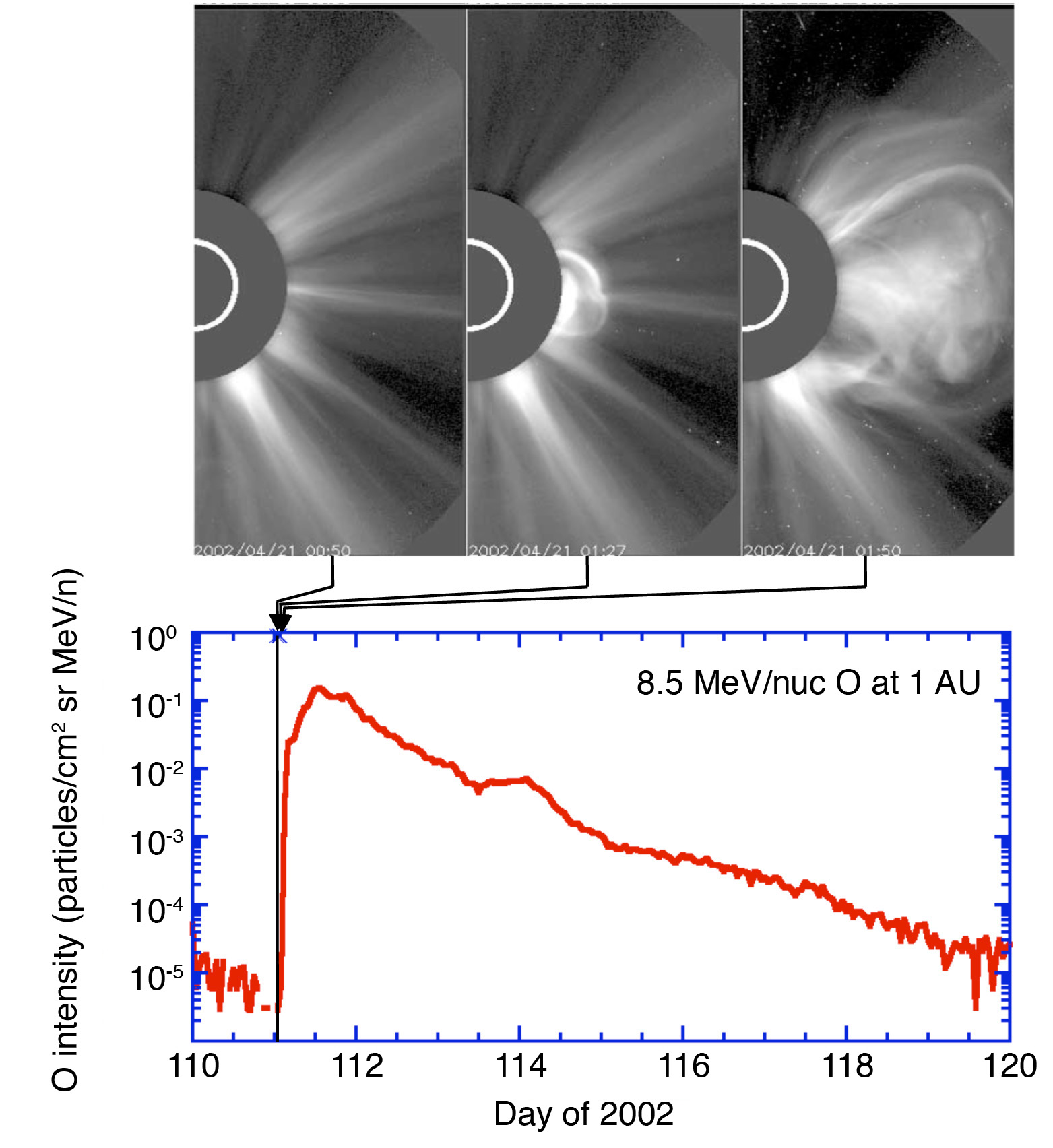}}
              \caption{Upper panel: SOHO LASCO observations of a CME erupting from the SunÕs western hemisphere, with exposure times at $00:50$, $01:27$ and $01:50$\,UT. The CME reached a speed of 2700\,km/s at 18\,R$_{\rm Sun}$, and the associated interplanetary shock passed Earth around $\sim 04:15$\,UT on 23 April 2002, about  51\,hours after the lift-off. Lower panel: ACE observations of high energy SEP O nuclei showing an increase in intensity of nearly 5 orders of magnitude beginning shortly after the CME lift-off. Note that, while the CME photos are all taken near the intensity onset, the ACE intensities remained elevated for days, long after the shock had passed the Earth. Solar Orbiter will provide much more accurate timing and particle distribution measurements because of its much shorter magnetic connection to the acceleration sites. This will pinpoint the acceleration mechanisms and determine the importance of interplanetary transport processes. (Adapted from \protect\opencite{Emslie:2004aa})
}
     \label{F-CME+Oxygen_in-situ}
   \end{figure}

\paragraph*{Present state of knowledge.} One of the two major physical mechanisms for energizing particles involves particles interacting with moving or turbulent magnetic fields, gaining small amounts of energy at each step and eventually reaching high energies. Called Fermi or stochastic acceleration, this mechanism is believed to operate in shock waves and in turbulent regions such as those associated with reconnecting magnetic fields or in heated coronal loops. The second major physical mechanism is a magnetic field whose strength or configuration changes in time, producing an electric field which can directly accelerate particles in a single step. At the Sun, such changes occur when large magnetic loop structures reconnect, or are explosively rearranged due to the stress from the motion of their footpoints on the solar surface (e.g., \opencite{Aschwanden:2006aa}; \opencite{Giacalone:2006aa}). Multiple processes may take place in SEP events, and while it is not possible to cleanly separate them, they can be split into two broad classes, the first being events associated with shock waves. Figure~\ref{F-SEP_gfx} shows a sketch wherein an instability in coronal magnetic loops has resulted in an eruption that launches a CME. As it moves into space, it drives a shock creating turbulence that accelerates SEPs from a seed population of ions filling the interplanetary medium (inset 2). Mixed into this may be particles from an associated solar flare (inset 1). CMEs often accelerate particles for hours as they move away from the Sun, and in some cases are still accelerating particles when they pass Earth orbit in a day or two (Figure~\ref{F-CME+Oxygen_in-situ}). Since CMEs can be huge as shown in the LASCO image, it is easy to see how they can fill a large portion of the heliosphere with SEPs. However, the correlation of the observed radiation intensities with CME properties is poor, indicating that additional aspects of the mechanism such as seed populations or shock geometry must play important roles that are not yet well understood \cite{Gopalswamy:2006aa,Desai:2006aa,Mewaldt:2006aa}.

\begin{figure}   
   \centerline{\includegraphics[width=\textwidth,clip=true]{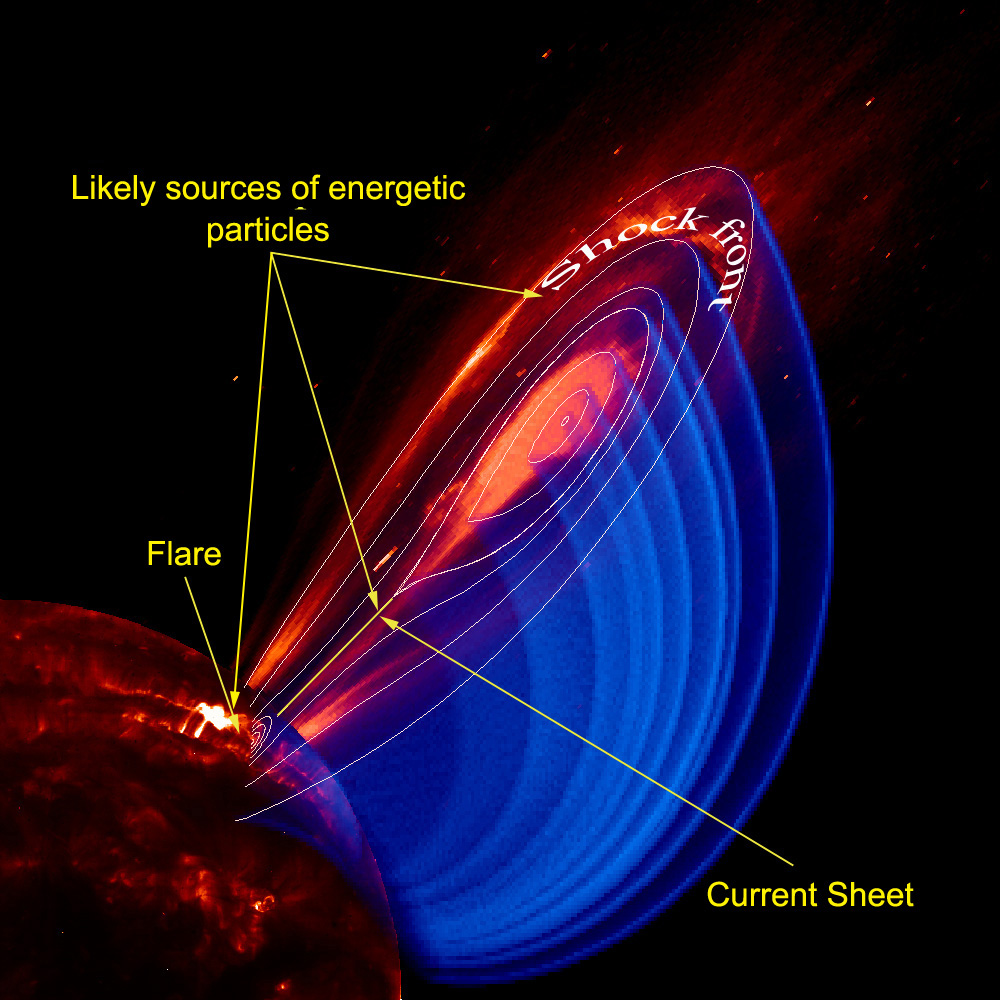}}
              \caption{Composite illustration of a unified flare/CME system showing potential solar energetic particle source regions. The coronagraph image (red image off the limb) shows the CME with a trailing current sheet seen nearly head-on. A cutaway of the modelled magnetic field structure is shown by the blue overlap. Post flare loops are shown on the UV disk image. By going closer to the Sun, Solar Orbiter will help to distinguish between the timing and release signatures from the shock front vs.\ the connection region at the current sheet. (From NASA's Solar Sentinels STDT report)
}
     \label{F-Flare+CME_composite}
   \end{figure}

The second class of events is associated with plasma and magnetic field processes in loops and active regions that accelerate particles. Reconnecting magnetic loops, and emerging magnetic flux regions provide sites for stochastic energetic particle acceleration or acceleration by electric fields. Because these regions are relatively small, the acceleration process is quick: on the order of seconds or minutes, but the resulting event is small and often difficult to observe. Since the energized particles are in the relatively high-density regions of the corona, they collide with coronal plasma, producing ultraviolet (UV) and X-ray signatures that make it possible to locate their acceleration sites and probe the local plasma density. Most of these particles remain trapped on their parent loops, travelling down the legs to the solar surface where they lose their energy to ambient material, producing X- and gamma-rays. A few escape on magnetic field lines leading to interplanetary space, traceable by their ('type III') radio signatures, electrons, and highly fractionated ion abundances where the rare $^3$He can be enhanced by $1000 - 10,000$ times more than in solar material. Figure~\ref{F-Flare+CME_composite} illustrates another site where reconnection can accelerate particles: in the current sheet behind a CME lift-off. In this case, particles can be accelerated for hours, and may 'leak' around the CME structure and become mixed with the shock accelerated particles \cite{Lin:2006aa,Cargill:2006aa,Drake:2009aa}.

The energetic particles from these events reach our detectors at Earth orbit after spiralling around the IMF, which is an Archimedes spiral on average. But since the IMF meanders and has many kinks, the length of the particle's path has a high degree of uncertainty, and the particles themselves scatter and mix, smearing and blurring signatures of the acceleration at the Sun. Although we can enumerate candidate mechanisms for producing SEPs, a critical question is: what actually happens in nature? Which processes dominate? How can shocks form fast enough to accelerate ions and electrons to relativistic energies in a matter of minutes, as happened in the January 20, 2005 SEP event? 

\paragraph*{How Solar Orbiter will address the question.}Solar Orbiter will make progress on the origins of SEPs by enabling precise determination of the sequences of events, along with comprehensive in-situ determination of the field and plasma properties and the suprathermal ion pool in the inner heliosphere. Recent progress at 1\,AU has relied on combining remote and in-situ observations from different missions such as ACE, SOHO, WIND, RHESSI, TRACE, Hinode, and SDO  --- where using multiple spacecraft is possible since they are all at virtually the same vantage point. But to do this close to the Sun, where there is an enormous observational advantage due to proximity, it is necessary to carry the whole suite on one spacecraft, since the probe's trajectory is not in synchronization with Earth. Almost the entire Solar Orbiter payload contributes to unravelling the question of SEP origins: visible, UV, and X-ray imaging of loops, flares, and CMEs with their location and timing; X-ray signatures of energetic particle interactions at loop footpoints, or in loops themselves; radio signatures of coronal shocks and escaping electrons; magnetic field, plasma wave and solar wind measurements to determine turbulence levels and identify shock passages; seed population specification from the heavy ion composition of solar wind and suprathermals in the inner heliosphere; finally, the accelerated energetic particles themselves: their timing, velocity distributions, scattering characteristics, and composition. 

\vspace{3mm}
\noindent
{\it(a) CME and shock associated SEPs.} Moving from the lower corona to the interplanetary medium, shocks evolve rapidly since the sound speed drops as plasma density and magnetic field strength decline as $\sim 1/r^2$. Solar Orbiter's coronagraphs will remotely identify shock front location, speed, and compression ratios through this critical region within $\sim 10$\,R$_{\rm Sun}$. Combining this information with local electron densities as well as coronal ion velocities given by Solar Orbiter radio and light polarization observations will provide critical constraints on shock evolution models in regions too close to the Sun for direct sampling. 

In the regions explored by Solar Orbiter close to the Sun, the IMF is almost radial with much less variation (uncertainty) in length than is the case at 1\,AU, so the knowledge of the actual path length improves by a factor of $3 - 5$ as the length shortens. Having observed the CMEs and their radio signatures in the corona and the X-ray signatures of the energetic particles near the Sun, Solar Orbiter will then determine subsequent arrival time of the particles in situ that can be accurately compared to CME position. As the shock then rolls past the spacecraft, Solar Orbiter will measure the shock speed and strength as well as the associated plasma turbulence, electric, and magnetic field fluctuations. This will give a complete description of the acceleration parameters in the inner heliosphere where much of the particle acceleration takes place. Indirect evidence from 1\,AU indicates that shock acceleration properties depend on the longitude of the shock compared to the observer; close to the Sun, Solar Orbiter can cleanly test this property since the IMF is nearly radial, the CME lift-off site is known, and the accelerated particles will have little chance to mix. In the high-latitude phase of the mission, Solar Orbiter will be able to look down on the longitudinal extent of CMEs in visible, UV, and hard X-rays, allowing first direct observations of the longitudinal size of the acceleration region. This will make it possible to test currently unconstrained acceleration and transport models by using measured CME size, speed, and shape to specify the accelerating shock.

\vspace{3mm}
\noindent
{\it(b) SEPs associated with coronal loops and reconnection regions.} As Solar Orbiter approaches the Sun, the photon and particle signatures from small events will increase by $1/r^2$, making it possible to observe events $15 - 20$ times smaller than ever before, in effect opening a new window for SEP processes. We may detect for the first time energetic particle populations from X-ray microflares, a candidate mechanism for coronal heating that cannot be studied further away from the Sun due to background problems. For the small flares that produce X-ray, electron, and $^3$He-enrichments we will observe with great accuracy events that at 1\,AU are not far above the level of detection: the timing of particle and radio signatures, the composition and spectra, etc., providing strong new constraints on the process operating in these events.  New insight into particle acceleration along coronal loops will be obtained since the $1/r^2$ sensitivity advantage and viewing geometry will make it possible to view the X-ray emission from the tops of loops in numerous cases where the much stronger footpoint sources are occulted behind the solar limb. These studies of faint coronal sources that are only rarely observable from 1\,AU will give crucial information about the location and plasma properties of suspected electron acceleration sites in the high corona. Furthermore, considering recent works of tracking of CMEs with heliospheric imagers (see \opencite{Harrison:2009aa} and references therein), it is likely that Solar Orbiter's heliospheric imager will be able to further constrain CME speed, shape and size and, in turn, SEP acceleration by imaging CMEs after they leave the coronagraph's field-of-view.

\vspace{3mm}
\noindent
\subsubsection{How are energetic particles released from their sources and distributed in space and time?}
\paragraph*{Present state of knowledge.} SEPs associated with CME driven shocks have been long known to often arrive at Earth orbit much later than would be expected based on their velocities (\opencite{Van-Hollebeke:1975aa}; \opencite{Tylka:2006aa}). There are two alternate processes that might cause this. (1) The acceleration may require significant time to energize the particles since they must repeatedly collide with the shock to gain energy in many small steps, so the process may continue for many hours as the shock moves well into the inner solar system. Or (2) the particle intensities near the shock may create strong turbulence that traps the particles in the vicinity of the shock, and their intensity observed at Earth orbit depends on the physics of the particles escaping from the trapping region. Once free of the vicinity of the shock, SEPs may spiral relatively freely on their way to Earth orbit, or more usually will be scattered repeatedly from kinks in the IMF, delaying their arrival further. The amount of scattering in the interplanetary space varies depending on other activity such as recent passage of other shocks or solar wind stream interactions. By the time the particles reach Earth orbit, they are so thoroughly mixed that these effects cannot be untangled \cite{Gopalswamy:2006aa,Cohen:2007aa}.

\begin{figure}   
   \centerline{\includegraphics[width=\textwidth,clip=true]{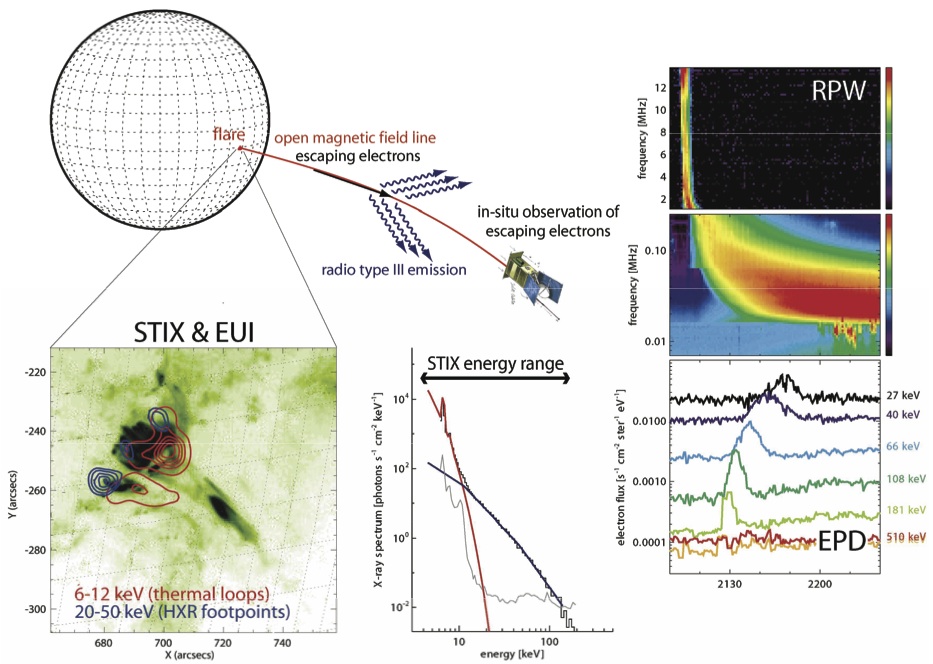}}
              \caption{Coordinated remote-sensing and in-situ observations of a flare source (lower left) producing a jet seen in UV and X-rays, which outline the loops and interactions at loop footpoints (blue). Escaping electrons produce a radio burst (upper right) whose frequency depends on coronal height of the emitting particles. At the Solar Orbiter spacecraft, the detection of energetic electrons times the arrival of escaping particles, and energetic ions provide signatures of extreme fractionation produced by the acceleration mechanism. The prompt arrival of the particles establishes that Solar Orbiter is magnetically connected to the X-ray source, allowing comparison with coronal magnetic field models in the region of the active region. (Adapted from A.\ Benz, $3^{\rm rd}$ Solar Orbiter Workshop, Sorrento) 
}
   \label{F-Benz_coord_obs_Sorrento}
   \end{figure}

Particles accelerated along magnetic loops can reach very high energies in seconds after the onset of flaring activity, and then collide with the solar surface where they emit gamma radiation. There is a poor correlation between the intensity of the gamma radiation and the SEP intensities observed at Earth orbit, so most particles from this powerful acceleration process do not escape. Much more common are flare events observed in UV and X-rays that produce sudden acceleration of electrons, sketched in Figure~\ref{F-Benz_coord_obs_Sorrento}. The electrons can escape from the corona, producing nonthermal radio emission as they interact with the local plasma. Moving from higher to lower frequencies as the local plasma density decreases with altitude, the (type III) radio emission makes it possible to track the energetic electron burst into interplanetary space where it may pass by the observer. Energetic ions, greatly enriched in $^3$He and heavy nuclei, accompany these electron bursts \cite{Lin:2006aa,Mason:2007aa}. Key open questions in shock associated events are whether particle arrival delays at 1\,AU are due to the length of time needed to accelerate the particles, or due to trapping in the turbulence near an accelerating shock, or a combination of both? For particles accelerated along loops, are the electrons and ions accelerated from sites low in the corona or at higher altitudes, and how are they related to the X- and gamma-ray signatures?

\paragraph*{How Solar Orbiter will address the question.} Solar Orbiter will advance our understanding of SEP acceleration associated with CME driven shocks by probing the inner heliospheric sites where particle acceleration and release take place. Solar Orbiter will observe how shocks evolve, and whether they are still accelerating particles as they pass by the spacecraft. If particle arrivals are controlled by the time it takes the shock to accelerate them, then the highest energy particles will be delayed since they require many more interactions with the shock. If trapping and release controls the timing, then the faster and slower particles will have similar intensity changes as the shock moves by. Since Solar Orbiter will simultaneously measure the turbulence properties in the shock acceleration region, it will be possible to construct a complete model of the acceleration process.

For SEPs accelerated along loops or in reconnection regions, Solar Orbiter will see the coronal location from UV and X-rays, and then trace the progress of released electrons by radio emission that will drift to the plasma frequency at the spacecraft for those bursts that pass by. This unambiguously establishes that the magnetic field line at Solar Orbiter connects to the coronal UV and X-ray emission site. Since Solar Orbiter can be connected to active regions for periods of days, this will provide multiple tracings between the heliospheric magnetic field and its origin in the corona. X-ray emission from the flaring sites can be used to derive the energetic electron spectrum at the flare site, which in turn can be compared with the escaping population to see if most of the accelerated electrons are released (usually most do not escape). Thanks to the $1/r^2$ intensity advantage, Solar Orbiter will observe thousands of these cases and thereby permit detailed mapping of coronal sources and the trapping properties of the acceleration sites. 

\subsubsection{What are the seed populations for energetic particles? }
\begin{figure}   
   \centerline{\includegraphics[width=\textwidth,clip=true]{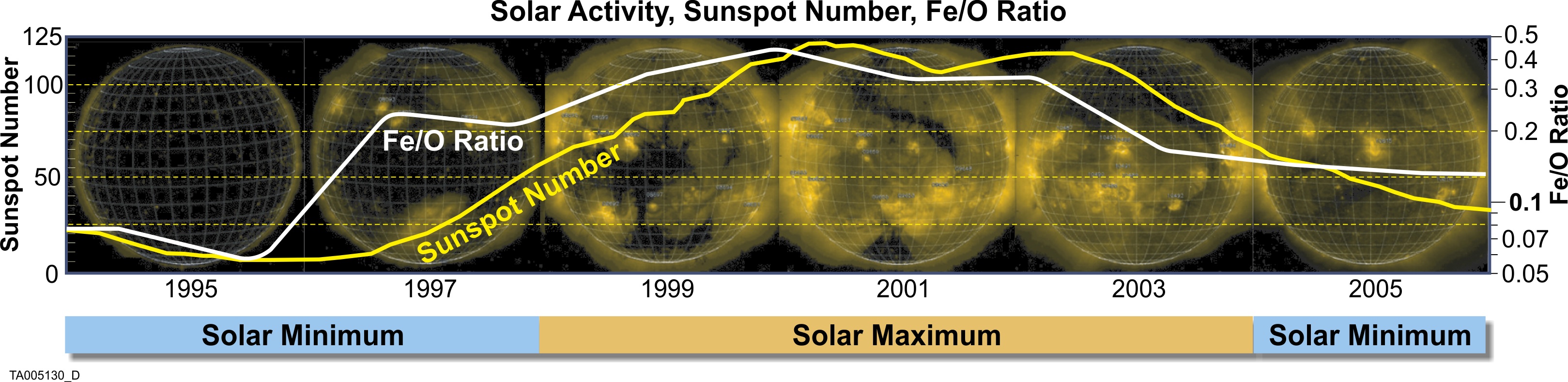}}
              \caption{The seed population for SEP shock accelerated particles is the suprathermal ion pool. At 1\,AU this pool has large variations over the solar cycle as shown above for the Fe/O ratio: the average pool composition goes from domination by flare activity at solar maximum (high Fe/O), to interplanetary sources at solar minimum (low Fe/O). Day-to-day variations in the poolÕs composition can be larger than shown here because of other transient activity such as interplanetary shock passages. Inside 1\,AU, where almost all shock acceleration of particles takes place, the suprathermal ion pool is completely unexplored, and will be mapped for the first time by Solar Orbiter. (From M.\,Desai, $2^{\rm nd}$ Solar Orbiter Workshop, Athens)
}
     \label{F-Desai_solar_cycle}
   \end{figure}

\paragraph*{Present state of knowledge.} The low-energy particles accelerated by CME-driven shocks to SEP energies are called the seed population. The observed ionization states of SEP ions show temperatures typical of the corona, ruling out hot material on flare loops as the seeds. But SEPs also show significant abundances of ions such as $^3$He and singly ionized He, which are virtually absent from the solar wind. The observed energetic particle abundances indicate that the suprathermal ion pool, composed of ions from a few times the speed of the solar wind to a few tens of it, is the likely source. At 1\,AU the suprathermal ion pool is $\sim100$ times more variable in intensity than the solar wind, and varies in composition depending on solar and interplanetary activity. The suprathermal ions are continuously present at 1\,AU (Figure~\ref{F-Desai_solar_cycle}), but it is not known if there is a continuous solar source, or if these ions are from other activities such as acceleration in association with fast- and slow-solar wind streams. Inside 1\,AU, the suprathermal ion pool is expected to show significant radial dependence due to the different processes that contribute to the mixture, but it is unexplored \cite{Desai:2006aa,Mewaldt:2007aa,Lee:2007aa,Fisk:2007aa}.
For SEPs accelerated along loops or in reconnection regions that give rise to electron and type-III radio bursts, ionization states are coronal-like at lower energies and change over to much hotter flare-like at high energies.  This may be evidence for a complex source, or, more likely, of energetic particle stripping as the ions escape from a low coronal source. For SEPs accelerated at reconnection sites behind CMEs (Figure~\ref{F-Flare+CME_composite}) abundances and ionization states would be coronal \cite{Klecker:2006aa}.

Critical questions in this area are: what is the suprathermal ion pool in the inner heliosphere, including its composition and temporal and spatial variations? What turbulence or stochastic mechanisms in the inner heliosphere accelerate particles to suprathermal energies? 

\paragraph*{How Solar Orbiter will address the question.} By systematically mapping the suprathermal ion pool in the inner heliosphere with spectroscopic and in-situ data, Solar Orbiter will provide the missing seed particle data for models of SEP acceleration associated with shocks. Since the suprathermal ion pool composition varies, different shock events will be expected to produce correspondingly different energetic particle populations that can be examined on a case-by-base basis. The high-latitude phase of the mission will add an important third dimension to the suprathermal pool mapping, since it will be more heavily influenced by, e.g., mid-latitude streamer belts, making it possible to probe the solar and interplanetary origins of the seed particle populations. Taken together, these observations will make it possible to significantly advance models of particle acceleration close to the Sun. 

For SEPs accelerated on loops or in reconnection regions the $1/r^2$ advantage of Solar Orbiter will again provide a decisive advantage since particle properties will be accurately measured and comparable with much more precise information on the coronal location. This will permit distinguishing between low coronal sources that result in stripping of escaping particles vs.\ higher sources which could mimic stripping properties. SEPs accelerated from reconnection regions behind CME lift-offs will be identified by comparing energetic particle timing with the location of the CME, and energetic particle composition with that determined spectroscopically for the remote coronal source. 

\subsection{How does the solar dynamo work and drive connections between the Sun and the heliosphere?}

The Sun's magnetic field dominates the solar atmosphere. It structures the coronal plasma, drives much of the coronal dynamics, and produces all the observed energetic phenomena. One of the most striking features of solar magnetism is its $\sim11$-year activity cycle, which is manifest in all the associated solar and heliospheric phenomena. Similar activity cycles are also observed in a broad range of stars in the right half of the Hertzsprung-Russell diagram, and the Sun is an important test case for dynamo models of stellar activity. 

The Sun's global magnetic field is generated by a dynamo generally believed to be seated in the tachocline, the shear layer at the base of the convection zone. According to flux-transport dynamo models (e.g., \opencite{Dikpati:2008aa}), meridional circulation, and other near-surface flows transport magnetic flux from decaying active regions to the poles. There subduction carries it to the tachocline to be reprocessed for the next cycle. This 'conveyor belt' scenario provides a natural explanation for the sunspot cycle, and characterizing the flows that drive it will provide a crucial test of our models and may also allow us to predict the length and amplitude of future cycles. However, current models fail miserably at predicting actual global solar behaviour. For example, the current sunspot minimum has been far deeper and longer than predicted by any solar modelling group, indicating that crucial elements are missing from current understanding.

A major weakness of current global dynamo models is poor constraint of the meridional circulation at high latitudes. The exact profile and nature of the turnover from poleward flow to subduction strongly affects behaviour of the resulting global dynamo (e.g., \opencite{Dikpati:1999aa}), but detecting and characterizing the solar flow is essentially impossible at shallow viewing angles in the ecliptic plane. In addition to the global dynamo, turbulent convection may drive a local dynamo that could be responsible for generating the observed weak, small-scale internetwork field, which is ubiquitous across the surface and appears to dominate the emergent unsigned flux there. 

A key objective of the Solar Orbiter mission is to measure and characterize the flows that transport the solar magnetic fields: complex near-surface flows, the meridional flow, and the differential rotation at all latitudes. Of particular and perhaps paramount importance for advancing our understanding of the solar dynamo and the polarity reversal of the global magnetic field is a detailed knowledge of magnetic flux transport near the poles. Hinode, peering over the Sun's limb from a heliographic latitude of 7$^{\circ}$, has provided a tantalizing glimpse of the Sun's high-latitude region above 70$^{\circ}$; however, observations from near the ecliptic lack the detail, coverage, and unambiguous interpretation needed to understand the properties and dynamics of the polar region. Thus, Solar Orbiter's imaging of the properties and dynamics of the polar region during the out-of-the-ecliptic phase of the mission (reaching heliographic latitudes of 25$^{\circ}$ during the nominal mission and as high as 34$^{\circ}$ during the extended mission) will provided urgently needed constraints on models of the solar dynamo.

Most of the open magnetic flux that extends into the heliosphere originates from the Sun's polar regions, from polar coronal holes. The current solar minimum activity period, which is deeper and more extended than previously measured minima, demonstrates the importance of this polar field to the solar wind and heliosphere. There is evidence that the solar wind dynamic pressure, composition and turbulence levels, as well as the strength of the heliospheric magnetic field, have all changed in the last few years in ways that are unprecedented in the space age. None of these changes were predicted, and current solar conditions present a challenge to our understanding of the solar dynamo and its effects on the solar system at large and the Earth in particular. 

Below we discuss in more detail three interrelated questions that flow down from this top-level question: How is magnetic flux transported to and reprocessed at high solar latitudes? What are the properties of the magnetic field at high solar latitudes? Are there separate dynamo processes in the Sun?

\subsubsection{How is magnetic flux transported to and reprocessed at high solar latitudes?}
\paragraph*{Present state of knowledge.} In the last decade, the mapping of surface and subsurface flow fields at low and middle latitudes has seen major advances, largely due to the availability of high-quality data from the SOHO's Michelson Doppler Imager (MDI) instrument and, more recently, SDO's Helioseismic and Magnetic Imager (HMI). These data have provided accurate knowledge of differential rotation, the low latitude, near-surface part of the meridional flows, and the near-surface torsional oscillations, which are rhythmic changes in the rotation speed that travel from mid-latitudes both equatorward and poleward \cite{Howe:2006aa}. Local helioseismic techniques have also reached a level of maturity that allows the three-dimensional structure of the shallow velocity field beneath the solar surface to be determined.

\begin{figure}   
   \centerline{\includegraphics[width=\textwidth,clip=true]{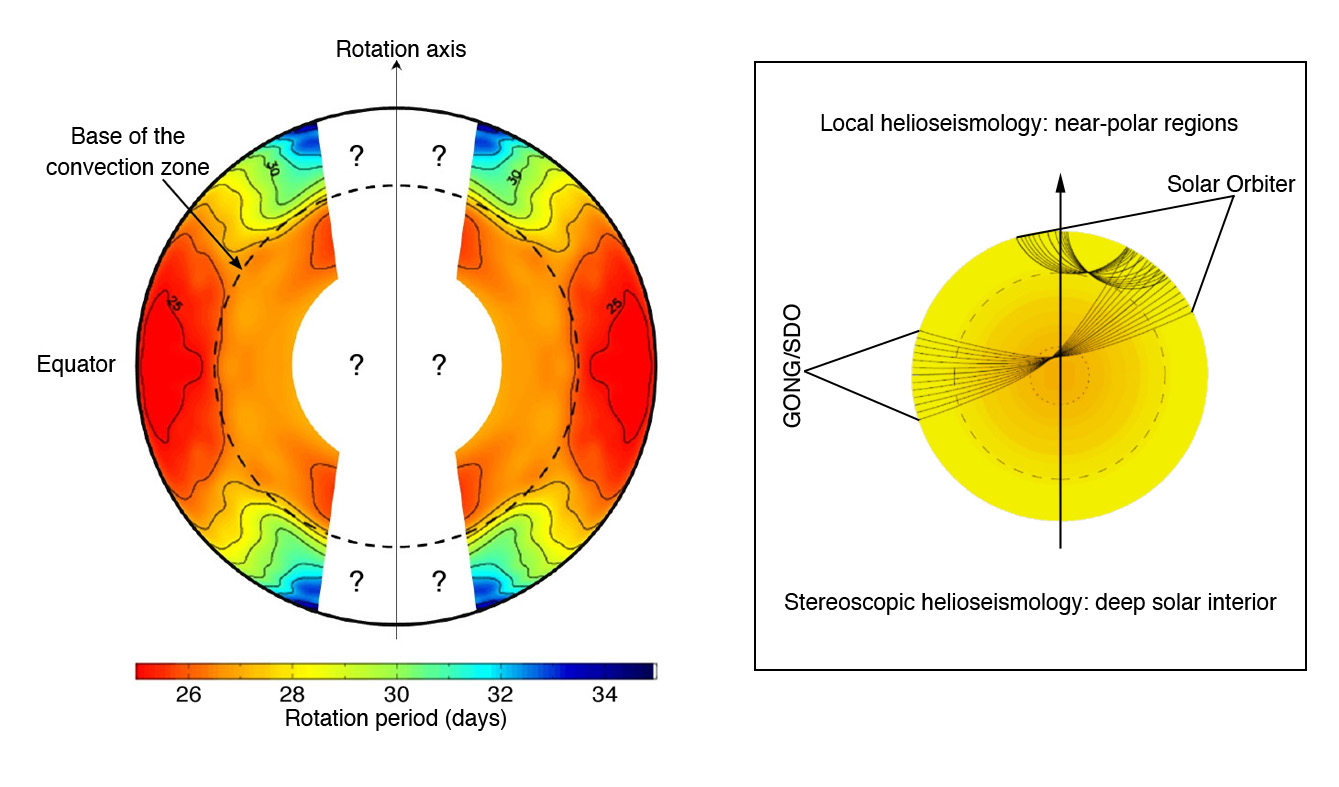}}
              \caption{Left: Rotation profile in the solar interior as derived from GONG and MDI data. By traveling to high latitudes, Solar Orbiter will use local helioseismology to determine the currently unknown properties of the solar interior below the poles \protect\cite{Corbard:1998aa}. Right: Solar OrbiterÕs helioseismology capabilities: (i) probing near polar regions with local helioseismology and (ii) probing the deep solar interior with stereoscopic helioseismology in combination with near-Earth observations, e.g.\ from GONG or SDO. (From \protect\opencite{Roth:2007aa})}
     \label{F-Helioseism_mroth}
   \end{figure}

Despite these advances, progress in understanding the operation of the solar dynamo depends on how well we understand differential rotation and the meridional flows near the poles of the Sun. However, because of the lack of out-of-the ecliptic observations, the near-polar flow fields remain poorly mapped, as does the differential rotation at high latitudes (see \opencite{Beck:2000aa,Thompson:2003aa} and Figure~\ref{F-Helioseism_mroth}). The meridional flow in particular, the very foundation of the flux transport dynamo, is not well characterized above $\sim 50^{\circ}$ latitude; it is not even certain that it consists of only one cell in each hemisphere. The return flow, believed to occur at the base of the convection zone, is entirely undetermined save for the requirement of mass conservation. All these flows must be better constrained observationally in order to help solve the puzzle of the solar cycle and to advance our understanding of the operation of the solar dynamo (and, more broadly, of stellar dynamos in general).

\paragraph*{How Solar Orbiter will address this question.} Solar Orbiter will measure or infer local and convective flows, rotation, and meridional circulation in the photosphere and in the subsurface convection zone at all heliographic latitudes including, during the later stages of the mission, at the critical near-polar latitudes. Solar Orbiter might be able to reveal patterns of differential rotation, the geometry of the meridional flow, the structure of subduction areas around the poles where the solar plasma dives back into the Sun, and the properties of convection cells below the solar surface. This will be achieved through correlation tracking of small features, direct imaging of Doppler shifts, and helioseismic observations. 

\begin{figure}   
   \centerline{\includegraphics[width=\textwidth,clip=true]{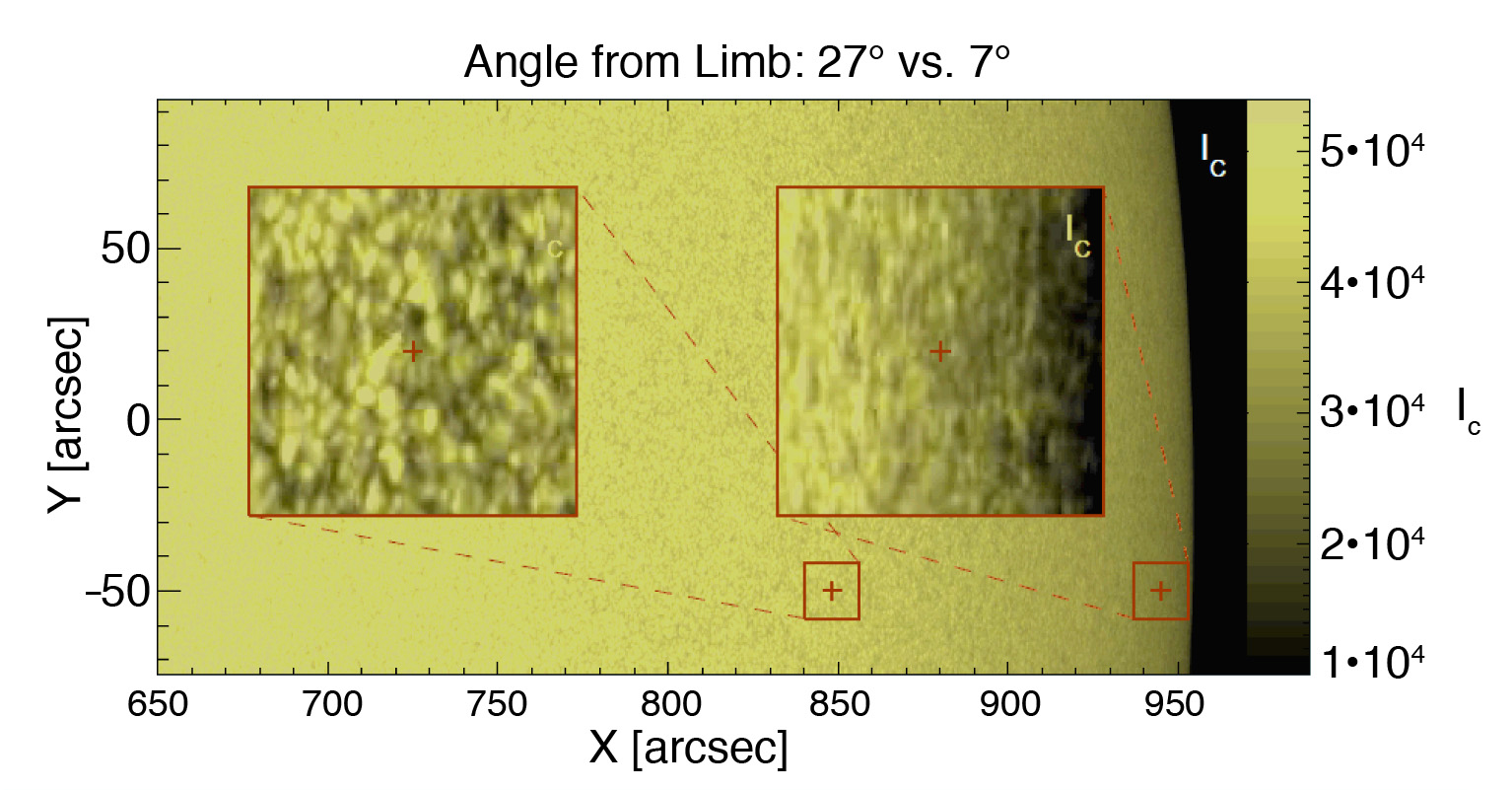}}
              \caption{Comparison of solar granulation at the poles as viewed from 27$^\circ$, where the fine scale structure can be resolved with much higher fidelity than from (7$^\circ$) inclination, obtained in the ecliptic plane twice a year. Hinode observations of the latter were used to obtain the first polar map of the vector magnetic field map \protect\cite{Tsuneta:2008aa}. Solar Orbiter will characterize the properties and dynamics of the polar regions in detail, including magnetic fields, plasma flows, and temperatures.
}
     \label{F-Granulation_angle}
   \end{figure}

Solar Orbiter will resolve small-scale magnetic features near the poles, even within the nominal mission phase (Figure~\ref{F-Granulation_angle}), and right up to the poles during the extended mission. It will determine the detailed surface flow field through tracking algorithms. Such algorithms provide only inconclusive results when applied to polar data obtained from near-Earth orbit due to the foreshortening. Doppler maps of the line-of-sight velocity component will complement the correlation tracking measurements and will also reveal convection, rotation, and meridional circulation flows. 

Time series of Doppler and intensity maps will be used to probe the three-dimensional mass flows in the upper layers of the convection zone, at high heliographic latitudes. The flows will be inferred using the methods of local helioseismology (e.g., \opencite{Gizon:2005aa}): time-distance helioseismology, ring diagram analysis, helioseismic holography, and direct modelling. Using SOHO/MDI Dopplergrams, it was demonstrated that even complex velocity fields can be derived with a single day of data (e.g., \opencite{Jackiewicz:2008aa}). The deeper layers of the convection zone will be studied using both local and the global methods of helioseismology. Moreover, Solar Orbiter will provide the first opportunity to implement the novel technique of stereoscopic helioseismology to probe flows and structural heterogeneities deep in the convection zone, even reaching down to the tachocline. 

Combining Solar Orbiter observations with ground- or space-based helioseismic observations from 1\,AU (e.g., GONG or SDO) will increase the observed fraction of the Sun's surface and thereby benefit global helioseismology because the modes of oscillation will be easier to disentangle (reduction of spatial leaks). With stereoscopic helioseismology, new acoustic ray paths can be taken into account to probe deeper layers in the interior (Figure~\ref{F-Helioseism_mroth}), including the bottom of the convection zone. 

\subsubsection{What are the properties of the magnetic field at high solar latitudes?}
\paragraph*{Present state of knowledge.} Meridional circulation transports the surface magnetic flux toward the poles, where a concentration of magnetic flux is expected to occur. However, because of the directional sensitivity of the Zeeman effect and magnetic polarity cancellation resulting from geometric foreshortening, present-day observations from the ecliptic at 1\,AU can provide only a poor representation of the polar magnetic field. The high resolution of Hinode's Solar Optical Telescope (SOT) can partly overcome the second disadvantage \cite{Tsuneta:2008aa}, but not the first. Consequently, an accurate quantitative estimate of the polar magnetic field remains a major and as yet unattained goal. 

The polar field is directly related to the dynamo process, presumably as a source of poloidal field that is wound up by the differential rotation in the shear layer at the base of the convection zone. The distribution of the magnetic field at the poles drives the formation and evolution of polar coronal holes, polar plumes, X-ray jets, and other events and structures that characterize the polar corona. Polar coronal holes have been intensively studied from the non-ideal vantage point offered by the ecliptic, but never imaged from outside the ecliptic. Consequently the distribution of the polar field and the origin of polar structures are only poorly determined. The fast solar wind is associated with open field lines inside coronal holes, whereas at least parts of the slow solar wind are thought to emanate from the coronal hole boundaries. Understanding the interaction of open and closed field lines across these boundaries is of paramount importance for elucidating the connection between the solar magnetic field and the heliosphere. 

As described in Section~\ref{CME_balance}, the magnetic flux in the heliosphere varies with the solar cycle and, while there is evidence that the heliospheric magnetic flux has increased substantially in the last hundred years, during the recent solar minimum the IMF strength was lower than at any time since the beginning of the space age. Models based on the injection of flux into the heliosphere by coronal mass ejections cannot explain this reduction, and it is becoming clear that the processes by which flux is added to and removed from the heliosphere are more complex than previously thought.

\paragraph*{How Solar Orbiter will address this question.} Solar Orbiter's suite of imaging instruments will characterize the properties and dynamics of the polar regions for the first time, including magnetic fields, plasma flows, and temperatures (Figure~\ref{F-Granulation_angle}). Solar Orbiter will measure the amount of polar magnetic flux, its spatial distribution and its evolution (by comparing results from different orbits), thereby providing an independent constraint on the strength and direction of the meridional flow near the pole. The evolution of Solar Orbiter's orbit to higher heliographic latitudes will make it possible to study the transport of magnetic flux from the activity belts toward the poles, which drives the polarity reversal of the global magnetic field (see \opencite{Wang:1989aa}; \opencite{Sheeley:1991aa}; \opencite{Makarov:2003aa}). From its viewpoint outside the ecliptic, Solar Orbiter will probe the cancellation processes that take place when flux elements of opposite polarity meet as part of the polarity reversal process. Joint observations from Solar Orbiter and spacecraft in the ecliptic will determine, with high accuracy, the transversal magnetic field, which is notoriously difficult to measure, along with derived quantities such as the electric current density.

\begin{figure}   
   \centerline{\includegraphics[width=\textwidth,clip=true]{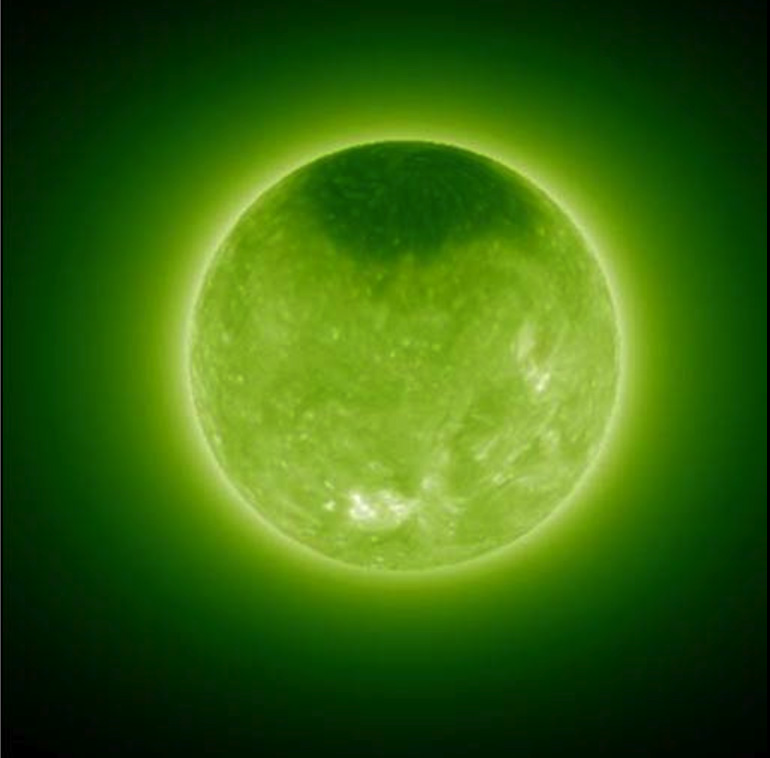}}
              \caption{Simulated view of the ultraviolet corona from 35$^\circ$ heliolatitude. Solar Orbiter's remote-sensing instruments and out-of-ecliptic vantage point will enable the first simultaneous measurements of the polar magnetic field and the associated structures in a polar coronal hole.}
     \label{F-EUI_polar}
   \end{figure}

Solar Orbiter will measure the photospheric magnetic field at the poles, while simultaneously imaging the coronal and heliospheric structure at visible and EUV wavelengths. In addition, as the spacecraft passes through the mid-latitude slow/fast wind boundary at around 0.5\,AU, the field and plasma properties of the solar wind will be measured. With the help of magnetic field extrapolation methods these observations will, for the first time, allow the photospheric and coronal magnetic field in polar coronal holes to be studied simultaneously and the evolution of polar coronal hole boundaries and other coronal structures to be investigated. The images are complementary to those from low-latitude instruments (see Figure~\ref{F-EUI_polar} for a simulated EUI image). 

Solar Orbiter's observations from progressively higher heliographic latitudes (25$^{\circ}$ by the end of the nominal mission) will enable coordinated investigation (jointly with spacecraft in the ecliptic) of the three-dimensional structure of the inner heliosphere. These observations will reveal the links between the Sun's polar regions and the properties of the solar wind and interplanetary magnetic field, in particular the heliospheric current sheet, which is used as a proxy for the tilt of the solar magnetic dipole. In addition, Solar Orbiter will pass both north and south of the solar equatorial plane in each orbit, with repeated transits through the equatorial streamer belt and through the slow/fast wind boundary at mid-latitudes into the polar wind, making it possible to follow the evolution of the solar wind and interplanetary magnetic field as well as of the sources in the polar coronal holes. Ulysses has shown that poleward of the edge of coronal holes the properties of the solar wind are relatively uniform, so that Solar Orbiter only needs to reach heliographic latitudes just above the coronal hole edge to enter the high-speed solar wind. The orbital inclination of 25$^{\circ}$ reached during the nominal mission is sufficiently high to satisfy this constraint. 

\subsubsection{Are there separate dynamo processes acting in the Sun?}
\paragraph*{Present state of knowledge.} MHD simulations indicate that a local turbulent dynamo should be acting in the Sun's turbulent convection zone \cite{Brun:2004aa} and even in the near-surface layers \cite{Vogler:2007aa}. Hinode/SOT has detected ubiquitous horizontal magnetic fields in quiet regions of the Sun \cite{Lites:2007aa}, which are possibly generated by local dynamo action \cite{Pietarila-Graham:2009aa}. These small, weak features (inter-network fields; \opencite{Zirin:1987aa}) bring 100 times more magnetic flux to the solar surface than the stronger features that are known to be the product of the global dynamo, and have themselves shown to be in cross-scale turbulent equilibrium \cite{Schrijver:1997aa}. Even the smallest observable features have been shown to be formed primarily by aggregation of yet smaller, yet more prevalent features too small to resolve with current instrumentation \cite{Lamb:2008aa,Lamb:2010aa}. It is, however, still uncertain whether a separate local, turbulent dynamo really is acting on the Sun and how strongly it contributes to the Sun's magnetic flux (and magnetic energy). In particular, all solar magnetic features, from the smallest observable intergranular flux concentrations to the largest active regions, have been shown \cite{Parnell:2009aa} to have a power law (scale free) probability distribution function, suggesting that a single turbulent mechanism may be responsible for all observable scales of magnetic activity.

\paragraph*{How Solar Orbiter will address this question.} One way to distinguish between the products of a global and a local dynamo is to study the distribution of small elements of freshly emerging magnetic flux over heliographic latitude. The global dynamo, presumably owing to the structure of the differential rotation and the meridional flow near the base of the convection zone, leads to the emergence of large bipolar magnetic regions (active regions) at the solar surface at latitudes between 5$^{\circ}$ and 30$^{\circ}$ and of smaller ephemeral active regions over a larger range of latitudes, but concentrated also at low latitudes. In contrast, a local turbulent dynamo is expected to enhance field more uniformly across the surface. 

Observations carried out from the ecliptic cannot quantitatively determine the latitudinal distribution of magnetic flux and in particular the emergence of small-scale magnetic features (inter-network fields) due to foreshortening and the different sensitivity of the Zeeman effect to longitudinal and transversal fields. Solar Orbiter, by flying to latitudes of 25$^{\circ}$ and higher above the ecliptic, will be able to measure weak magnetic features equally well at low and high latitudes \cite{MartinezPillet:2007aa}. If the number and size (i.e., magnetic flux) distributions of such features are significantly different at high latitudes, then even the weak features are probably due to the global dynamo. If, however, they are evenly distributed, then the evidence for a significant role of a local dynamo will be greatly strengthened. Current work is confounded by viewing angle restrictions near the poles, by the ubiquitous seething horizontal field (e.g., \opencite{Harvey:2007aa}), and by small deflections in near-vertical fields, which dominate observed feature distributions near the limb of the Sun.


\section{Spacecraft}
\label{S-Spacecraft}
\begin{figure}   
   \centerline{\includegraphics[width=\textwidth,clip=true]{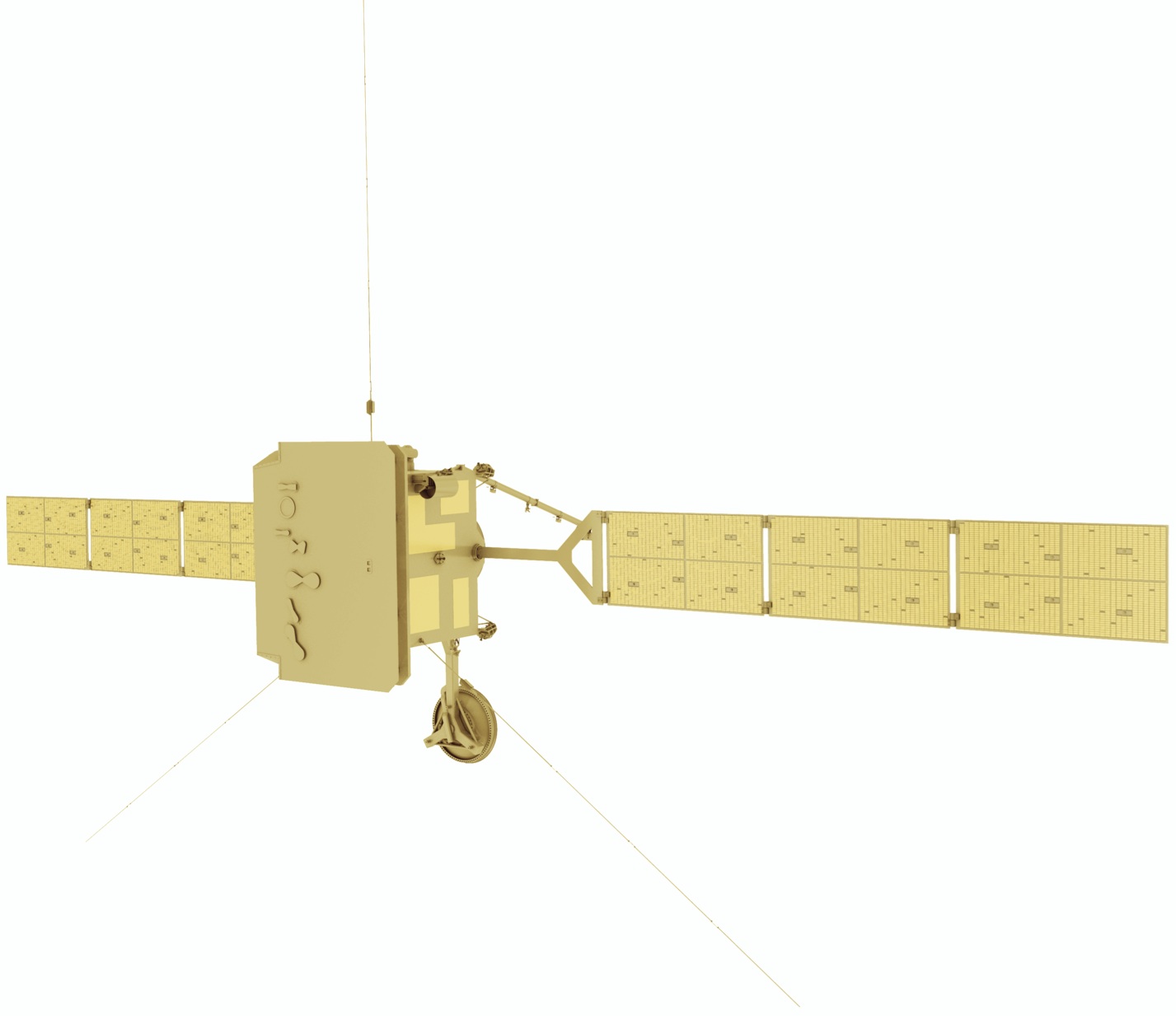}}
              \caption{Front view of the spacecraft with the three RPW antennas,  high-gain antenna, instrument boom and solar arrays deployed. The heat shield has feed-throughs with doors for the remote-sensing instruments.
}
   \label{F-SO_front}
   \end{figure}

The Solar Orbiter spacecraft (Figure \ref{F-SO_front}) is a Sun-pointed, 3-axis stabilized platform, with a dedicated heat shield to provide protection from the high levels of solar flux near perihelion. Feed-throughs in the heat shield (with individual doors) provide the remote-sensing instruments with their required fields-of-view to the Sun. Single-sided solar arrays provide the required power throughout the mission over the wide range of distances from the Sun and can be rotated about their longitudinal axis to manage the array temperature, particularly important during closest approach to the Sun. 

An articulated high-temperature high-gain antenna provides nominal communication with the ground station, and a medium-gain antenna and two low-gain antennas are included for use as back-up. The design drivers for the Solar Orbiter spacecraft come not only from the need to satisfy the mission's technical and performance requirements, but also from the need to minimize the total cost of the mission. The adopted philosophy is therefore to avoid technology development as far as possible, in order to maintain the cost-cap of the mission in keeping with its Cosmic Vision M-class status. The design of Solar Orbiter has therefore incorporated technology items from ESA's BepiColombo mercury mission \cite{2010P&SS...58....2B} where appropriate. Furthermore, design heritage from ESA's "Express" series of missions, with their goal of rapid and streamlined development, has also featured heavily in the Solar Orbiter spacecraft design.

\section{Science Operations}
\label{S-Operations}
\begin{figure}   
   \centerline{\includegraphics[width=\textwidth,clip=true]{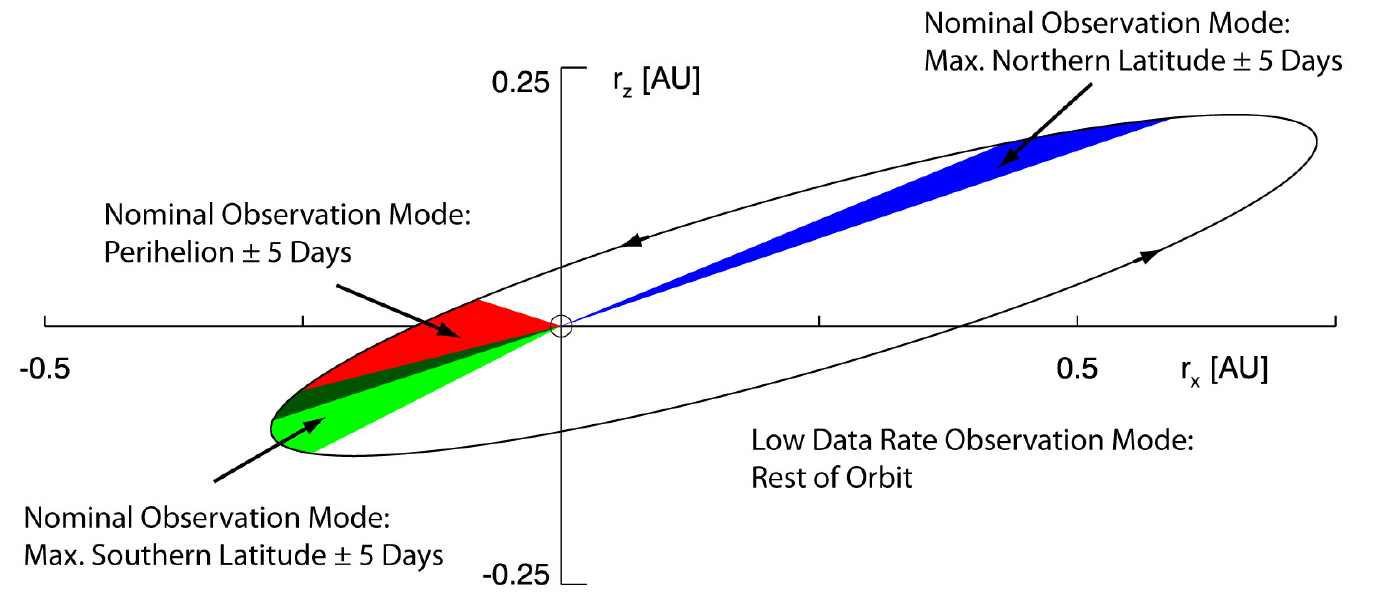}}
              \caption{Observation modes of Solar Orbiter. Science data is collected at high rate during three 10-day windows of each orbit, centered around the perihelion and the extrema in solar latitude, and at a lower rate during the remainder of each orbit. Shown here is the first orbit that exceeds a heliographic latitude of 25$^\circ$. The area shaded in dark green indicates an overlap between two nominal observation windows. The $r_{\rm x}$ axis lies in the ecliptic plane and the $r_{\rm z}$ axis is perpendicular to it.
}
   \label{F-RS-windows}
   \end{figure}

One of the strengths of the Solar Orbiter mission is the synergy between in-situ and remote-sensing observations, and each science objective requires coordinated observations between several in-situ and remote-sensing instruments. Another unique aspect of Solar Orbiter, in contrast to near-Earth observatory missions like SOHO, is that Solar Orbiter will operate much like a planetary encounter mission, with the main scientific activity and planning taking place during the near-Sun encounter part of each orbit. Specifically, observations with the remote-sensing instruments will be organized into three 10-day intervals centered around perihelion and either maximum latitude or maximum corotation passages (Figure \ref{F-RS-windows}). As a baseline, the in-situ instruments will operate continuously during normal operations. Another important aspect of this mission, from a science operations standpoint, is that every science orbit is different, with different orbital characteristics (Sun--spacecraft distance, Earth--spacecraft distance, etc.). Science and operations planning for each orbit is therefore critical, with specific orbits expected to be dedicated to specific science problems. This will be similar to what has been used successfully in the ESA/NASA SOHO mission's Joint Observation Programs (JOPs). 

\section{International Cooperation}
Solar Orbiter is an ESA-led mission with strong NASA participation. Specifically, NASA will provide the launch vehicle and parts of the scientific payload (the SoloHI instrument and the HIS sensor of SWA). The mission also has important synergies with NASA's Solar Probe Plus mission, and coordinated observations are expected to enhance greatly the scientific return of both missions. 

Solar Probe Plus, which has entered Phase B in February 2012, is scheduled for launch in September 2018. Due to its exceptional launch characteristics, Solar Probe Plus will reach its first perihelion of 0.16\,AU already three months into the mission. If launched in January 2017, Solar Orbiter will be in its cruise phase at this time, during which the in-situ instruments are fully operational for heliodistances less than 1.2\,AU, i.e.\ throughout most of the orbit. It is envisaged that the remote-sensing instruments will be switched on for dedicated observing campaigns with Solar Probe Plus.

In the overall international context, Solar Orbiter is ESA's primary contribution to the International Living With a Star (ILWS) initiative.

\section{Conclusions}

Understanding the connections and the coupling between the Sun and the heliosphere is of fundamental importance to understanding how our solar system works. To reach this goal, Solar Orbiter will make in-situ measurements of the solar wind plasma, fields, waves, and energetic particles as close as 0.28\,AU from the Sun, simultaneously with high-resolution imaging and spectroscopic observations of the Sun in and out of the ecliptic plane. The combination of in-situ and remote sensing instruments on the same spacecraft, together with the new, inner-heliospheric perspective, distinguishes Solar Orbiter from all previous and current missions, enabling breakthrough science which can be achieved in no other way. 

In addition to delivering ground-breaking science in its own right, Solar Orbiter also has important synergies with NASA's Solar Probe Plus mission. Coordinated observations with this mission, combined with data from other missions operating in the inner heliosphere (or providing remote-sensing observations of the near-Sun environment), will contribute greatly to our understanding of the Sun and its environment.

\begin{acks}
Contributions to this paper were provided by the PIs and Co-PIs, the ESA Solar Orbiter Project Team, the NASA Solar Orbiter Collaboration Project Team, E.~Marsch (MPS Lindau), M.~Velli (JPL/U.\ Firenze), C.~DeForest (SwRI Boulder), D.~Hassler (SwRI Boulder) and W.~Lewis (SwRI San Antonio). The authors would like to thank the referee, guest editor and journal editors for comments and suggestions, which helped to improve the quality of this work.
\end{acks}

\bibliographystyle{spr-mp-sola}
\bibliography{aamnem99,loops}

\begin{thebibliography}{124}
\ifx \bisbn   \undefined \def \bisbn  #1{ISBN #1}\fi
\ifx \binits  \undefined \def \binits#1{#1}\fi
\ifx \bauthor  \undefined \def \bauthor#1{#1}\fi
\ifx \batitle  \undefined \def \batitle#1{#1}\fi
\ifx \bjtitle  \undefined \def \bjtitle#1{\textit{#1}}\fi
\ifx \bvolume  \undefined \def \bvolume#1{\textbf{#1}}\fi
\ifx \byear  \undefined \def \byear#1{#1}\fi
\ifx \bissue  \undefined \def \bissue#1{#1}\fi
\ifx \bfpage  \undefined \def \bfpage#1{#1}\fi
\ifx \blpage  \undefined \def \blpage #1{#1}\fi
\ifx \burl  \undefined \def \burl#1{\textsf{#1}}\fi
\ifx \href  \undefined \def \href#1#2{\textsf{#2}}\fi
\ifx \doiurl  \undefined \def
  \doiurl#1{\href{http://dx.doi.org/#1}{\textsf{#1}}}\fi
\ifx \betal  \undefined \def \betal{\textit{et al.}}\fi
\ifx \binstitute  \undefined \def \binstitute#1{#1}\fi
\ifx \bctitle  \undefined \def \bctitle#1{#1}\fi
\ifx \beditor  \undefined \def \beditor#1{#1}\fi
\ifx \bpublisher  \undefined \def \bpublisher#1{#1}\fi
\ifx \bbtitle  \undefined \def \bbtitle#1{\textit{#1}}\fi
\ifx \bedition  \undefined \def \bedition#1{#1}\fi
\ifx \bseriesno  \undefined \def \bseriesno#1{\textbf{#1}}\fi
\ifx \blocation  \undefined \def \blocation#1{#1}\fi
\ifx \bsertitle  \undefined \def \bsertitle#1{\textit{#1}}\fi
\ifx \bsnm \undefined \def \bsnm#1{#1}\fi
\ifx \bsuffix \undefined \def \bsuffix#1{#1}\fi
\ifx \bparticle \undefined \def \bparticle#1{#1}\fi
\ifx \barticle \undefined \def \barticle#1{}\fi
\ifx \botherref \undefined \def \botherref#1{}\fi
\ifx \url \undefined \def \url#1{\textsf{#1}}\fi
\ifx \bchapter \undefined \def \bchapter#1{}\fi
\ifx \bbook \undefined \def \bbook#1{}\fi
\ifx \bcomment \undefined \def \bcomment#1{#1}\fi
\ifx \oauthor \undefined \def \oauthor#1{#1}\fi
\ifx \citeauthoryear \undefined \def \citeauthoryear#1{#1}\fi
\def \endbibitem {}
\ifx \bconflocation  \undefined \def \bconflocation#1{#1} \fi

\bibitem[\protect\citeauthoryear{{Acton} \textit{et~al.}}{1992}]{Acton:1992aa}
\begin{barticle}
\bauthor{\bsnm{{Acton}}, \binits{L.}},
\bauthor{\bsnm{{Tsuneta}}, \binits{S.}},
\bauthor{\bsnm{{Ogawara}}, \binits{Y.}},
\bauthor{\bsnm{{Bentley}}, \binits{R.}},
\bauthor{\bsnm{{Bruner}}, \binits{M.}},
\bauthor{\bsnm{{Canfield}}, \binits{R.}},
\bauthor{\bsnm{{Culhane}}, \binits{L.}},
\bauthor{\bsnm{{Doschek}}, \binits{G.}},
\bauthor{\bsnm{{Hiei}}, \binits{E.}},
\bauthor{\bsnm{{Hirayama}}, \binits{T.}}:
\byear{1992},
\batitle{{The Yohkoh mission for high-energy solar physics}}.
\bjtitle{Science}
\bvolume{258},
\bfpage{618}\,--\,\blpage{625}.
doi:\doiurl{10.1126/science.258.5082.618}.
\end{barticle}
\endbibitem

\bibitem[\protect\citeauthoryear{{Antiochos}
  \textit{et~al.}}{2011}]{Antiochos:2011aa}
\begin{barticle}
\bauthor{\bsnm{{Antiochos}}, \binits{S.K.}},
\bauthor{\bsnm{{Miki{\'c}}}, \binits{Z.}},
\bauthor{\bsnm{{Titov}}, \binits{V.S.}},
\bauthor{\bsnm{{Lionello}}, \binits{R.}},
\bauthor{\bsnm{{Linker}}, \binits{J.A.}}:
\byear{2011},
\batitle{{A Model for the Sources of the Slow Solar Wind}}.
\bjtitle{\apj}
\bvolume{731},
\bfpage{112}.
doi:\doiurl{10.1088/0004-637X/731/2/112}.
\end{barticle}
\endbibitem

\bibitem[\protect\citeauthoryear{{Antonucci}, {Abbo}, and
  {Dodero}}{2005}]{Antonucci:2005aa}
\begin{barticle}
\bauthor{\bsnm{{Antonucci}}, \binits{E.}},
\bauthor{\bsnm{{Abbo}}, \binits{L.}},
\bauthor{\bsnm{{Dodero}}, \binits{M.A.}}:
\byear{2005},
\batitle{{Slow wind and magnetic topology in the solar minimum corona in
  1996-1997}}.
\bjtitle{\aap}
\bvolume{435},
\bfpage{699}\,--\,\blpage{711}.
doi:\doiurl{10.1051/0004-6361:20047126}.
\end{barticle}
\endbibitem

\bibitem[\protect\citeauthoryear{{Aschwanden}}{2006}]{Aschwanden:2006aa}
\begin{barticle}
\bauthor{\bsnm{{Aschwanden}}, \binits{M.J.}}:
\byear{2006},
\batitle{{The Localization of Particle Acceleration Sites in Solar Flares and
  CMES}}.
\bjtitle{\ssr}
\bvolume{124},
\bfpage{361}\,--\,\blpage{372}.
doi:\doiurl{10.1007/s11214-006-9095-9}.
\end{barticle}
\endbibitem

\bibitem[\protect\citeauthoryear{{Axford} and {McKenzie}}{1992}]{Axford:1992aa}
\begin{bchapter}
\bauthor{\bsnm{{Axford}}, \binits{W.I.}},
\bauthor{\bsnm{{McKenzie}}, \binits{J.F.}}:
\byear{1992},
\bctitle{{The origin of high speed solar wind streams}}.
In: \beditor{\bsnm{{E.~Marsch \& R.~Schwenn}}} (ed.)
\bbtitle{Solar Wind 7 - Proceedings of the 3rd COSPAR Colloquium, COSPAR
  Colloquia Series Vol. 3},
\bpublisher{Pergamon Press},
\blocation{Oxford},
\bfpage{1}\,--\,\blpage{5}.
\end{bchapter}
\endbibitem

\bibitem[\protect\citeauthoryear{{Beck}}{2000}]{Beck:2000aa}
\begin{barticle}
\bauthor{\bsnm{{Beck}}, \binits{J.G.}}:
\byear{2000},
\batitle{{A comparison of differential rotation measurements - (Invited
  Review)}}.
\bjtitle{\solphys}
\bvolume{191},
\bfpage{47}\,--\,\blpage{70}.
\end{barticle}
\endbibitem

\bibitem[\protect\citeauthoryear{{Benkhoff}
  \textit{et~al.}}{2010}]{2010P&SS...58....2B}
\begin{barticle}
\bauthor{\bsnm{{Benkhoff}}, \binits{J.}},
\bauthor{\bsnm{{van Casteren}}, \binits{J.}},
\bauthor{\bsnm{{Hayakawa}}, \binits{H.}},
\bauthor{\bsnm{{Fujimoto}}, \binits{M.}},
\bauthor{\bsnm{{Laakso}}, \binits{H.}},
\bauthor{\bsnm{{Novara}}, \binits{M.}},
\bauthor{\bsnm{{Ferri}}, \binits{P.}},
\bauthor{\bsnm{{Middleton}}, \binits{H.R.}},
\bauthor{\bsnm{{Ziethe}}, \binits{R.}}:
\byear{2010},
\batitle{{BepiColombo -- Comprehensive exploration of Mercury: Mission overview
  and science goals}}.
\bjtitle{\planss}
\bvolume{58},
\bfpage{2}\,--\,\blpage{20}.
doi:\doiurl{10.1016/j.pss.2009.09.020}.
\end{barticle}
\endbibitem

\bibitem[\protect\citeauthoryear{{Borovsky}}{2008}]{Borovsky:2008aa}
\begin{barticle}
\bauthor{\bsnm{{Borovsky}}, \binits{J.E.}}:
\byear{2008},
\batitle{{Flux tube texture of the solar wind: Strands of the magnetic carpet
  at 1 AU?}}
\bjtitle{Journal of Geophysical Research (Space Physics)}
\bvolume{113},
\bfpage{8110}.
doi:\doiurl{10.1029/2007JA012684}.
\end{barticle}
\endbibitem

\bibitem[\protect\citeauthoryear{{Breech}
  \textit{et~al.}}{2008}]{Breech:2008aa}
\begin{barticle}
\bauthor{\bsnm{{Breech}}, \binits{B.}},
\bauthor{\bsnm{{Matthaeus}}, \binits{W.H.}},
\bauthor{\bsnm{{Minnie}}, \binits{J.}},
\bauthor{\bsnm{{Bieber}}, \binits{J.W.}},
\bauthor{\bsnm{{Oughton}}, \binits{S.}},
\bauthor{\bsnm{{Smith}}, \binits{C.W.}},
\bauthor{\bsnm{{Isenberg}}, \binits{P.A.}}:
\byear{2008},
\batitle{{Turbulence transport throughout the heliosphere}}.
\bjtitle{Journal of Geophysical Research (Space Physics)}
\bvolume{113},
\bfpage{8105}.
doi:\doiurl{10.1029/2007JA012711}.
\end{barticle}
\endbibitem

\bibitem[\protect\citeauthoryear{{Brun}, {Miesch}, and
  {Toomre}}{2004}]{Brun:2004aa}
\begin{barticle}
\bauthor{\bsnm{{Brun}}, \binits{A.S.}},
\bauthor{\bsnm{{Miesch}}, \binits{M.S.}},
\bauthor{\bsnm{{Toomre}}, \binits{J.}}:
\byear{2004},
\batitle{{Global-Scale Turbulent Convection and Magnetic Dynamo Action in the
  Solar Envelope}}.
\bjtitle{\apj}
\bvolume{614},
\bfpage{1073}\,--\,\blpage{1098}.
doi:\doiurl{10.1086/423835}.
\end{barticle}
\endbibitem

\bibitem[\protect\citeauthoryear{{Bruno} \textit{et~al.}}{2001}]{Bruno:2001aa}
\begin{barticle}
\bauthor{\bsnm{{Bruno}}, \binits{R.}},
\bauthor{\bsnm{{Carbone}}, \binits{V.}},
\bauthor{\bsnm{{Veltri}}, \binits{P.}},
\bauthor{\bsnm{{Pietropaolo}}, \binits{E.}},
\bauthor{\bsnm{{Bavassano}}, \binits{B.}}:
\byear{2001},
\batitle{{Identifying intermittency events in the solar wind}}.
\bjtitle{\planss}
\bvolume{49},
\bfpage{1201}\,--\,\blpage{1210}.
doi:\doiurl{10.1016/S0032-0633(01)00061-7}.
\end{barticle}
\endbibitem

\bibitem[\protect\citeauthoryear{{Cargill}
  \textit{et~al.}}{2006}]{Cargill:2006aa}
\begin{barticle}
\bauthor{\bsnm{{Cargill}}, \binits{P.J.}},
\bauthor{\bsnm{{Vlahos}}, \binits{L.}},
\bauthor{\bsnm{{Turkmani}}, \binits{R.}},
\bauthor{\bsnm{{Galsgaard}}, \binits{K.}},
\bauthor{\bsnm{{Isliker}}, \binits{H.}}:
\byear{2006},
\batitle{{Particle Acceleration in a Three-Dimensional Model of Reconnecting
  Coronal Magnetic Fields}}.
\bjtitle{\ssr}
\bvolume{124},
\bfpage{249}\,--\,\blpage{259}.
doi:\doiurl{10.1007/s11214-006-9108-8}.
\end{barticle}
\endbibitem

\bibitem[\protect\citeauthoryear{{Cirtain}
  \textit{et~al.}}{2007}]{Cirtain:2007aa}
\begin{barticle}
\bauthor{\bsnm{{Cirtain}}, \binits{J.W.}},
\bauthor{\bsnm{{Golub}}, \binits{L.}},
\bauthor{\bsnm{{Lundquist}}, \binits{L.}},
\bauthor{\bsnm{{van Ballegooijen}}, \binits{A.}},
\bauthor{\bsnm{{Savcheva}}, \binits{A.}},
\bauthor{\bsnm{{Shimojo}}, \binits{M.}},
\bauthor{\bsnm{{DeLuca}}, \binits{E.}},
\bauthor{\bsnm{{Tsuneta}}, \binits{S.}},
\bauthor{\bsnm{{Sakao}}, \binits{T.}},
\bauthor{\bsnm{{Reeves}}, \binits{K.}},
\bauthor{\bsnm{{Weber}}, \binits{M.}},
\bauthor{\bsnm{{Kano}}, \binits{R.}},
\bauthor{\bsnm{{Narukage}}, \binits{N.}},
\bauthor{\bsnm{{Shibasaki}}, \binits{K.}}:
\byear{2007},
\batitle{{Evidence for Alfv{\'e}n Waves in Solar X-ray Jets}}.
\bjtitle{Science}
\bvolume{318},
\bfpage{1580}\,--\,\blpage{1582}.
doi:\doiurl{10.1126/science.1147050}.
\end{barticle}
\endbibitem

\bibitem[\protect\citeauthoryear{{Cohen} \textit{et~al.}}{2007}]{Cohen:2007aa}
\begin{barticle}
\bauthor{\bsnm{{Cohen}}, \binits{C.M.S.}},
\bauthor{\bsnm{{Mewaldt}}, \binits{R.A.}},
\bauthor{\bsnm{{Leske}}, \binits{R.A.}},
\bauthor{\bsnm{{Cummings}}, \binits{A.C.}},
\bauthor{\bsnm{{Stone}}, \binits{E.C.}},
\bauthor{\bsnm{{Wiedenbeck}}, \binits{M.E.}},
\bauthor{\bsnm{{von Rosenvinge}}, \binits{T.T.}},
\bauthor{\bsnm{{Mason}}, \binits{G.M.}}:
\byear{2007},
\batitle{{Solar Elemental Composition Based on Studies of Solar Energetic
  Particles}}.
\bjtitle{\ssr}
\bvolume{130},
\bfpage{183}\,--\,\blpage{194}.
doi:\doiurl{10.1007/s11214-007-9218-y}.
\end{barticle}
\endbibitem

\bibitem[\protect\citeauthoryear{Corbard}{1998}]{Corbard:1998aa}
\begin{botherref}
\oauthor{\bsnm{Corbard}, \binits{T.}}:
1998,
Inversion des mesures heliosismologiques: la rotation interne du soleil.
PhD thesis,
Universit{\'e} de Nice.
\end{botherref}
\endbibitem

\bibitem[\protect\citeauthoryear{{Cranmer}, {van Ballegooijen}, and
  {Edgar}}{2007}]{Cranmer:2007aa}
\begin{barticle}
\bauthor{\bsnm{{Cranmer}}, \binits{S.R.}},
\bauthor{\bsnm{{van Ballegooijen}}, \binits{A.A.}},
\bauthor{\bsnm{{Edgar}}, \binits{R.J.}}:
\byear{2007},
\batitle{{Self-consistent Coronal Heating and Solar Wind Acceleration from
  Anisotropic Magnetohydrodynamic Turbulence}}.
\bjtitle{\apjs}
\bvolume{171},
\bfpage{520}\,--\,\blpage{551}.
doi:\doiurl{10.1086/518001}.
\end{barticle}
\endbibitem

\bibitem[\protect\citeauthoryear{{De Pontieu}
  \textit{et~al.}}{2009}]{De-Pontieu:2009aa}
\begin{barticle}
\bauthor{\bsnm{{De Pontieu}}, \binits{B.}},
\bauthor{\bsnm{{McIntosh}}, \binits{S.W.}},
\bauthor{\bsnm{{Hansteen}}, \binits{V.H.}},
\bauthor{\bsnm{{Schrijver}}, \binits{C.J.}}:
\byear{2009},
\batitle{{Observing the Roots of Solar Coronal Heating - in the Chromosphere}}.
\bjtitle{\apjl}
\bvolume{701},
\bfpage{L1}\,--\,\blpage{L6}.
doi:\doiurl{10.1088/0004-637X/701/1/L1}.
\end{barticle}
\endbibitem

\bibitem[\protect\citeauthoryear{{De Pontieu}
  \textit{et~al.}}{2011}]{De-Pontieu:2011aa}
\begin{barticle}
\bauthor{\bsnm{{De Pontieu}}, \binits{B.}},
\bauthor{\bsnm{{McIntosh}}, \binits{S.W.}},
\bauthor{\bsnm{{Carlsson}}, \binits{M.}},
\bauthor{\bsnm{{Hansteen}}, \binits{V.H.}},
\bauthor{\bsnm{{Tarbell}}, \binits{T.D.}},
\bauthor{\bsnm{{Boerner}}, \binits{P.}},
\bauthor{\bsnm{{Martinez-Sykora}}, \binits{J.}},
\bauthor{\bsnm{{Schrijver}}, \binits{C.J.}},
\bauthor{\bsnm{{Title}}, \binits{A.M.}}:
\byear{2011},
\batitle{{The Origins of Hot Plasma in the Solar Corona}}.
\bjtitle{Science}
\bvolume{331},
\bfpage{55}\,--\,\blpage{58}.
doi:\doiurl{10.1126/science.1197738}.
\end{barticle}
\endbibitem

\bibitem[\protect\citeauthoryear{{Desai} \textit{et~al.}}{2006}]{Desai:2006aa}
\begin{barticle}
\bauthor{\bsnm{{Desai}}, \binits{M.I.}},
\bauthor{\bsnm{{Mason}}, \binits{G.M.}},
\bauthor{\bsnm{{Mazur}}, \binits{J.E.}},
\bauthor{\bsnm{{Dwyer}}, \binits{J.R.}}:
\byear{2006},
\batitle{{The Seed Population for Energetic Particles Accelerated by CME-Driven
  Shocks}}.
\bjtitle{\ssr}
\bvolume{124},
\bfpage{261}\,--\,\blpage{275}.
doi:\doiurl{10.1007/s11214-006-9109-7}.
\end{barticle}
\endbibitem

\bibitem[\protect\citeauthoryear{{Dikpati} and
  {Charbonneau}}{1999}]{Dikpati:1999aa}
\begin{barticle}
\bauthor{\bsnm{{Dikpati}}, \binits{M.}},
\bauthor{\bsnm{{Charbonneau}}, \binits{P.}}:
\byear{1999},
\batitle{{A Babcock-Leighton Flux Transport Dynamo with Solar-like Differential
  Rotation}}.
\bjtitle{\apj}
\bvolume{518},
\bfpage{508}\,--\,\blpage{520}.
doi:\doiurl{10.1086/307269}.
\end{barticle}
\endbibitem

\bibitem[\protect\citeauthoryear{{Dikpati} and {Gilman}}{2008}]{Dikpati:2008aa}
\begin{barticle}
\bauthor{\bsnm{{Dikpati}}, \binits{M.}},
\bauthor{\bsnm{{Gilman}}, \binits{P.A.}}:
\byear{2008},
\batitle{{Global solar dynamo models: Simulations and predictions}}.
\bjtitle{Journal of Astrophysics and Astronomy}
\bvolume{29},
\bfpage{29}\,--\,\blpage{39}.
doi:\doiurl{10.1007/s12036-008-0004-3}.
\end{barticle}
\endbibitem

\bibitem[\protect\citeauthoryear{{Dodero}
  \textit{et~al.}}{1998}]{Dodero:1998aa}
\begin{barticle}
\bauthor{\bsnm{{Dodero}}, \binits{M.A.}},
\bauthor{\bsnm{{Antonucci}}, \binits{E.}},
\bauthor{\bsnm{{Giordano}}, \binits{S.}},
\bauthor{\bsnm{{Martin}}, \binits{R.}}:
\byear{1998},
\batitle{{Solar Wind Velocity and Anisotropic Coronal Kinetic Temperature
  Measured with the O VI Doublet Ratio}}.
\bjtitle{\solphys}
\bvolume{183},
\bfpage{77}\,--\,\blpage{90}.
\end{barticle}
\endbibitem

\bibitem[\protect\citeauthoryear{{Domingo}, {Fleck}, and
  {Poland}}{1995}]{Domingo:1995aa}
\begin{barticle}
\bauthor{\bsnm{{Domingo}}, \binits{V.}},
\bauthor{\bsnm{{Fleck}}, \binits{B.}},
\bauthor{\bsnm{{Poland}}, \binits{A.I.}}:
\byear{1995},
\batitle{{The SOHO Mission: an Overview}}.
\bjtitle{\solphys}
\bvolume{162},
\bfpage{1}\,--\,\blpage{37}.
doi:\doiurl{10.1007/BF00733425}.
\end{barticle}
\endbibitem

\bibitem[\protect\citeauthoryear{{Drake} \textit{et~al.}}{2009}]{Drake:2009aa}
\begin{barticle}
\bauthor{\bsnm{{Drake}}, \binits{J.F.}},
\bauthor{\bsnm{{Cassak}}, \binits{P.A.}},
\bauthor{\bsnm{{Shay}}, \binits{M.A.}},
\bauthor{\bsnm{{Swisdak}}, \binits{M.}},
\bauthor{\bsnm{{Quataert}}, \binits{E.}}:
\byear{2009},
\batitle{{A Magnetic Reconnection Mechanism for Ion Acceleration and Abundance
  Enhancements in Impulsive Flares}}.
\bjtitle{\apjl}
\bvolume{700},
\bfpage{L16}\,--\,\blpage{L20}.
doi:\doiurl{10.1088/0004-637X/700/1/L16}.
\end{barticle}
\endbibitem

\bibitem[\protect\citeauthoryear{{Emslie}
  \textit{et~al.}}{2004}]{Emslie:2004aa}
\begin{barticle}
\bauthor{\bsnm{{Emslie}}, \binits{A.G.}},
\bauthor{\bsnm{{Kucharek}}, \binits{H.}},
\bauthor{\bsnm{{Dennis}}, \binits{B.R.}},
\bauthor{\bsnm{{Gopalswamy}}, \binits{N.}},
\bauthor{\bsnm{{Holman}}, \binits{G.D.}},
\bauthor{\bsnm{{Share}}, \binits{G.H.}},
\bauthor{\bsnm{{Vourlidas}}, \binits{A.}},
\bauthor{\bsnm{{Forbes}}, \binits{T.G.}},
\bauthor{\bsnm{{Gallagher}}, \binits{P.T.}},
\bauthor{\bsnm{{Mason}}, \binits{G.M.}},
\bauthor{\bsnm{{Metcalf}}, \binits{T.R.}},
\bauthor{\bsnm{{Mewaldt}}, \binits{R.A.}},
\bauthor{\bsnm{{Murphy}}, \binits{R.J.}},
\bauthor{\bsnm{{Schwartz}}, \binits{R.A.}},
\bauthor{\bsnm{{Zurbuchen}}, \binits{T.H.}}:
\byear{2004},
\batitle{{Energy partition in two solar flare/CME events}}.
\bjtitle{Journal of Geophysical Research (Space Physics)}
\bvolume{109},
\bfpage{10104}.
doi:\doiurl{10.1029/2004JA010571}.
\end{barticle}
\endbibitem

\bibitem[\protect\citeauthoryear{{Fisk}}{2003}]{Fisk:2003aa}
\begin{barticle}
\bauthor{\bsnm{{Fisk}}, \binits{L.A.}}:
\byear{2003},
\batitle{{Acceleration of the solar wind as a result of the reconnection of
  open magnetic flux with coronal loops}}.
\bjtitle{Journal of Geophysical Research (Space Physics)}
\bvolume{108},
\bfpage{1157}.
doi:\doiurl{10.1029/2002JA009284}.
\end{barticle}
\endbibitem

\bibitem[\protect\citeauthoryear{{Fisk} and {Gloeckler}}{2007}]{Fisk:2007aa}
\begin{barticle}
\bauthor{\bsnm{{Fisk}}, \binits{L.A.}},
\bauthor{\bsnm{{Gloeckler}}, \binits{G.}}:
\byear{2007},
\batitle{{Acceleration and Composition of Solar Wind Suprathermal Tails}}.
\bjtitle{\ssr}
\bvolume{130},
\bfpage{153}\,--\,\blpage{160}.
doi:\doiurl{10.1007/s11214-007-9180-8}.
\end{barticle}
\endbibitem

\bibitem[\protect\citeauthoryear{{Fisk} and {Schwadron}}{2001}]{Fisk:2001aa}
\begin{barticle}
\bauthor{\bsnm{{Fisk}}, \binits{L.A.}},
\bauthor{\bsnm{{Schwadron}}, \binits{N.A.}}:
\byear{2001},
\batitle{{The Behavior of the Open Magnetic Field of the Sun}}.
\bjtitle{\apj}
\bvolume{560},
\bfpage{425}\,--\,\blpage{438}.
doi:\doiurl{10.1086/322503}.
\end{barticle}
\endbibitem

\bibitem[\protect\citeauthoryear{{Fisk} and {Zhao}}{2009}]{Fisk:2009aa}
\begin{bchapter}
\bauthor{\bsnm{{Fisk}}, \binits{L.A.}},
\bauthor{\bsnm{{Zhao}}, \binits{L.}}:
\byear{2009},
\bctitle{{The heliospheric magnetic field and the solar wind during the solar
  cycle}}.
In: \beditor{\bsnm{{Gopalswamy}}, \binits{N.}},
\beditor{\bsnm{{Webb}}, \binits{D.F.}} (eds.)
\bbtitle{IAU Symposium},
\bsertitle{IAU Symposium}
\bseriesno{257},
\bfpage{109}\,--\,\blpage{120}.
doi:\doiurl{10.1017/S1743921309029160}.
\end{bchapter}
\endbibitem

\bibitem[\protect\citeauthoryear{{Fisk} and {Zurbuchen}}{2006}]{Fisk:2006aa}
\begin{barticle}
\bauthor{\bsnm{{Fisk}}, \binits{L.A.}},
\bauthor{\bsnm{{Zurbuchen}}, \binits{T.H.}}:
\byear{2006},
\batitle{{Distribution and properties of open magnetic flux outside of coronal
  holes}}.
\bjtitle{Journal of Geophysical Research (Space Physics)}
\bvolume{111},
\bfpage{9115}.
doi:\doiurl{10.1029/2005JA011575}.
\end{barticle}
\endbibitem

\bibitem[\protect\citeauthoryear{{Fisk}, {Schwadron}, and
  {Zurbuchen}}{1998}]{Fisk:1998aa}
\begin{barticle}
\bauthor{\bsnm{{Fisk}}, \binits{L.A.}},
\bauthor{\bsnm{{Schwadron}}, \binits{N.A.}},
\bauthor{\bsnm{{Zurbuchen}}, \binits{T.H.}}:
\byear{1998},
\batitle{{On the Slow Solar Wind}}.
\bjtitle{\ssr}
\bvolume{86},
\bfpage{51}\,--\,\blpage{60}.
doi:\doiurl{10.1023/A:1005015527146}.
\end{barticle}
\endbibitem

\bibitem[\protect\citeauthoryear{{Fisk}, {Schwadron}, and
  {Zurbuchen}}{1999}]{Fisk:1999aa}
\begin{barticle}
\bauthor{\bsnm{{Fisk}}, \binits{L.A.}},
\bauthor{\bsnm{{Schwadron}}, \binits{N.A.}},
\bauthor{\bsnm{{Zurbuchen}}, \binits{T.H.}}:
\byear{1999},
\batitle{{Acceleration of the fast solar wind by the emergence of new magnetic
  flux}}.
\bjtitle{\jgr}
\bvolume{104},
\bfpage{19765}\,--\,\blpage{19772}.
doi:\doiurl{10.1029/1999JA900256}.
\end{barticle}
\endbibitem

\bibitem[\protect\citeauthoryear{{Geiss}}{1982}]{Geiss1982SSRv}
\begin{barticle}
\bauthor{\bsnm{{Geiss}}, \binits{J.}}:
\byear{1982},
\batitle{{Processes affecting abundances in the solar wind}}.
\bjtitle{\ssr}
\bvolume{33},
\bfpage{201}\,--\,\blpage{217}.
doi:\doiurl{10.1007/BF00213254}.
\end{barticle}
\endbibitem

\bibitem[\protect\citeauthoryear{{Geiss} \textit{et~al.}}{1995}]{Geiss:1995aa}
\begin{barticle}
\bauthor{\bsnm{{Geiss}}, \binits{J.}},
\bauthor{\bsnm{{Gloeckler}}, \binits{G.}},
\bauthor{\bsnm{{von Steiger}}, \binits{R.}},
\bauthor{\bsnm{{Balsiger}}, \binits{H.}},
\bauthor{\bsnm{{Fisk}}, \binits{L.A.}},
\bauthor{\bsnm{{Galvin}}, \binits{A.B.}},
\bauthor{\bsnm{{Ipavich}}, \binits{F.M.}},
\bauthor{\bsnm{{Livi}}, \binits{S.}},
\bauthor{\bsnm{{McKenzie}}, \binits{J.F.}},
\bauthor{\bsnm{{Ogilvie}}, \binits{K.W.}},
\bauthor{\bsnm{{Wilken}}, \binits{B.}}:
\byear{1995},
\batitle{{The Southern High-Speed Stream: Results from the SWICS Instrument on
  Ulysses}}.
\bjtitle{Science}
\bvolume{268},
\bfpage{1033}\,--\,\blpage{1036}.
doi:\doiurl{10.1126/science.7754380}.
\end{barticle}
\endbibitem

\bibitem[\protect\citeauthoryear{{Getman}
  \textit{et~al.}}{2008}]{Getman:2008aa}
\begin{barticle}
\bauthor{\bsnm{{Getman}}, \binits{K.V.}},
\bauthor{\bsnm{{Feigelson}}, \binits{E.D.}},
\bauthor{\bsnm{{Broos}}, \binits{P.S.}},
\bauthor{\bsnm{{Micela}}, \binits{G.}},
\bauthor{\bsnm{{Garmire}}, \binits{G.P.}}:
\byear{2008},
\batitle{{X-Ray Flares in Orion Young Stars. I. Flare Characteristics}}.
\bjtitle{\apj}
\bvolume{688},
\bfpage{418}\,--\,\blpage{436}.
doi:\doiurl{10.1086/592033}.
\end{barticle}
\endbibitem

\bibitem[\protect\citeauthoryear{{Giacalone} and
  {K{\'o}ta}}{2006}]{Giacalone:2006aa}
\begin{barticle}
\bauthor{\bsnm{{Giacalone}}, \binits{J.}},
\bauthor{\bsnm{{K{\'o}ta}}, \binits{J.}}:
\byear{2006},
\batitle{{Acceleration of Solar-Energetic Particles by Shocks}}.
\bjtitle{\ssr}
\bvolume{124},
\bfpage{277}\,--\,\blpage{288}.
doi:\doiurl{10.1007/s11214-006-9110-1}.
\end{barticle}
\endbibitem

\bibitem[\protect\citeauthoryear{{Gizon} and {Birch}}{2005}]{Gizon:2005aa}
\begin{barticle}
\bauthor{\bsnm{{Gizon}}, \binits{L.}},
\bauthor{\bsnm{{Birch}}, \binits{A.C.}}:
\byear{2005},
\batitle{{Local Helioseismology}}.
\bjtitle{Living Reviews in Solar Physics}
\bvolume{2},
\bfpage{6}.
\end{barticle}
\endbibitem

\bibitem[\protect\citeauthoryear{{Gopalswamy}}{2006}]{Gopalswamy:2006aa}
\begin{barticle}
\bauthor{\bsnm{{Gopalswamy}}, \binits{N.}}:
\byear{2006},
\batitle{{Properties of Interplanetary Coronal Mass Ejections}}.
\bjtitle{\ssr}
\bvolume{124},
\bfpage{145}\,--\,\blpage{168}.
doi:\doiurl{10.1007/s11214-006-9102-1}.
\end{barticle}
\endbibitem

\bibitem[\protect\citeauthoryear{{Gopalswamy}
  \textit{et~al.}}{2001}]{Gopalswamy:2001aa}
\begin{barticle}
\bauthor{\bsnm{{Gopalswamy}}, \binits{N.}},
\bauthor{\bsnm{{Yashiro}}, \binits{S.}},
\bauthor{\bsnm{{Kaiser}}, \binits{M.L.}},
\bauthor{\bsnm{{Howard}}, \binits{R.A.}},
\bauthor{\bsnm{{Bougeret}}, \binits{J.-L.}}:
\byear{2001},
\batitle{{Radio Signatures of Coronal Mass Ejection Interaction: Coronal Mass
  Ejection Cannibalism?}}
\bjtitle{\apjl}
\bvolume{548},
\bfpage{L91}\,--\,\blpage{L94}.
doi:\doiurl{10.1086/318939}.
\end{barticle}
\endbibitem

\bibitem[\protect\citeauthoryear{{Gopalswamy}
  \textit{et~al.}}{2002}]{Gopalswamy:2002aa}
\begin{barticle}
\bauthor{\bsnm{{Gopalswamy}}, \binits{N.}},
\bauthor{\bsnm{{Yashiro}}, \binits{S.}},
\bauthor{\bsnm{{Kaiser}}, \binits{M.L.}},
\bauthor{\bsnm{{Howard}}, \binits{R.A.}},
\bauthor{\bsnm{{Bougeret}}, \binits{J.-L.}}:
\byear{2002},
\batitle{{Interplanetary radio emission due to interaction between two coronal
  mass ejections}}.
\bjtitle{\grl}
\bvolume{29}(\bissue{8}),
\bfpage{080000}\,--\,\blpage{1}.
doi:\doiurl{10.1029/2001GL013606}.
\end{barticle}
\endbibitem

\bibitem[\protect\citeauthoryear{{Gopalswamy}
  \textit{et~al.}}{2008}]{Gopalswamy:2008aa}
\begin{barticle}
\bauthor{\bsnm{{Gopalswamy}}, \binits{N.}},
\bauthor{\bsnm{{Yashiro}}, \binits{S.}},
\bauthor{\bsnm{{Xie}}, \binits{H.}},
\bauthor{\bsnm{{Akiyama}}, \binits{S.}},
\bauthor{\bsnm{{Aguilar-Rodriguez}}, \binits{E.}},
\bauthor{\bsnm{{Kaiser}}, \binits{M.L.}},
\bauthor{\bsnm{{Howard}}, \binits{R.A.}},
\bauthor{\bsnm{{Bougeret}}, \binits{J.-L.}}:
\byear{2008},
\batitle{{Radio-Quiet Fast and Wide Coronal Mass Ejections}}.
\bjtitle{\apj}
\bvolume{674},
\bfpage{560}\,--\,\blpage{569}.
doi:\doiurl{10.1086/524765}.
\end{barticle}
\endbibitem

\bibitem[\protect\citeauthoryear{{Handy} \textit{et~al.}}{1999}]{Handy:1999aa}
\begin{barticle}
\bauthor{\bsnm{{Handy}}, \binits{B.N.}},
\bauthor{\bsnm{{Acton}}, \binits{L.W.}},
\bauthor{\bsnm{{Kankelborg}}, \binits{C.C.}},
\bauthor{\bsnm{{Wolfson}}, \binits{C.J.}},
\bauthor{\bsnm{{Akin}}, \binits{D.J.}},
\bauthor{\bsnm{{Bruner}}, \binits{M.E.}},
\bauthor{\bsnm{{Caravalho}}, \binits{R.}},
\bauthor{\bsnm{{Catura}}, \binits{R.C.}},
\bauthor{\bsnm{{Chevalier}}, \binits{R.}},
\bauthor{\bsnm{{Duncan}}, \binits{D.W.}},
\bauthor{\bsnm{{Edwards}}, \binits{C.G.}},
\bauthor{\bsnm{{Feinstein}}, \binits{C.N.}},
\bauthor{\bsnm{{Freeland}}, \binits{S.L.}},
\bauthor{\bsnm{{Friedlaender}}, \binits{F.M.}},
\bauthor{\bsnm{{Hoffmann}}, \binits{C.H.}},
\bauthor{\bsnm{{Hurlburt}}, \binits{N.E.}},
\bauthor{\bsnm{{Jurcevich}}, \binits{B.K.}},
\bauthor{\bsnm{{Katz}}, \binits{N.L.}},
\bauthor{\bsnm{{Kelly}}, \binits{G.A.}},
\bauthor{\bsnm{{Lemen}}, \binits{J.R.}},
\bauthor{\bsnm{{Levay}}, \binits{M.}},
\bauthor{\bsnm{{Lindgren}}, \binits{R.W.}},
\bauthor{\bsnm{{Mathur}}, \binits{D.P.}},
\bauthor{\bsnm{{Meyer}}, \binits{S.B.}},
\bauthor{\bsnm{{Morrison}}, \binits{S.J.}},
\bauthor{\bsnm{{Morrison}}, \binits{M.D.}},
\bauthor{\bsnm{{Nightingale}}, \binits{R.W.}},
\bauthor{\bsnm{{Pope}}, \binits{T.P.}},
\bauthor{\bsnm{{Rehse}}, \binits{R.A.}},
\bauthor{\bsnm{{Schrijver}}, \binits{C.J.}},
\bauthor{\bsnm{{Shine}}, \binits{R.A.}},
\bauthor{\bsnm{{Shing}}, \binits{L.}},
\bauthor{\bsnm{{Strong}}, \binits{K.T.}},
\bauthor{\bsnm{{Tarbell}}, \binits{T.D.}},
\bauthor{\bsnm{{Title}}, \binits{A.M.}},
\bauthor{\bsnm{{Torgerson}}, \binits{D.D.}},
\bauthor{\bsnm{{Golub}}, \binits{L.}},
\bauthor{\bsnm{{Bookbinder}}, \binits{J.A.}},
\bauthor{\bsnm{{Caldwell}}, \binits{D.}},
\bauthor{\bsnm{{Cheimets}}, \binits{P.N.}},
\bauthor{\bsnm{{Davis}}, \binits{W.N.}},
\bauthor{\bsnm{{Deluca}}, \binits{E.E.}},
\bauthor{\bsnm{{McMullen}}, \binits{R.A.}},
\bauthor{\bsnm{{Warren}}, \binits{H.P.}},
\bauthor{\bsnm{{Amato}}, \binits{D.}},
\bauthor{\bsnm{{Fisher}}, \binits{R.}},
\bauthor{\bsnm{{Maldonado}}, \binits{H.}},
\bauthor{\bsnm{{Parkinson}}, \binits{C.}}:
\byear{1999},
\batitle{{The transition region and coronal explorer}}.
\bjtitle{\solphys}
\bvolume{187},
\bfpage{229}\,--\,\blpage{260}.
doi:\doiurl{10.1023/A:1005166902804}.
\end{barticle}
\endbibitem

\bibitem[\protect\citeauthoryear{{Hansteen} and {Leer}}{1995}]{Hansteen:1995aa}
\begin{barticle}
\bauthor{\bsnm{{Hansteen}}, \binits{V.H.}},
\bauthor{\bsnm{{Leer}}, \binits{E.}}:
\byear{1995},
\batitle{{Coronal heating, densities, and temperatures and solar wind
  acceleration}}.
\bjtitle{\jgr}
\bvolume{100},
\bfpage{21577}\,--\,\blpage{21594}.
doi:\doiurl{10.1029/95JA02300}.
\end{barticle}
\endbibitem

\bibitem[\protect\citeauthoryear{{Harrison}
  \textit{et~al.}}{2009}]{Harrison:2009aa}
\begin{barticle}
\bauthor{\bsnm{{Harrison}}, \binits{R.A.}},
\bauthor{\bsnm{{Davies}}, \binits{J.A.}},
\bauthor{\bsnm{{Rouillard}}, \binits{A.P.}},
\bauthor{\bsnm{{Davis}}, \binits{C.J.}},
\bauthor{\bsnm{{Eyles}}, \binits{C.J.}},
\bauthor{\bsnm{{Bewsher}}, \binits{D.}},
\bauthor{\bsnm{{Crothers}}, \binits{S.R.}},
\bauthor{\bsnm{{Howard}}, \binits{R.A.}},
\bauthor{\bsnm{{Sheeley}}, \binits{N.R.}},
\bauthor{\bsnm{{Vourlidas}}, \binits{A.}},
\bauthor{\bsnm{{Webb}}, \binits{D.F.}},
\bauthor{\bsnm{{Brown}}, \binits{D.S.}},
\bauthor{\bsnm{{Dorrian}}, \binits{G.D.}}:
\byear{2009},
\batitle{{Two Years of the STEREO Heliospheric Imagers. Invited Review}}.
\bjtitle{\solphys}
\bvolume{256},
\bfpage{219}\,--\,\blpage{237}.
doi:\doiurl{10.1007/s11207-009-9352-7}.
\end{barticle}
\endbibitem

\bibitem[\protect\citeauthoryear{{Harvey}
  \textit{et~al.}}{2007}]{Harvey:2007aa}
\begin{barticle}
\bauthor{\bsnm{{Harvey}}, \binits{J.W.}},
\bauthor{\bsnm{{Branston}}, \binits{D.}},
\bauthor{\bsnm{{Henney}}, \binits{C.J.}},
\bauthor{\bsnm{{Keller}}, \binits{C.U.}},
\bauthor{\bsnm{{SOLIS and GONG Teams}}}:
\byear{2007},
\batitle{{Seething Horizontal Magnetic Fields in the Quiet Solar Photosphere}}.
\bjtitle{\apjl}
\bvolume{659},
\bfpage{L177}\,--\,\blpage{L180}.
doi:\doiurl{10.1086/518036}.
\end{barticle}
\endbibitem

\bibitem[\protect\citeauthoryear{{Horbury}, {Forman}, and
  {Oughton}}{2008}]{Horbury:2008aa}
\begin{barticle}
\bauthor{\bsnm{{Horbury}}, \binits{T.S.}},
\bauthor{\bsnm{{Forman}}, \binits{M.}},
\bauthor{\bsnm{{Oughton}}, \binits{S.}}:
\byear{2008},
\batitle{{Anisotropic Scaling of Magnetohydrodynamic Turbulence}}.
\bjtitle{Physical Review Letters}
\bvolume{101}(\bissue{17}),
\bfpage{175005}.
doi:\doiurl{10.1103/PhysRevLett.101.175005}.
\end{barticle}
\endbibitem

\bibitem[\protect\citeauthoryear{{Howe} \textit{et~al.}}{2006}]{Howe:2006aa}
\begin{barticle}
\bauthor{\bsnm{{Howe}}, \binits{R.}},
\bauthor{\bsnm{{Komm}}, \binits{R.}},
\bauthor{\bsnm{{Hill}}, \binits{F.}},
\bauthor{\bsnm{{Ulrich}}, \binits{R.}},
\bauthor{\bsnm{{Haber}}, \binits{D.A.}},
\bauthor{\bsnm{{Hindman}}, \binits{B.W.}},
\bauthor{\bsnm{{Schou}}, \binits{J.}},
\bauthor{\bsnm{{Thompson}}, \binits{M.J.}}:
\byear{2006},
\batitle{{Large-Scale Zonal Flows Near the Solar Surface}}.
\bjtitle{\solphys}
\bvolume{235},
\bfpage{1}\,--\,\blpage{15}.
doi:\doiurl{10.1007/s11207-006-0117-2}.
\end{barticle}
\endbibitem

\bibitem[\protect\citeauthoryear{{Jackiewicz}, {Gizon}, and
  {Birch}}{2008}]{Jackiewicz:2008aa}
\begin{barticle}
\bauthor{\bsnm{{Jackiewicz}}, \binits{J.}},
\bauthor{\bsnm{{Gizon}}, \binits{L.}},
\bauthor{\bsnm{{Birch}}, \binits{A.C.}}:
\byear{2008},
\batitle{{High-Resolution Mapping of Flows in the Solar Interior: Fully
  Consistent OLA Inversion of Helioseismic Travel Times}}.
\bjtitle{\solphys}
\bvolume{251},
\bfpage{381}\,--\,\blpage{415}.
doi:\doiurl{10.1007/s11207-008-9158-z}.
\end{barticle}
\endbibitem

\bibitem[\protect\citeauthoryear{{Kaiser}
  \textit{et~al.}}{2008}]{Kaiser:2008aa}
\begin{barticle}
\bauthor{\bsnm{{Kaiser}}, \binits{M.L.}},
\bauthor{\bsnm{{Kucera}}, \binits{T.A.}},
\bauthor{\bsnm{{Davila}}, \binits{J.M.}},
\bauthor{\bsnm{{St.~Cyr}}, \binits{O.C.}},
\bauthor{\bsnm{{Guhathakurta}}, \binits{M.}},
\bauthor{\bsnm{{Christian}}, \binits{E.}}:
\byear{2008},
\batitle{{The STEREO Mission: An Introduction}}.
\bjtitle{\ssr}
\bvolume{136},
\bfpage{5}\,--\,\blpage{16}.
doi:\doiurl{10.1007/s11214-007-9277-0}.
\end{barticle}
\endbibitem

\bibitem[\protect\citeauthoryear{{Kilpua}
  \textit{et~al.}}{2011}]{Kilpua:2011aa}
\begin{barticle}
\bauthor{\bsnm{{Kilpua}}, \binits{E.K.J.}},
\bauthor{\bsnm{{Jian}}, \binits{L.K.}},
\bauthor{\bsnm{{Li}}, \binits{Y.}},
\bauthor{\bsnm{{Luhmann}}, \binits{J.G.}},
\bauthor{\bsnm{{Russell}}, \binits{C.T.}}:
\byear{2011},
\batitle{{Multipoint ICME encounters: Pre-STEREO and STEREO observations}}.
\bjtitle{Journal of Atmospheric and Solar-Terrestrial Physics}
\bvolume{73},
\bfpage{1228}\,--\,\blpage{1241}.
doi:\doiurl{10.1016/j.jastp.2010.10.012}.
\end{barticle}
\endbibitem

\bibitem[\protect\citeauthoryear{{Klecker}, {M{\"o}bius}, and
  {Popecki}}{2006}]{Klecker:2006aa}
\begin{barticle}
\bauthor{\bsnm{{Klecker}}, \binits{B.}},
\bauthor{\bsnm{{M{\"o}bius}}, \binits{E.}},
\bauthor{\bsnm{{Popecki}}, \binits{M.A.}}:
\byear{2006},
\batitle{{Solar Energetic Particle Charge States: An Overview}}.
\bjtitle{\ssr}
\bvolume{124},
\bfpage{289}\,--\,\blpage{301}.
doi:\doiurl{10.1007/s11214-006-9111-0}.
\end{barticle}
\endbibitem

\bibitem[\protect\citeauthoryear{{Klimchuk}}{2006}]{Klimchuk:2006aa}
\begin{barticle}
\bauthor{\bsnm{{Klimchuk}}, \binits{J.A.}}:
\byear{2006},
\batitle{{On Solving the Coronal Heating Problem}}.
\bjtitle{\solphys}
\bvolume{234},
\bfpage{41}\,--\,\blpage{77}.
doi:\doiurl{10.1007/s11207-006-0055-z}.
\end{barticle}
\endbibitem

\bibitem[\protect\citeauthoryear{{Kohl} \textit{et~al.}}{1997}]{Kohl:1997aa}
\begin{barticle}
\bauthor{\bsnm{{Kohl}}, \binits{J.L.}},
\bauthor{\bsnm{{Noci}}, \binits{G.}},
\bauthor{\bsnm{{Antonucci}}, \binits{E.}},
\bauthor{\bsnm{{Tondello}}, \binits{G.}},
\bauthor{\bsnm{{Huber}}, \binits{M.C.E.}},
\bauthor{\bsnm{{Gardner}}, \binits{L.D.}},
\bauthor{\bsnm{{Nicolosi}}, \binits{P.}},
\bauthor{\bsnm{{Strachan}}, \binits{L.}},
\bauthor{\bsnm{{Fineschi}}, \binits{S.}},
\bauthor{\bsnm{{Raymond}}, \binits{J.C.}},
\bauthor{\bsnm{{Romoli}}, \binits{M.}},
\bauthor{\bsnm{{Spadaro}}, \binits{D.}},
\bauthor{\bsnm{{Panasyuk}}, \binits{A.}},
\bauthor{\bsnm{{Siegmund}}, \binits{O.H.W.}},
\bauthor{\bsnm{{Benna}}, \binits{C.}},
\bauthor{\bsnm{{Ciaravella}}, \binits{A.}},
\bauthor{\bsnm{{Cranmer}}, \binits{S.R.}},
\bauthor{\bsnm{{Giordano}}, \binits{S.}},
\bauthor{\bsnm{{Karovska}}, \binits{M.}},
\bauthor{\bsnm{{Martin}}, \binits{R.}},
\bauthor{\bsnm{{Michels}}, \binits{J.}},
\bauthor{\bsnm{{Modigliani}}, \binits{A.}},
\bauthor{\bsnm{{Naletto}}, \binits{G.}},
\bauthor{\bsnm{{Pernechele}}, \binits{C.}},
\bauthor{\bsnm{{Poletto}}, \binits{G.}},
\bauthor{\bsnm{{Smith}}, \binits{P.L.}}:
\byear{1997},
\batitle{{First Results from the SOHO Ultraviolet Coronagraph Spectrometer}}.
\bjtitle{\solphys}
\bvolume{175},
\bfpage{613}\,--\,\blpage{644}.
doi:\doiurl{10.1023/A:1004903206467}.
\end{barticle}
\endbibitem

\bibitem[\protect\citeauthoryear{{Kohl} \textit{et~al.}}{1998}]{Kohl:1998aa}
\begin{barticle}
\bauthor{\bsnm{{Kohl}}, \binits{J.L.}},
\bauthor{\bsnm{{Noci}}, \binits{G.}},
\bauthor{\bsnm{{Antonucci}}, \binits{E.}},
\bauthor{\bsnm{{Tondello}}, \binits{G.}},
\bauthor{\bsnm{{Huber}}, \binits{M.C.E.}},
\bauthor{\bsnm{{Cranmer}}, \binits{S.R.}},
\bauthor{\bsnm{{Strachan}}, \binits{L.}},
\bauthor{\bsnm{{Panasyuk}}, \binits{A.V.}},
\bauthor{\bsnm{{Gardner}}, \binits{L.D.}},
\bauthor{\bsnm{{Romoli}}, \binits{M.}},
\bauthor{\bsnm{{Fineschi}}, \binits{S.}},
\bauthor{\bsnm{{Dobrzycka}}, \binits{D.}},
\bauthor{\bsnm{{Raymond}}, \binits{J.C.}},
\bauthor{\bsnm{{Nicolosi}}, \binits{P.}},
\bauthor{\bsnm{{Siegmund}}, \binits{O.H.W.}},
\bauthor{\bsnm{{Spadaro}}, \binits{D.}},
\bauthor{\bsnm{{Benna}}, \binits{C.}},
\bauthor{\bsnm{{Ciaravella}}, \binits{A.}},
\bauthor{\bsnm{{Giordano}}, \binits{S.}},
\bauthor{\bsnm{{Habbal}}, \binits{S.R.}},
\bauthor{\bsnm{{Karovska}}, \binits{M.}},
\bauthor{\bsnm{{Li}}, \binits{X.}},
\bauthor{\bsnm{{Martin}}, \binits{R.}},
\bauthor{\bsnm{{Michels}}, \binits{J.G.}},
\bauthor{\bsnm{{Modigliani}}, \binits{A.}},
\bauthor{\bsnm{{Naletto}}, \binits{G.}},
\bauthor{\bsnm{{O'Neal}}, \binits{R.H.}},
\bauthor{\bsnm{{Pernechele}}, \binits{C.}},
\bauthor{\bsnm{{Poletto}}, \binits{G.}},
\bauthor{\bsnm{{Smith}}, \binits{P.L.}},
\bauthor{\bsnm{{Suleiman}}, \binits{R.M.}}:
\byear{1998},
\batitle{{UVCS/SOHO Empirical Determinations of Anisotropic Velocity
  Distributions in the Solar Corona}}.
\bjtitle{\apjl}
\bvolume{501},
\bfpage{L127}.
doi:\doiurl{10.1086/311434}.
\end{barticle}
\endbibitem

\bibitem[\protect\citeauthoryear{{Kohl} \textit{et~al.}}{2006}]{Kohl:2006aa}
\begin{barticle}
\bauthor{\bsnm{{Kohl}}, \binits{J.L.}},
\bauthor{\bsnm{{Noci}}, \binits{G.}},
\bauthor{\bsnm{{Cranmer}}, \binits{S.R.}},
\bauthor{\bsnm{{Raymond}}, \binits{J.C.}}:
\byear{2006},
\batitle{{Ultraviolet spectroscopy of the extended solar corona}}.
\bjtitle{\aapr}
\bvolume{13},
\bfpage{31}\,--\,\blpage{157}.
doi:\doiurl{10.1007/s00159-005-0026-7}.
\end{barticle}
\endbibitem

\bibitem[\protect\citeauthoryear{{Kosugi}
  \textit{et~al.}}{2007}]{Kosugi:2007aa}
\begin{barticle}
\bauthor{\bsnm{{Kosugi}}, \binits{T.}},
\bauthor{\bsnm{{Matsuzaki}}, \binits{K.}},
\bauthor{\bsnm{{Sakao}}, \binits{T.}},
\bauthor{\bsnm{{Shimizu}}, \binits{T.}},
\bauthor{\bsnm{{Sone}}, \binits{Y.}},
\bauthor{\bsnm{{Tachikawa}}, \binits{S.}},
\bauthor{\bsnm{{Hashimoto}}, \binits{T.}},
\bauthor{\bsnm{{Minesugi}}, \binits{K.}},
\bauthor{\bsnm{{Ohnishi}}, \binits{A.}},
\bauthor{\bsnm{{Yamada}}, \binits{T.}},
\bauthor{\bsnm{{Tsuneta}}, \binits{S.}},
\bauthor{\bsnm{{Hara}}, \binits{H.}},
\bauthor{\bsnm{{Ichimoto}}, \binits{K.}},
\bauthor{\bsnm{{Suematsu}}, \binits{Y.}},
\bauthor{\bsnm{{Shimojo}}, \binits{M.}},
\bauthor{\bsnm{{Watanabe}}, \binits{T.}},
\bauthor{\bsnm{{Shimada}}, \binits{S.}},
\bauthor{\bsnm{{Davis}}, \binits{J.M.}},
\bauthor{\bsnm{{Hill}}, \binits{L.D.}},
\bauthor{\bsnm{{Owens}}, \binits{J.K.}},
\bauthor{\bsnm{{Title}}, \binits{A.M.}},
\bauthor{\bsnm{{Culhane}}, \binits{J.L.}},
\bauthor{\bsnm{{Harra}}, \binits{L.K.}},
\bauthor{\bsnm{{Doschek}}, \binits{G.A.}},
\bauthor{\bsnm{{Golub}}, \binits{L.}}:
\byear{2007},
\batitle{{The Hinode (Solar-B) Mission: An Overview}}.
\bjtitle{\solphys}
\bvolume{243},
\bfpage{3}\,--\,\blpage{17}.
doi:\doiurl{10.1007/s11207-007-9014-6}.
\end{barticle}
\endbibitem

\bibitem[\protect\citeauthoryear{{Lamb} \textit{et~al.}}{2008}]{Lamb:2008aa}
\begin{barticle}
\bauthor{\bsnm{{Lamb}}, \binits{D.A.}},
\bauthor{\bsnm{{DeForest}}, \binits{C.E.}},
\bauthor{\bsnm{{Hagenaar}}, \binits{H.J.}},
\bauthor{\bsnm{{Parnell}}, \binits{C.E.}},
\bauthor{\bsnm{{Welsch}}, \binits{B.T.}}:
\byear{2008},
\batitle{{Solar Magnetic Tracking. II. The Apparent Unipolar Origin of
  Quiet-Sun Flux}}.
\bjtitle{\apj}
\bvolume{674},
\bfpage{520}\,--\,\blpage{529}.
doi:\doiurl{10.1086/524372}.
\end{barticle}
\endbibitem

\bibitem[\protect\citeauthoryear{{Lamb} \textit{et~al.}}{2010}]{Lamb:2010aa}
\begin{barticle}
\bauthor{\bsnm{{Lamb}}, \binits{D.A.}},
\bauthor{\bsnm{{DeForest}}, \binits{C.E.}},
\bauthor{\bsnm{{Hagenaar}}, \binits{H.J.}},
\bauthor{\bsnm{{Parnell}}, \binits{C.E.}},
\bauthor{\bsnm{{Welsch}}, \binits{B.T.}}:
\byear{2010},
\batitle{{Solar Magnetic Tracking. III. Apparent Unipolar Flux Emergence in
  High-resolution Observations}}.
\bjtitle{\apj}
\bvolume{720},
\bfpage{1405}\,--\,\blpage{1416}.
doi:\doiurl{10.1088/0004-637X/720/2/1405}.
\end{barticle}
\endbibitem

\bibitem[\protect\citeauthoryear{{Lee}}{2007}]{Lee:2007aa}
\begin{barticle}
\bauthor{\bsnm{{Lee}}, \binits{M.A.}}:
\byear{2007},
\batitle{{What Determines the Composition of SEPs in Gradual Events?}}
\bjtitle{\ssr}
\bvolume{130},
\bfpage{221}\,--\,\blpage{229}.
doi:\doiurl{10.1007/s11214-007-9188-0}.
\end{barticle}
\endbibitem

\bibitem[\protect\citeauthoryear{{Li} \textit{et~al.}}{1998}]{Li:1998aa}
\begin{barticle}
\bauthor{\bsnm{{Li}}, \binits{X.}},
\bauthor{\bsnm{{Habbal}}, \binits{S.R.}},
\bauthor{\bsnm{{Kohl}}, \binits{J.}},
\bauthor{\bsnm{{Noci}}, \binits{G.}}:
\byear{1998},
\batitle{{The Effect of Temperature Anisotropy on Observations of Doppler
  Dimming and Pumping in the Inner Corona}}.
\bjtitle{\apjl}
\bvolume{501},
\bfpage{L133}.
doi:\doiurl{10.1086/311428}.
\end{barticle}
\endbibitem

\bibitem[\protect\citeauthoryear{{Lin} and {Forbes}}{2000}]{Lin:2000aa}
\begin{barticle}
\bauthor{\bsnm{{Lin}}, \binits{J.}},
\bauthor{\bsnm{{Forbes}}, \binits{T.G.}}:
\byear{2000},
\batitle{{Effects of reconnection on the coronal mass ejection process}}.
\bjtitle{\jgr}
\bvolume{105},
\bfpage{2375}\,--\,\blpage{2392}.
doi:\doiurl{10.1029/1999JA900477}.
\end{barticle}
\endbibitem

\bibitem[\protect\citeauthoryear{{Lin}}{2006}]{Lin:2006aa}
\begin{barticle}
\bauthor{\bsnm{{Lin}}, \binits{R.P.}}:
\byear{2006},
\batitle{{Particle Acceleration by the Sun: Electrons, Hard
  X-rays/Gamma-rays}}.
\bjtitle{\ssr}
\bvolume{124},
\bfpage{233}\,--\,\blpage{248}.
doi:\doiurl{10.1007/s11214-006-9107-9}.
\end{barticle}
\endbibitem

\bibitem[\protect\citeauthoryear{{Lin} \textit{et~al.}}{2002}]{Lin:2002aa}
\begin{barticle}
\bauthor{\bsnm{{Lin}}, \binits{R.P.}},
\bauthor{\bsnm{{Dennis}}, \binits{B.R.}},
\bauthor{\bsnm{{Hurford}}, \binits{G.J.}},
\bauthor{\bsnm{{Smith}}, \binits{D.M.}},
\bauthor{\bsnm{{Zehnder}}, \binits{A.}},
\bauthor{\bsnm{{Harvey}}, \binits{P.R.}},
\bauthor{\bsnm{{Curtis}}, \binits{D.W.}},
\bauthor{\bsnm{{Pankow}}, \binits{D.}},
\bauthor{\bsnm{{Turin}}, \binits{P.}},
\bauthor{\bsnm{{Bester}}, \binits{M.}},
\bauthor{\bsnm{{Csillaghy}}, \binits{A.}},
\bauthor{\bsnm{{Lewis}}, \binits{M.}},
\bauthor{\bsnm{{Madden}}, \binits{N.}},
\bauthor{\bsnm{{van Beek}}, \binits{H.F.}},
\bauthor{\bsnm{{Appleby}}, \binits{M.}},
\bauthor{\bsnm{{Raudorf}}, \binits{T.}},
\bauthor{\bsnm{{McTiernan}}, \binits{J.}},
\bauthor{\bsnm{{Ramaty}}, \binits{R.}},
\bauthor{\bsnm{{Schmahl}}, \binits{E.}},
\bauthor{\bsnm{{Schwartz}}, \binits{R.}},
\bauthor{\bsnm{{Krucker}}, \binits{S.}},
\bauthor{\bsnm{{Abiad}}, \binits{R.}},
\bauthor{\bsnm{{Quinn}}, \binits{T.}},
\bauthor{\bsnm{{Berg}}, \binits{P.}},
\bauthor{\bsnm{{Hashii}}, \binits{M.}},
\bauthor{\bsnm{{Sterling}}, \binits{R.}},
\bauthor{\bsnm{{Jackson}}, \binits{R.}},
\bauthor{\bsnm{{Pratt}}, \binits{R.}},
\bauthor{\bsnm{{Campbell}}, \binits{R.D.}},
\bauthor{\bsnm{{Malone}}, \binits{D.}},
\bauthor{\bsnm{{Landis}}, \binits{D.}},
\bauthor{\bsnm{{Barrington-Leigh}}, \binits{C.P.}},
\bauthor{\bsnm{{Slassi-Sennou}}, \binits{S.}},
\bauthor{\bsnm{{Cork}}, \binits{C.}},
\bauthor{\bsnm{{Clark}}, \binits{D.}},
\bauthor{\bsnm{{Amato}}, \binits{D.}},
\bauthor{\bsnm{{Orwig}}, \binits{L.}},
\bauthor{\bsnm{{Boyle}}, \binits{R.}},
\bauthor{\bsnm{{Banks}}, \binits{I.S.}},
\bauthor{\bsnm{{Shirey}}, \binits{K.}},
\bauthor{\bsnm{{Tolbert}}, \binits{A.K.}},
\bauthor{\bsnm{{Zarro}}, \binits{D.}},
\bauthor{\bsnm{{Snow}}, \binits{F.}},
\bauthor{\bsnm{{Thomsen}}, \binits{K.}},
\bauthor{\bsnm{{Henneck}}, \binits{R.}},
\bauthor{\bsnm{{McHedlishvili}}, \binits{A.}},
\bauthor{\bsnm{{Ming}}, \binits{P.}},
\bauthor{\bsnm{{Fivian}}, \binits{M.}},
\bauthor{\bsnm{{Jordan}}, \binits{J.}},
\bauthor{\bsnm{{Wanner}}, \binits{R.}},
\bauthor{\bsnm{{Crubb}}, \binits{J.}},
\bauthor{\bsnm{{Preble}}, \binits{J.}},
\bauthor{\bsnm{{Matranga}}, \binits{M.}},
\bauthor{\bsnm{{Benz}}, \binits{A.}},
\bauthor{\bsnm{{Hudson}}, \binits{H.}},
\bauthor{\bsnm{{Canfield}}, \binits{R.C.}},
\bauthor{\bsnm{{Holman}}, \binits{G.D.}},
\bauthor{\bsnm{{Crannell}}, \binits{C.}},
\bauthor{\bsnm{{Kosugi}}, \binits{T.}},
\bauthor{\bsnm{{Emslie}}, \binits{A.G.}},
\bauthor{\bsnm{{Vilmer}}, \binits{N.}},
\bauthor{\bsnm{{Brown}}, \binits{J.C.}},
\bauthor{\bsnm{{Johns-Krull}}, \binits{C.}},
\bauthor{\bsnm{{Aschwanden}}, \binits{M.}},
\bauthor{\bsnm{{Metcalf}}, \binits{T.}},
\bauthor{\bsnm{{Conway}}, \binits{A.}}:
\byear{2002},
\batitle{{The Reuven Ramaty High-Energy Solar Spectroscopic Imager (RHESSI)}}.
\bjtitle{\solphys}
\bvolume{210},
\bfpage{3}\,--\,\blpage{32}.
doi:\doiurl{10.1023/A:1022428818870}.
\end{barticle}
\endbibitem

\bibitem[\protect\citeauthoryear{{Lites} \textit{et~al.}}{2007}]{Lites:2007aa}
\begin{barticle}
\bauthor{\bsnm{{Lites}}, \binits{B.}},
\bauthor{\bsnm{{Socas-Navarro}}, \binits{H.}},
\bauthor{\bsnm{{Kubo}}, \binits{M.}},
\bauthor{\bsnm{{Berger}}, \binits{T.}},
\bauthor{\bsnm{{Frank}}, \binits{Z.}},
\bauthor{\bsnm{{Shine}}, \binits{R.A.}},
\bauthor{\bsnm{{Tarbell}}, \binits{T.D.}},
\bauthor{\bsnm{{Title}}, \binits{A.M.}},
\bauthor{\bsnm{{Ichimoto}}, \binits{K.}},
\bauthor{\bsnm{{Katsukawa}}, \binits{Y.}},
\bauthor{\bsnm{{Tsuneta}}, \binits{S.}},
\bauthor{\bsnm{{Suematsu}}, \binits{Y.}},
\bauthor{\bsnm{{Shimizu}}, \binits{T.}}:
\byear{2007},
\batitle{{Hinode Observations of Horizontal Quiet Sun Magnetic Flux and the
  ``Hidden Turbulent Magnetic Flux''}}.
\bjtitle{\pasj}
\bvolume{59},
\bfpage{571}.
\end{barticle}
\endbibitem

\bibitem[\protect\citeauthoryear{{Liu} \textit{et~al.}}{2010}]{Liu:2010ab}
\begin{barticle}
\bauthor{\bsnm{{Liu}}, \binits{R.}},
\bauthor{\bsnm{{Liu}}, \binits{C.}},
\bauthor{\bsnm{{Park}}, \binits{S.-H.}},
\bauthor{\bsnm{{Wang}}, \binits{H.}}:
\byear{2010},
\batitle{{Gradual Inflation of Active-region Coronal Arcades Building up to
  Coronal Mass Ejections}}.
\bjtitle{\apj}
\bvolume{723},
\bfpage{229}\,--\,\blpage{240}.
doi:\doiurl{10.1088/0004-637X/723/1/229}.
\end{barticle}
\endbibitem

\bibitem[\protect\citeauthoryear{{Lockwood}, {Stamper}, and
  {Wild}}{1999}]{Lockwood:1999aa}
\begin{barticle}
\bauthor{\bsnm{{Lockwood}}, \binits{M.}},
\bauthor{\bsnm{{Stamper}}, \binits{R.}},
\bauthor{\bsnm{{Wild}}, \binits{M.N.}}:
\byear{1999},
\batitle{{A doubling of the Sun's coronal magnetic field during the past 100
  years}}.
\bjtitle{\nat}
\bvolume{399},
\bfpage{437}\,--\,\blpage{439}.
doi:\doiurl{10.1038/20867}.
\end{barticle}
\endbibitem

\bibitem[\protect\citeauthoryear{{Lugaz}, {Manchester}, and
  {Gombosi}}{2005}]{Lugaz:2005aa}
\begin{barticle}
\bauthor{\bsnm{{Lugaz}}, \binits{N.}},
\bauthor{\bsnm{{Manchester}}, \binits{W.B.} \bsuffix{IV}},
\bauthor{\bsnm{{Gombosi}}, \binits{T.I.}}:
\byear{2005},
\batitle{{Numerical Simulation of the Interaction of Two Coronal Mass Ejections
  from Sun to Earth}}.
\bjtitle{\apj}
\bvolume{634},
\bfpage{651}\,--\,\blpage{662}.
doi:\doiurl{10.1086/491782}.
\end{barticle}
\endbibitem

\bibitem[\protect\citeauthoryear{{Lynch} \textit{et~al.}}{2004}]{Lynch:2004aa}
\begin{barticle}
\bauthor{\bsnm{{Lynch}}, \binits{B.J.}},
\bauthor{\bsnm{{Antiochos}}, \binits{S.K.}},
\bauthor{\bsnm{{MacNeice}}, \binits{P.J.}},
\bauthor{\bsnm{{Zurbuchen}}, \binits{T.H.}},
\bauthor{\bsnm{{Fisk}}, \binits{L.A.}}:
\byear{2004},
\batitle{{Observable Properties of the Breakout Model for Coronal Mass
  Ejections}}.
\bjtitle{\apj}
\bvolume{617},
\bfpage{589}\,--\,\blpage{599}.
doi:\doiurl{10.1086/424564}.
\end{barticle}
\endbibitem

\bibitem[\protect\citeauthoryear{{Makarov}, {Tlatov}, and
  {Sivaraman}}{2003}]{Makarov:2003aa}
\begin{barticle}
\bauthor{\bsnm{{Makarov}}, \binits{V.I.}},
\bauthor{\bsnm{{Tlatov}}, \binits{A.G.}},
\bauthor{\bsnm{{Sivaraman}}, \binits{K.R.}}:
\byear{2003},
\batitle{{Duration of Polar Activity Cycles and Their Relation to Sunspot
  Activity}}.
\bjtitle{\solphys}
\bvolume{214},
\bfpage{41}\,--\,\blpage{54}.
\end{barticle}
\endbibitem

\bibitem[\protect\citeauthoryear{{Mann} \textit{et~al.}}{2003}]{Mann:2003aa}
\begin{barticle}
\bauthor{\bsnm{{Mann}}, \binits{G.}},
\bauthor{\bsnm{{Klassen}}, \binits{A.}},
\bauthor{\bsnm{{Aurass}}, \binits{H.}},
\bauthor{\bsnm{{Classen}}, \binits{H.-T.}}:
\byear{2003},
\batitle{{Formation and development of shock waves in the solar corona and the
  near-Sun interplanetary space}}.
\bjtitle{\aap}
\bvolume{400},
\bfpage{329}\,--\,\blpage{336}.
doi:\doiurl{10.1051/0004-6361:20021593}.
\end{barticle}
\endbibitem

\bibitem[\protect\citeauthoryear{{Marino}
  \textit{et~al.}}{2008}]{Marino:2008aa}
\begin{barticle}
\bauthor{\bsnm{{Marino}}, \binits{R.}},
\bauthor{\bsnm{{Sorriso-Valvo}}, \binits{L.}},
\bauthor{\bsnm{{Carbone}}, \binits{V.}},
\bauthor{\bsnm{{Noullez}}, \binits{A.}},
\bauthor{\bsnm{{Bruno}}, \binits{R.}},
\bauthor{\bsnm{{Bavassano}}, \binits{B.}}:
\byear{2008},
\batitle{{Heating the Solar Wind by a Magnetohydrodynamic Turbulent Energy
  Cascade}}.
\bjtitle{\apjl}
\bvolume{677},
\bfpage{L71}\,--\,\blpage{L74}.
doi:\doiurl{10.1086/587957}.
\end{barticle}
\endbibitem

\bibitem[\protect\citeauthoryear{{Marsch}}{2006}]{Marsch:2006ab}
\begin{barticle}
\bauthor{\bsnm{{Marsch}}, \binits{E.}}:
\byear{2006},
\batitle{{Kinetic Physics of the Solar Corona and Solar Wind}}.
\bjtitle{Living Reviews in Solar Physics}
\bvolume{3},
\bfpage{1}.
\end{barticle}
\endbibitem

\bibitem[\protect\citeauthoryear{{Marsch}
  \textit{et~al.}}{2006}]{Marsch:2006aa}
\begin{barticle}
\bauthor{\bsnm{{Marsch}}, \binits{E.}},
\bauthor{\bsnm{{Zhou}}, \binits{G.-Q.}},
\bauthor{\bsnm{{He}}, \binits{J.-S.}},
\bauthor{\bsnm{{Tu}}, \binits{C.-Y.}}:
\byear{2006},
\batitle{{Magnetic structure of the solar transition region as observed in
  various ultraviolet lines emitted at different temperatures}}.
\bjtitle{\aap}
\bvolume{457},
\bfpage{699}\,--\,\blpage{706}.
doi:\doiurl{10.1051/0004-6361:20065665}.
\end{barticle}
\endbibitem

\bibitem[\protect\citeauthoryear{{Marsden} and
  {M\"uller}}{2011}]{Marsden:2011aa}
\begin{botherref}
\oauthor{\bsnm{{Marsden}}, \binits{R.G.}},
\oauthor{\bsnm{{M\"uller}}, \binits{D.}}:
2011,
\textit{{\rm Solar Orbiter Definition Study Report, ESA/SRE(2011)14}},
http://sci.esa.int/science-e/www/object/index.cfm?fobjectid=48985.
\end{botherref}
\endbibitem

\bibitem[\protect\citeauthoryear{Mart{\'{i}}nez~Pillet}{2007}]{MartinezPillet:2007aa}
\begin{bchapter}
\bauthor{\bsnm{Mart{\'{i}}nez~Pillet}, \binits{V.}}:
\byear{2007},
\bctitle{Instrumental approaches to magnetic and velocity measurements in and
  out of the ecliptic plane}.
In: \beditor{\bsnm{{Marsch}}, \binits{E.}},
\beditor{\bsnm{{Tsinganos}}, \binits{K.}},
\beditor{\bsnm{{Marsden}}, \binits{R.}},
\beditor{\bsnm{{Conroy}}, \binits{L.}} (eds.)
\bbtitle{Proceedings of the 2$^{nd}$ Solar Orbiter Workshop},
\bsertitle{ESA Special Publication}
\bseriesno{641},
\bfpage{1}\,--\,\blpage{6}.
\end{bchapter}
\endbibitem

\bibitem[\protect\citeauthoryear{{Mason}}{2007}]{Mason:2007aa}
\begin{barticle}
\bauthor{\bsnm{{Mason}}, \binits{G.M.}}:
\byear{2007},
\batitle{{$^{3}$He-Rich Solar Energetic Particle Events}}.
\bjtitle{\ssr}
\bvolume{130},
\bfpage{231}\,--\,\blpage{242}.
doi:\doiurl{10.1007/s11214-007-9156-8}.
\end{barticle}
\endbibitem

\bibitem[\protect\citeauthoryear{{Matteini}
  \textit{et~al.}}{2007}]{Matteini:2007aa}
\begin{barticle}
\bauthor{\bsnm{{Matteini}}, \binits{L.}},
\bauthor{\bsnm{{Landi}}, \binits{S.}},
\bauthor{\bsnm{{Hellinger}}, \binits{P.}},
\bauthor{\bsnm{{Pantellini}}, \binits{F.}},
\bauthor{\bsnm{{Maksimovic}}, \binits{M.}},
\bauthor{\bsnm{{Velli}}, \binits{M.}},
\bauthor{\bsnm{{Goldstein}}, \binits{B.E.}},
\bauthor{\bsnm{{Marsch}}, \binits{E.}}:
\byear{2007},
\batitle{{Evolution of the solar wind proton temperature anisotropy from 0.3 to
  2.5 AU}}.
\bjtitle{\grl}
\bvolume{34},
\bfpage{20105}.
doi:\doiurl{10.1029/2007GL030920}.
\end{barticle}
\endbibitem

\bibitem[\protect\citeauthoryear{{McComas}
  \textit{et~al.}}{2008}]{McComas:2008aa}
\begin{barticle}
\bauthor{\bsnm{{McComas}}, \binits{D.J.}},
\bauthor{\bsnm{{Ebert}}, \binits{R.W.}},
\bauthor{\bsnm{{Elliott}}, \binits{H.A.}},
\bauthor{\bsnm{{Goldstein}}, \binits{B.E.}},
\bauthor{\bsnm{{Gosling}}, \binits{J.T.}},
\bauthor{\bsnm{{Schwadron}}, \binits{N.A.}},
\bauthor{\bsnm{{Skoug}}, \binits{R.M.}}:
\byear{2008},
\batitle{{Weaker solar wind from the polar coronal holes and the whole Sun}}.
\bjtitle{\grl}
\bvolume{35},
\bfpage{18103}.
doi:\doiurl{10.1029/2008GL034896}.
\end{barticle}
\endbibitem

\bibitem[\protect\citeauthoryear{{McIntosh}, {Davey}, and
  {Hassler}}{2006}]{McIntosh:2006aa}
\begin{barticle}
\bauthor{\bsnm{{McIntosh}}, \binits{S.W.}},
\bauthor{\bsnm{{Davey}}, \binits{A.R.}},
\bauthor{\bsnm{{Hassler}}, \binits{D.M.}}:
\byear{2006},
\batitle{{Simple Magnetic Flux Balance as an Indicator of Ne VIII Doppler
  Velocity Partitioning in an Equatorial Coronal Hole}}.
\bjtitle{\apjl}
\bvolume{644},
\bfpage{L87}\,--\,\blpage{L91}.
doi:\doiurl{10.1086/505488}.
\end{barticle}
\endbibitem

\bibitem[\protect\citeauthoryear{{Mewaldt}}{2006}]{Mewaldt:2006aa}
\begin{barticle}
\bauthor{\bsnm{{Mewaldt}}, \binits{R.A.}}:
\byear{2006},
\batitle{{Solar Energetic Particle Composition, Energy Spectra, and Space
  Weather}}.
\bjtitle{\ssr}
\bvolume{124},
\bfpage{303}\,--\,\blpage{316}.
doi:\doiurl{10.1007/s11214-006-9091-0}.
\end{barticle}
\endbibitem

\bibitem[\protect\citeauthoryear{{Mewaldt}
  \textit{et~al.}}{2007}]{Mewaldt:2007aa}
\begin{barticle}
\bauthor{\bsnm{{Mewaldt}}, \binits{R.A.}},
\bauthor{\bsnm{{Cohen}}, \binits{C.M.S.}},
\bauthor{\bsnm{{Mason}}, \binits{G.M.}},
\bauthor{\bsnm{{Cummings}}, \binits{A.C.}},
\bauthor{\bsnm{{Desai}}, \binits{M.I.}},
\bauthor{\bsnm{{Leske}}, \binits{R.A.}},
\bauthor{\bsnm{{Raines}}, \binits{J.}},
\bauthor{\bsnm{{Stone}}, \binits{E.C.}},
\bauthor{\bsnm{{Wiedenbeck}}, \binits{M.E.}},
\bauthor{\bsnm{{von Rosenvinge}}, \binits{T.T.}},
\bauthor{\bsnm{{Zurbuchen}}, \binits{T.H.}}:
\byear{2007},
\batitle{{On the Differences in Composition between Solar Energetic Particles
  and Solar Wind}}.
\bjtitle{\ssr}
\bvolume{130},
\bfpage{207}\,--\,\blpage{219}.
doi:\doiurl{10.1007/s11214-007-9187-1}.
\end{barticle}
\endbibitem

\bibitem[\protect\citeauthoryear{{Neugebauer}
  \textit{et~al.}}{1995}]{Neugebauer:1995aa}
\begin{barticle}
\bauthor{\bsnm{{Neugebauer}}, \binits{M.}},
\bauthor{\bsnm{{Goldstein}}, \binits{B.E.}},
\bauthor{\bsnm{{McComas}}, \binits{D.J.}},
\bauthor{\bsnm{{Suess}}, \binits{S.T.}},
\bauthor{\bsnm{{Balogh}}, \binits{A.}}:
\byear{1995},
\batitle{{Ulysses observations of microstreams in the solar wind from coronal
  holes}}.
\bjtitle{\jgr}
\bvolume{100},
\bfpage{23389}\,--\,\blpage{23396}.
doi:\doiurl{10.1029/95JA02723}.
\end{barticle}
\endbibitem

\bibitem[\protect\citeauthoryear{{Ontiveros} and
  {Vourlidas}}{2009}]{Ontiveros:2009aa}
\begin{barticle}
\bauthor{\bsnm{{Ontiveros}}, \binits{V.}},
\bauthor{\bsnm{{Vourlidas}}, \binits{A.}}:
\byear{2009},
\batitle{{Quantitative Measurements of Coronal Mass Ejection-Driven Shocks from
  LASCO Observations}}.
\bjtitle{\apj}
\bvolume{693},
\bfpage{267}\,--\,\blpage{275}.
doi:\doiurl{10.1088/0004-637X/693/1/267}.
\end{barticle}
\endbibitem

\bibitem[\protect\citeauthoryear{{Owens} and {Crooker}}{2006}]{Owens:2006aa}
\begin{barticle}
\bauthor{\bsnm{{Owens}}, \binits{M.J.}},
\bauthor{\bsnm{{Crooker}}, \binits{N.U.}}:
\byear{2006},
\batitle{{Coronal mass ejections and magnetic flux buildup in the
  heliosphere}}.
\bjtitle{Journal of Geophysical Research (Space Physics)}
\bvolume{111},
\bfpage{10104}.
doi:\doiurl{10.1029/2006JA011641}.
\end{barticle}
\endbibitem

\bibitem[\protect\citeauthoryear{{Owens} \textit{et~al.}}{2008}]{Owens:2008aa}
\begin{barticle}
\bauthor{\bsnm{{Owens}}, \binits{M.J.}},
\bauthor{\bsnm{{Crooker}}, \binits{N.U.}},
\bauthor{\bsnm{{Schwadron}}, \binits{N.A.}},
\bauthor{\bsnm{{Horbury}}, \binits{T.S.}},
\bauthor{\bsnm{{Yashiro}}, \binits{S.}},
\bauthor{\bsnm{{Xie}}, \binits{H.}},
\bauthor{\bsnm{{St.~Cyr}}, \binits{O.C.}},
\bauthor{\bsnm{{Gopalswamy}}, \binits{N.}}:
\byear{2008},
\batitle{{Conservation of open solar magnetic flux and the floor in the
  heliospheric magnetic field}}.
\bjtitle{\grl}
\bvolume{35},
\bfpage{20108}.
doi:\doiurl{10.1029/2008GL035813}.
\end{barticle}
\endbibitem

\bibitem[\protect\citeauthoryear{{Parnell}
  \textit{et~al.}}{2009}]{Parnell:2009aa}
\begin{barticle}
\bauthor{\bsnm{{Parnell}}, \binits{C.E.}},
\bauthor{\bsnm{{DeForest}}, \binits{C.E.}},
\bauthor{\bsnm{{Hagenaar}}, \binits{H.J.}},
\bauthor{\bsnm{{Johnston}}, \binits{B.A.}},
\bauthor{\bsnm{{Lamb}}, \binits{D.A.}},
\bauthor{\bsnm{{Welsch}}, \binits{B.T.}}:
\byear{2009},
\batitle{{A Power-Law Distribution of Solar Magnetic Fields Over More Than Five
  Decades in Flux}}.
\bjtitle{\apj}
\bvolume{698},
\bfpage{75}\,--\,\blpage{82}.
doi:\doiurl{10.1088/0004-637X/698/1/75}.
\end{barticle}
\endbibitem

\bibitem[\protect\citeauthoryear{{Patsourakos} and
  {Vourlidas}}{2009}]{Patsourakos:2009aa}
\begin{barticle}
\bauthor{\bsnm{{Patsourakos}}, \binits{S.}},
\bauthor{\bsnm{{Vourlidas}}, \binits{A.}}:
\byear{2009},
\batitle{{''Extreme Ultraviolet Waves'' are Waves: First Quadrature
  Observations of an Extreme Ultraviolet Wave from STEREO}}.
\bjtitle{\apjl}
\bvolume{700},
\bfpage{L182}\,--\,\blpage{L186}.
doi:\doiurl{10.1088/0004-637X/700/2/L182}.
\end{barticle}
\endbibitem

\bibitem[\protect\citeauthoryear{{Pesnell}, {Thompson}, and
  {Chamberlin}}{2012}]{Pesnell:2012aa}
\begin{barticle}
\bauthor{\bsnm{{Pesnell}}, \binits{W.D.}},
\bauthor{\bsnm{{Thompson}}, \binits{B.J.}},
\bauthor{\bsnm{{Chamberlin}}, \binits{P.C.}}:
\byear{2012},
\batitle{{The Solar Dynamics Observatory (SDO)}}.
\bjtitle{\solphys}
\bvolume{275},
\bfpage{3}\,--\,\blpage{15}.
doi:\doiurl{10.1007/s11207-011-9841-3}.
\end{barticle}
\endbibitem

\bibitem[\protect\citeauthoryear{{Pietarila Graham}, {Danilovic}, and
  {Sch{\"u}ssler}}{2009}]{Pietarila-Graham:2009aa}
\begin{barticle}
\bauthor{\bsnm{{Pietarila Graham}}, \binits{J.}},
\bauthor{\bsnm{{Danilovic}}, \binits{S.}},
\bauthor{\bsnm{{Sch{\"u}ssler}}, \binits{M.}}:
\byear{2009},
\batitle{{Turbulent Magnetic Fields in the Quiet Sun: Implications of Hinode
  Observations and Small-Scale Dynamo Simulations}}.
\bjtitle{\apj}
\bvolume{693},
\bfpage{1728}\,--\,\blpage{1735}.
doi:\doiurl{10.1088/0004-637X/693/2/1728}.
\end{barticle}
\endbibitem

\bibitem[\protect\citeauthoryear{{Porsche}}{1977}]{Porsche:1977aa}
\begin{barticle}
\bauthor{\bsnm{{Porsche}}, \binits{H.}}:
\byear{1977},
\batitle{{General aspects of the mission Helios 1 and 2. Introduction to a
  special issue on initial scientific results of the Helios mission.}}
\bjtitle{Journal of Geophysics/Zeitschrift f{\"u}r Geophysik}
\bvolume{42},
\bfpage{551}\,--\,\blpage{559}.
\end{barticle}
\endbibitem

\bibitem[\protect\citeauthoryear{{Reale}}{2010}]{Reale:2010ab}
\begin{barticle}
\bauthor{\bsnm{{Reale}}, \binits{F.}}:
\byear{2010},
\batitle{{Coronal Loops: Observations and Modeling of Confined Plasma}}.
\bjtitle{Living Reviews in Solar Physics}
\bvolume{7},
\bfpage{5}.
\end{barticle}
\endbibitem

\bibitem[\protect\citeauthoryear{{Richardson} and
  {Cane}}{2004}]{Richardson:2004aa}
\begin{barticle}
\bauthor{\bsnm{{Richardson}}, \binits{I.G.}},
\bauthor{\bsnm{{Cane}}, \binits{H.V.}}:
\byear{2004},
\batitle{{The fraction of interplanetary coronal mass ejections that are
  magnetic clouds: Evidence for a solar cycle variation}}.
\bjtitle{\grl}
\bvolume{31},
\bfpage{18804}.
doi:\doiurl{10.1029/2004GL020958}.
\end{barticle}
\endbibitem

\bibitem[\protect\citeauthoryear{{Richardson} and
  {Cane}}{2010}]{Richardson:2010aa}
\begin{barticle}
\bauthor{\bsnm{{Richardson}}, \binits{I.G.}},
\bauthor{\bsnm{{Cane}}, \binits{H.V.}}:
\byear{2010},
\batitle{{Near-Earth Interplanetary Coronal Mass Ejections During Solar Cycle
  23 (1996 - 2009): Catalog and Summary of Properties}}.
\bjtitle{\solphys}
\bvolume{264},
\bfpage{189}\,--\,\blpage{237}.
doi:\doiurl{10.1007/s11207-010-9568-6}.
\end{barticle}
\endbibitem

\bibitem[\protect\citeauthoryear{Roth}{2007}]{Roth:2007aa}
\begin{botherref}
\oauthor{\bsnm{Roth}, \binits{M.}}:
2007,
In: Kneer, F., Puschmann, K.G., Wittmann, A.D. (eds.)
\textit{Modern Solar Facilities - Advanced Solar Science},
\textit{Proceedings of a Workshop Held at G{\"o}ttingen, September 27-29,
  2006},
Universit{\"a}tsverlag G{\"o}ttingen.
9781931968782.
\end{botherref}
\endbibitem

\bibitem[\protect\citeauthoryear{{Rouillard}, {Lockwood}, and
  {Finch}}{2007}]{Rouillard:2007aa}
\begin{barticle}
\bauthor{\bsnm{{Rouillard}}, \binits{A.P.}},
\bauthor{\bsnm{{Lockwood}}, \binits{M.}},
\bauthor{\bsnm{{Finch}}, \binits{I.}}:
\byear{2007},
\batitle{{Centennial changes in the solar wind speed and in the open solar
  flux}}.
\bjtitle{Journal of Geophysical Research (Space Physics)}
\bvolume{112},
\bfpage{5103}.
doi:\doiurl{10.1029/2006JA012130}.
\end{barticle}
\endbibitem

\bibitem[\protect\citeauthoryear{{Schrijver}
  \textit{et~al.}}{1997}]{Schrijver:1997aa}
\begin{barticle}
\bauthor{\bsnm{{Schrijver}}, \binits{C.J.}},
\bauthor{\bsnm{{Title}}, \binits{A.M.}},
\bauthor{\bsnm{{van Ballegooijen}}, \binits{A.A.}},
\bauthor{\bsnm{{Hagenaar}}, \binits{H.J.}},
\bauthor{\bsnm{{Shine}}, \binits{R.A.}}:
\byear{1997},
\batitle{{Sustaining the Quiet Photospheric Network: The Balance of Flux
  Emergence, Fragmentation, Merging, and Cancellation}}.
\bjtitle{\apj}
\bvolume{487},
\bfpage{424}.
doi:\doiurl{10.1086/304581}.
\end{barticle}
\endbibitem

\bibitem[\protect\citeauthoryear{{Schwadron} and
  {McComas}}{2003}]{Schwadron:2003aa}
\begin{barticle}
\bauthor{\bsnm{{Schwadron}}, \binits{N.A.}},
\bauthor{\bsnm{{McComas}}, \binits{D.J.}}:
\byear{2003},
\batitle{{Solar Wind Scaling Law}}.
\bjtitle{\apj}
\bvolume{599},
\bfpage{1395}\,--\,\blpage{1403}.
doi:\doiurl{10.1086/379541}.
\end{barticle}
\endbibitem

\bibitem[\protect\citeauthoryear{{Schwadron} and
  {McComas}}{2008}]{Schwadron:2008aa}
\begin{barticle}
\bauthor{\bsnm{{Schwadron}}, \binits{N.A.}},
\bauthor{\bsnm{{McComas}}, \binits{D.J.}}:
\byear{2008},
\batitle{{The Solar Wind Power from Magnetic Flux}}.
\bjtitle{\apjl}
\bvolume{686},
\bfpage{L33}\,--\,\blpage{L36}.
doi:\doiurl{10.1086/592877}.
\end{barticle}
\endbibitem

\bibitem[\protect\citeauthoryear{{Schwenn} and {Marsch}}{1990}]{Schwenn:1990aa}
\begin{bbook}
\bauthor{\bsnm{{Schwenn}}, \binits{R.}},
\bauthor{\bsnm{{Marsch}}, \binits{E.}}:
\byear{1990},
\bbtitle{{Physics of the Inner Heliosphere I. Large-Scale Phenomena}},
\bsertitle{Physics and Chemistry in Space}
\bseriesno{20},
\bpublisher{Springer},
\blocation{Berlin}.
\end{bbook}
\endbibitem

\bibitem[\protect\citeauthoryear{{Schwenn} and {Marsch}}{1991}]{Schwenn:1991ab}
\begin{bbook}
\bauthor{\bsnm{{Schwenn}}, \binits{R.}},
\bauthor{\bsnm{{Marsch}}, \binits{E.}}:
\byear{1991},
\bbtitle{{Physics of the Inner Heliosphere II. Particles, Waves and
  Turbulence}},
\bsertitle{Physics and Chemistry in Space}
\bseriesno{21},
\bpublisher{Springer},
\blocation{Berlin}.
\end{bbook}
\endbibitem

\bibitem[\protect\citeauthoryear{{Sheeley}}{1991}]{Sheeley:1991aa}
\begin{barticle}
\bauthor{\bsnm{{Sheeley}}, \binits{N.R.} \bsuffix{Jr.}}:
\byear{1991},
\batitle{{Polar faculae - 1906-1990}}.
\bjtitle{\apj}
\bvolume{374},
\bfpage{386}\,--\,\blpage{389}.
doi:\doiurl{10.1086/170129}.
\end{barticle}
\endbibitem

\bibitem[\protect\citeauthoryear{{Smith} \textit{et~al.}}{2001}]{Smith:2001aa}
\begin{barticle}
\bauthor{\bsnm{{Smith}}, \binits{C.W.}},
\bauthor{\bsnm{{Mullan}}, \binits{D.J.}},
\bauthor{\bsnm{{Ness}}, \binits{N.F.}},
\bauthor{\bsnm{{Skoug}}, \binits{R.M.}},
\bauthor{\bsnm{{Steinberg}}, \binits{J.}}:
\byear{2001},
\batitle{{Day the solar wind almost disappeared: Magnetic field fluctuations,
  wave refraction and dissipation}}.
\bjtitle{\jgr}
\bvolume{106},
\bfpage{18625}\,--\,\blpage{18634}.
doi:\doiurl{10.1029/2001JA000022}.
\end{barticle}
\endbibitem

\bibitem[\protect\citeauthoryear{{Smith} \textit{et~al.}}{2000}]{Smith:2000aa}
\begin{barticle}
\bauthor{\bsnm{{Smith}}, \binits{E.J.}},
\bauthor{\bsnm{{Jokipii}}, \binits{J.R.}},
\bauthor{\bsnm{{K{\'o}ta}}, \binits{J.}},
\bauthor{\bsnm{{Lepping}}, \binits{R.P.}},
\bauthor{\bsnm{{Szabo}}, \binits{A.}}:
\byear{2000},
\batitle{{Evidence of a North-South Asymmetry in the Heliosphere Associated
  with a Southward Displacement of the Heliospheric Current Sheet}}.
\bjtitle{\apj}
\bvolume{533},
\bfpage{1084}\,--\,\blpage{1089}.
doi:\doiurl{10.1086/308685}.
\end{barticle}
\endbibitem

\bibitem[\protect\citeauthoryear{{Stone}}{1977}]{Stone:1977aa}
\begin{barticle}
\bauthor{\bsnm{{Stone}}, \binits{E.C.}}:
\byear{1977},
\batitle{{The Voyager Missions to the Outer System}}.
\bjtitle{\ssr}
\bvolume{21},
\bfpage{75}.
doi:\doiurl{10.1007/BF00200845}.
\end{barticle}
\endbibitem

\bibitem[\protect\citeauthoryear{{Telloni}, {Antonucci}, and
  {Dodero}}{2007}]{Telloni:2007aa}
\begin{barticle}
\bauthor{\bsnm{{Telloni}}, \binits{D.}},
\bauthor{\bsnm{{Antonucci}}, \binits{E.}},
\bauthor{\bsnm{{Dodero}}, \binits{M.A.}}:
\byear{2007},
\batitle{{Oxygen temperature anisotropy and solar wind heating above coronal
  holes out to 5\,R$_{Sun}$}}.
\bjtitle{\aap}
\bvolume{476},
\bfpage{1341}\,--\,\blpage{1346}.
doi:\doiurl{10.1051/0004-6361:20077660}.
\end{barticle}
\endbibitem

\bibitem[\protect\citeauthoryear{{Thieme}, {Marsch}, and
  {Schwenn}}{1990}]{Thieme:1990aa}
\begin{barticle}
\bauthor{\bsnm{{Thieme}}, \binits{K.M.}},
\bauthor{\bsnm{{Marsch}}, \binits{E.}},
\bauthor{\bsnm{{Schwenn}}, \binits{R.}}:
\byear{1990},
\batitle{{Spatial structures in high-speed streams as signatures of fine
  structures in coronal holes}}.
\bjtitle{Annales Geophysicae}
\bvolume{8},
\bfpage{713}\,--\,\blpage{723}.
\end{barticle}
\endbibitem

\bibitem[\protect\citeauthoryear{{Thompson}
  \textit{et~al.}}{2003}]{Thompson:2003aa}
\begin{barticle}
\bauthor{\bsnm{{Thompson}}, \binits{M.J.}},
\bauthor{\bsnm{{Christensen-Dalsgaard}}, \binits{J.}},
\bauthor{\bsnm{{Miesch}}, \binits{M.S.}},
\bauthor{\bsnm{{Toomre}}, \binits{J.}}:
\byear{2003},
\batitle{{The Internal Rotation of the Sun}}.
\bjtitle{\araa}
\bvolume{41},
\bfpage{599}\,--\,\blpage{643}.
doi:\doiurl{10.1146/annurev.astro.41.011802.094848}.
\end{barticle}
\endbibitem

\bibitem[\protect\citeauthoryear{{Tsuneta}
  \textit{et~al.}}{2008}]{Tsuneta:2008aa}
\begin{barticle}
\bauthor{\bsnm{{Tsuneta}}, \binits{S.}},
\bauthor{\bsnm{{Ichimoto}}, \binits{K.}},
\bauthor{\bsnm{{Katsukawa}}, \binits{Y.}},
\bauthor{\bsnm{{Lites}}, \binits{B.W.}},
\bauthor{\bsnm{{Matsuzaki}}, \binits{K.}},
\bauthor{\bsnm{{Nagata}}, \binits{S.}},
\bauthor{\bsnm{{Orozco Su{\'a}rez}}, \binits{D.}},
\bauthor{\bsnm{{Shimizu}}, \binits{T.}},
\bauthor{\bsnm{{Shimojo}}, \binits{M.}},
\bauthor{\bsnm{{Shine}}, \binits{R.A.}},
\bauthor{\bsnm{{Suematsu}}, \binits{Y.}},
\bauthor{\bsnm{{Suzuki}}, \binits{T.K.}},
\bauthor{\bsnm{{Tarbell}}, \binits{T.D.}},
\bauthor{\bsnm{{Title}}, \binits{A.M.}}:
\byear{2008},
\batitle{{The Magnetic Landscape of the Sun's Polar Region}}.
\bjtitle{\apj}
\bvolume{688},
\bfpage{1374}\,--\,\blpage{1381}.
doi:\doiurl{10.1086/592226}.
\end{barticle}
\endbibitem

\bibitem[\protect\citeauthoryear{{Tu} and {Marsch}}{1990}]{Tu:1990aa}
\begin{barticle}
\bauthor{\bsnm{{Tu}}, \binits{C.-Y.}},
\bauthor{\bsnm{{Marsch}}, \binits{E.}}:
\byear{1990},
\batitle{{Evidence for a 'background' spectrum of solar wind turbulence in the
  inner heliosphere}}.
\bjtitle{\jgr}
\bvolume{95},
\bfpage{4337}\,--\,\blpage{4341}.
doi:\doiurl{10.1029/JA095iA04p04337}.
\end{barticle}
\endbibitem

\bibitem[\protect\citeauthoryear{{Tu} \textit{et~al.}}{2005}]{Tu:2005aa}
\begin{barticle}
\bauthor{\bsnm{{Tu}}, \binits{C.-Y.}},
\bauthor{\bsnm{{Zhou}}, \binits{C.}},
\bauthor{\bsnm{{Marsch}}, \binits{E.}},
\bauthor{\bsnm{{Xia}}, \binits{L.-D.}},
\bauthor{\bsnm{{Zhao}}, \binits{L.}},
\bauthor{\bsnm{{Wang}}, \binits{J.-X.}},
\bauthor{\bsnm{{Wilhelm}}, \binits{K.}}:
\byear{2005},
\batitle{{Solar Wind Origin in Coronal Funnels}}.
\bjtitle{Science}
\bvolume{308},
\bfpage{519}\,--\,\blpage{523}.
doi:\doiurl{10.1126/science.1109447}.
\end{barticle}
\endbibitem

\bibitem[\protect\citeauthoryear{{Tylka} \textit{et~al.}}{2006}]{Tylka:2006aa}
\begin{barticle}
\bauthor{\bsnm{{Tylka}}, \binits{A.J.}},
\bauthor{\bsnm{{Cohen}}, \binits{C.M.S.}},
\bauthor{\bsnm{{Dietrich}}, \binits{W.F.}},
\bauthor{\bsnm{{Lee}}, \binits{M.A.}},
\bauthor{\bsnm{{Maclennan}}, \binits{C.G.}},
\bauthor{\bsnm{{Mewaldt}}, \binits{R.A.}},
\bauthor{\bsnm{{Ng}}, \binits{C.K.}},
\bauthor{\bsnm{{Reames}}, \binits{D.V.}}:
\byear{2006},
\batitle{{A Comparative Study of Ion Characteristics in the Large Gradual Solar
  Energetic Particle Events of 2002 April 21 and 2002 August 24}}.
\bjtitle{\apjs}
\bvolume{164},
\bfpage{536}\,--\,\blpage{551}.
doi:\doiurl{10.1086/503203}.
\end{barticle}
\endbibitem

\bibitem[\protect\citeauthoryear{{Van Hollebeke}, {Ma Sung}, and
  {McDonald}}{1975}]{Van-Hollebeke:1975aa}
\begin{barticle}
\bauthor{\bsnm{{Van Hollebeke}}, \binits{M.A.I.}},
\bauthor{\bsnm{{Ma Sung}}, \binits{L.S.}},
\bauthor{\bsnm{{McDonald}}, \binits{F.B.}}:
\byear{1975},
\batitle{{The variation of solar proton energy spectra and size distribution
  with heliolongitude}}.
\bjtitle{\solphys}
\bvolume{41},
\bfpage{189}\,--\,\blpage{223}.
doi:\doiurl{10.1007/BF00152967}.
\end{barticle}
\endbibitem

\bibitem[\protect\citeauthoryear{{V{\"o}gler} and
  {Sch{\"u}ssler}}{2007}]{Vogler:2007aa}
\begin{barticle}
\bauthor{\bsnm{{V{\"o}gler}}, \binits{A.}},
\bauthor{\bsnm{{Sch{\"u}ssler}}, \binits{M.}}:
\byear{2007},
\batitle{{A solar surface dynamo}}.
\bjtitle{\aap}
\bvolume{465},
\bfpage{L43}\,--\,\blpage{L46}.
doi:\doiurl{10.1051/0004-6361:20077253}.
\end{barticle}
\endbibitem

\bibitem[\protect\citeauthoryear{{von Steiger}, {Geiss}, and
  {Gloeckler}}{1997}]{von-Steiger:1997aa}
\begin{bchapter}
\bauthor{\bsnm{{von Steiger}}, \binits{R.}},
\bauthor{\bsnm{{Geiss}}, \binits{J.}},
\bauthor{\bsnm{{Gloeckler}}, \binits{G.}}:
\byear{1997},
\bctitle{{Composition of the Solar Wind}}.
In: \beditor{\bsnm{{J.~R.~Jokipii, C.~P.~Sonett, \& M.~S.~Giampapa}}} (ed.)
\bbtitle{Cosmic Winds and the Heliosphere},
\bfpage{581}.
\end{bchapter}
\endbibitem

\bibitem[\protect\citeauthoryear{{Vourlidas}
  \textit{et~al.}}{2003}]{Vourlidas:2003aa}
\begin{barticle}
\bauthor{\bsnm{{Vourlidas}}, \binits{A.}},
\bauthor{\bsnm{{Wu}}, \binits{S.T.}},
\bauthor{\bsnm{{Wang}}, \binits{A.H.}},
\bauthor{\bsnm{{Subramanian}}, \binits{P.}},
\bauthor{\bsnm{{Howard}}, \binits{R.A.}}:
\byear{2003},
\batitle{{Direct Detection of a Coronal Mass Ejection-Associated Shock in Large
  Angle and Spectrometric Coronagraph Experiment White-Light Images}}.
\bjtitle{\apj}
\bvolume{598},
\bfpage{1392}\,--\,\blpage{1402}.
doi:\doiurl{10.1086/379098}.
\end{barticle}
\endbibitem

\bibitem[\protect\citeauthoryear{{Vr{\v s}nak} and
  {Cliver}}{2008}]{Vrsnak:2008aa}
\begin{barticle}
\bauthor{\bsnm{{Vr{\v s}nak}}, \binits{B.}},
\bauthor{\bsnm{{Cliver}}, \binits{E.W.}}:
\byear{2008},
\batitle{{Origin of Coronal Shock Waves. Invited Review}}.
\bjtitle{\solphys}
\bvolume{253},
\bfpage{215}\,--\,\blpage{235}.
doi:\doiurl{10.1007/s11207-008-9241-5}.
\end{barticle}
\endbibitem

\bibitem[\protect\citeauthoryear{{Wang} and {Robbrecht}}{2011}]{Wang:2011aa}
\begin{barticle}
\bauthor{\bsnm{{Wang}}, \binits{Y.-M.}},
\bauthor{\bsnm{{Robbrecht}}, \binits{E.}}:
\byear{2011},
\batitle{{Asymmetric Sunspot Activity and the Southward Displacement of the
  Heliospheric Current Sheet}}.
\bjtitle{\apj}
\bvolume{736},
\bfpage{136}.
doi:\doiurl{10.1088/0004-637X/736/2/136}.
\end{barticle}
\endbibitem

\bibitem[\protect\citeauthoryear{{Wang} and {Sheeley}}{2006}]{Wang:2006aa}
\begin{barticle}
\bauthor{\bsnm{{Wang}}, \binits{Y.-M.}},
\bauthor{\bsnm{{Sheeley}}, \binits{N.R.} \bsuffix{Jr.}}:
\byear{2006},
\batitle{{Sources of the Solar Wind at Ulysses during 1990-2006}}.
\bjtitle{\apj}
\bvolume{653},
\bfpage{708}\,--\,\blpage{718}.
doi:\doiurl{10.1086/508929}.
\end{barticle}
\endbibitem

\bibitem[\protect\citeauthoryear{{Wang}, {Lean}, and
  {Sheeley}}{2000}]{Wang:2000aa}
\begin{barticle}
\bauthor{\bsnm{{Wang}}, \binits{Y.-M.}},
\bauthor{\bsnm{{Lean}}, \binits{J.}},
\bauthor{\bsnm{{Sheeley}}, \binits{N.R.}}:
\byear{2000},
\batitle{{The long-term variation of the Sun's open magnetic flux}}.
\bjtitle{\grl}
\bvolume{27},
\bfpage{505}\,--\,\blpage{508}.
doi:\doiurl{10.1029/1999GL010744}.
\end{barticle}
\endbibitem

\bibitem[\protect\citeauthoryear{{Wang}, {Nash}, and
  {Sheeley}}{1989}]{Wang:1989aa}
\begin{barticle}
\bauthor{\bsnm{{Wang}}, \binits{Y.-M.}},
\bauthor{\bsnm{{Nash}}, \binits{A.G.}},
\bauthor{\bsnm{{Sheeley}}, \binits{N.R.} \bsuffix{Jr.}}:
\byear{1989},
\batitle{{Evolution of the sun's polar fields during sunspot cycle 21 -
  Poleward surges and long-term behavior}}.
\bjtitle{\apj}
\bvolume{347},
\bfpage{529}\,--\,\blpage{539}.
doi:\doiurl{10.1086/168143}.
\end{barticle}
\endbibitem

\bibitem[\protect\citeauthoryear{{Wang} \textit{et~al.}}{2007}]{Wang:2007aa}
\begin{barticle}
\bauthor{\bsnm{{Wang}}, \binits{Y.-M.}},
\bauthor{\bsnm{{Biersteker}}, \binits{J.B.}},
\bauthor{\bsnm{{Sheeley}}, \binits{N.R.} \bsuffix{Jr.}},
\bauthor{\bsnm{{Koutchmy}}, \binits{S.}},
\bauthor{\bsnm{{Mouette}}, \binits{J.}},
\bauthor{\bsnm{{Druckm{\"u}ller}}, \binits{M.}}:
\byear{2007},
\batitle{{The Solar Eclipse of 2006 and the Origin of Raylike Features in the
  White-Light Corona}}.
\bjtitle{\apj}
\bvolume{660},
\bfpage{882}\,--\,\blpage{892}.
doi:\doiurl{10.1086/512480}.
\end{barticle}
\endbibitem

\bibitem[\protect\citeauthoryear{{Wenzel}
  \textit{et~al.}}{1992}]{Wenzel:1992aa}
\begin{barticle}
\bauthor{\bsnm{{Wenzel}}, \binits{K.P.}},
\bauthor{\bsnm{{Marsden}}, \binits{R.G.}},
\bauthor{\bsnm{{Page}}, \binits{D.E.}},
\bauthor{\bsnm{{Smith}}, \binits{E.J.}}:
\byear{1992},
\batitle{{The Ulysses Mission}}.
\bjtitle{\aaps}
\bvolume{92},
\bfpage{207}.
\end{barticle}
\endbibitem

\bibitem[\protect\citeauthoryear{{Zhang} and {Dere}}{2006}]{Zhang:2006aa}
\begin{barticle}
\bauthor{\bsnm{{Zhang}}, \binits{J.}},
\bauthor{\bsnm{{Dere}}, \binits{K.P.}}:
\byear{2006},
\batitle{{A Statistical Study of Main and Residual Accelerations of Coronal
  Mass Ejections}}.
\bjtitle{\apj}
\bvolume{649},
\bfpage{1100}\,--\,\blpage{1109}.
doi:\doiurl{10.1086/506903}.
\end{barticle}
\endbibitem

\bibitem[\protect\citeauthoryear{{Zirin}}{1987}]{Zirin:1987aa}
\begin{barticle}
\bauthor{\bsnm{{Zirin}}, \binits{H.}}:
\byear{1987},
\batitle{{Weak solar fields and their connection to the solar cycle}}.
\bjtitle{\solphys}
\bvolume{110},
\bfpage{101}\,--\,\blpage{107}.
doi:\doiurl{10.1007/BF00148205}.
\end{barticle}
\endbibitem

\end{thebibliography}

\end{article} 

\end{document}